\documentclass[12pt]{article}

\usepackage[dvipdfmx]{graphicx}
\usepackage{bmpsize}
\usepackage{bbding}
\usepackage{multirow}

\usepackage{newtxtext}
\usepackage{tikz-feynman}
\usepackage[utf8]{inputenc}

\usepackage[square,comma,numbers,sort&compress]{natbib}
\usepackage[left=20mm,right=20mm,top=20mm,bottom=20mm]{geometry}
\usepackage{ascmac}
\usepackage{bm}

\usepackage{epsf}
\usepackage{xcolor}
\usepackage{amsmath}
\usepackage{amssymb}
\usepackage{mathrsfs}
\usepackage{latexsym}
\usepackage{braket}
\usepackage{slashed}
\usepackage{float}
\usepackage{cases}
\usepackage{bm}
\usepackage{arydshln}
\usepackage{hyperref}
\hypersetup{ 
 setpagesize=false,
 bookmarksnumbered=true,%
 bookmarksopen=true,%
 colorlinks=true,%
 linkcolor=blue,
 citecolor=red,
}
\usepackage{comment}
\usepackage[utf8]{inputenc}
\usepackage{braket}
\usepackage{multicol}
\usepackage{lipsum}

\usepackage{ulem}
\usepackage{fancybox}
 \usepackage{cancel}

\newcommand{\vep}{\varepsilon}

\newcommand{\A}{\mathcal{A}}

\newcommand{\F}{\mathcal{F}}

\renewcommand{\L}{\mathcal{L}}

\newcommand{\al}[1]{\begin{align}#1\end{align}}
\newcommand{\als}[1]{\begin{align*}#1\end{align*}}
\newcommand{\bp}{\begin{pmatrix}}
\newcommand{\ep}{\end{pmatrix}}
\newcommand{\bb}{\begin{bmatrix}}
\newcommand{\eb}{\end{bmatrix}}

\newcommand{\paren}[1]{\left(#1\right)}
\newcommand{\sqbr}[1]{\left[#1\right]}
\newcommand{\ab}[1]{\left|#1\right|}
\newcommand{\fn}[1]{\!\left(#1\right)}
\newcommand{\br}[1]{\left\{#1\right\}}
\newcommand{\tx}[1]{\text{#1}}

\newcommand{\wh}{\widehat}

\newcommand{\ol}[1]{\overline{#1}}

\newcommand{\ov}{\over}
\newcommand{\dg}{\dagger}

\newcommand{\cred}[1]{{\color{black}#1}}

\usepackage{ulem}

\newcommand{\pal}{\partial}

\newcommand{\fulltoday}{\number\day\space \ifcase\month\or
    January\or February\or March\or April\or May\or June\or
    July\or August\or September\or October\or November\or December\fi
    \space\number\year}

 \makeatletter
    
    \@addtoreset{equation}{section}
  \makeatother

\allowdisplaybreaks[3]

\usepackage{calc}
\newcounter{hours}\newcounter{minutes}
\makeatletter                   
\renewcommand*{\thehours}{\two@digits\c@hours}
\renewcommand*{\theminutes}{\two@digits\c@minutes}
\makeatother

\begin{document}

\allowdisplaybreaks[2]
\renewcommand{\thefootnote}{*}

\newlength{\mylength}

\title{
%
$Z_4$ scotogenic model with a Higgs portal
}

\author{
Noel Jonathan Jobu\thanks{E-mail: \tt nj702@snu.edu.in}, \mbox{} 
Kenji Nishiwaki\thanks{E-mail: \tt kenji.nishiwaki@snu.edu.in}
\bigskip\\
\it\normalsize
Department of Physics, School of Natural Sciences, \\ 
\it\normalsize
Shiv Nadar Institution of Eminence (Deemed to be University), \\
\it\normalsize
Tehsil Dadri, Gautam Buddha Nagar, Uttar Pradesh,
201314, India
}
\maketitle
\begin{abstract}
\noindent
We propose a new Scotogenic type model based on a global $Z_4$ symmetry involving dark matter candidates.
After the symmetry breaking as $Z_4$ to $Z_2$ via the singlet scalar vacuum expectation value~(VEV), the lightest Majorana fermion works as a viable thermal freeze-out dark matter~(DM) candidate, and the mass terms for active neutrinos are generated as a finite quantum correction at the 1-loop level.
A key point of realising our Scotogenic structure is to introduce two types of Majorana fermions (heavy right-handed neutrinos) and inert Higgs doublets with opposite $Z_4$ parities. 
Since a large VEV for the singlet scalar is not so harmful in an appropriate realisation of the Higgs mechanism for the SM gauge symmetry, we can naturally realise a TeV-scale fermionic DM candidate, where constraints via direct detection experiments are less than those for sub-TeV DM.
Our scenario involves the Higgs-portal DM interactions, which help the realisation of the correct DM relic abundance.
Relying on the structure of the model, it is possible to find a natural partner for coannihilation.
Our scenario can be investigated via the measurement of the Higgs trilinear self-coupling at the Large Hadron Collider.
The simplest way to evade the domain-wall problem by adding a tiny soft $Z_2$ breaking term works, keeping a sufficient longevity of the decaying DM lifetime. 
\end{abstract}

\newpage

\renewcommand\thefootnote{\arabic{footnote}}
\setcounter{footnote}{0}


\section{Introduction}

It is well known that the original form of the Standard Model~(SM) should be upgraded, at least in the sector associated with neutrinos, due to the observations of neutrino oscillations, which originate from minuscule but nonzero masses of the neutrinos in the SM.
Over the last two decades or more after the confirmation of the neutrino oscillations~\cite{Super-Kamiokande:1998kpq,SNO:2002tuh} (refer to Review~\cite{Giganti:2017fhf}), the experimental developments in neutrino oscillation have been significant,
and now {provide} relevant constraints on the CP-violating phase of the Pontecorvo-Maki-Nakagawa-Sakata~(PMNS) matrix~\cite{Pontecorvo:1957cp,Pontecorvo:1957qd,Maki:1962mu}.
The measurements of the three mixing angles and the CP phase will be significantly upgraded in the coming next-generation experiments,
e.g., DUNE~\cite{DUNE:2020jqi} and Hyper-Kamiokande~\cite{Hyper-Kamiokande:2016srs}.
Given this situation, it is becoming increasingly important to consider the structure of the neutrino sector in depth.
Furthermore, it should be noted that the neutrino sector significantly impacts the possible scenarios in the early stages of the universe in conjunction with dark matter and other matters.

A key question is: What is the origin of the tiny masses of the neutrinos?
The simplest scenario relies on the Yukawa terms with the Higgs doublet, the doublet lepton fields and the newly added
right-handed singlet neutrinos, where the Dirac mass terms for the neutrinos are successfully generated via the Higgs mechanism,
but we should swallow to introduce tiny Yukawa couplings, at least ${\cal O}\fn{10^{-12}}$ or {smaller}.
A relaxation in the tuning of Yukawa couplings is realised by adopting the see-saw mechanism, where the Yukawa structure of the sub-eV neutrino masses (and also mixings) involves connections with other heavy particles (see Review~\cite{Xing:2020ijf}).
Depending on how the left-handed active neutrinos couple with heavy particles, see-saw scenarios are classified as
Canonical/Type-I~\cite{Minkowski:1977sc,Yanagida:1979as,Gell-Mann:1979vob,Glashow:1979nm,Mohapatra:1979ia,Ramond:1979py},
Type-I\!I~\cite{Schechter:1980gr,Lazarides:1980nt,Mohapatra:1980yp,Wetterich:1981bx,Schechter:1981cv},
Type-I\!I\!I~\cite{Foot:1988aq},
Inverse~\cite{Mohapatra:1986bd,Wyler:1982dd},
Linear~\cite{Wyler:1982dd,Akhmedov:1995ip,Akhmedov:1995vm}, and others.
Note that a seminal application of the right-handed neutrinos, which can appear in the see-saw texture, is the Baryogenesis via Leptogenesis~\cite{Fukugita:1986hr}.

There are interesting variations of the see-saw scenario, suggested by the quantum nature of neutrinos.
The above scenarios considered the case where neutrino mass terms exist at the lowest order of perturbation,
but it is also possible that these are produced quantumly via loops of heavy new particles.
As suggested by the results of the operator analysis~\cite{Weinberg:1979sa,Bonnet:2012kz},
if active neutrino mass terms are represented as higher-dimensional operators of the SM fields in a gauge-invariant way
(before the electroweak symmetry breaking), no divergences emerge in such loop-induced neutrino mass terms.
We list early-stage concrete models based on this idea~\cite{Zee:1980ai,Cheng:1980qt,Zee:1985id,Babu:1988ki,Ma:1998dn,Krauss:2002px}
(see also Review on this kind of models~\cite{Cai:2017jrq}).
A branch of such scenarios called the {\it Scotogenic} model is a fascinating arena in particle physics, where
additional symmetry/symmetries (imposed on such a model) ensure the loop origin of the active neutrino mass terms but also the stability of a new particle, which behaves as a dark matter~(DM)~\cite{Ma:2006km}.
In this type of scenario, neutrino physics is mutually connected with dark matter physics, and the model predictability will be enhanced.
Lots of studies have been done on such ideas in many contexts based on various kinds of symmetries;
{We can find various examples}
based on global $Z_2$ symmetry~\cite{Ma:2008cu,Farzan:2009ji,Chen:2009gd,Farzan:2010mr,Aoki:2011yk,Parida:2011wh,Cai:2011qr,Chao:2012sz,Farzan:2012sa,Okada:2012np,Hehn:2012kz,Kajiyama:2013zla,Hirsch:2013ola,Ma:2013nga,Brdar:2013iea,Law:2013saa,Okada:2014qsa,Patra:2014sua,Fraser:2014yha,Okada:2014nea,Baek:2015mna,Chowdhury:2015sla,Diaz:2016udz,Ferreira:2016sbb,Ahriche:2016cio,vonderPahlen:2016cbw,Lu:2016dbc,Merle:2016scw,Rocha-Moran:2016enp,Nomura:2016dnf,Cheung:2016frv,Chowdhury:2016mtl,Cheung:2017efc,Lee:2017ekw,Fortes:2017ndr,Tang:2017rhv,Guo:2018iix,Rojas:2018wym,Aranda:2018lif,Han:2019lux,Suematsu:2019kst,Pramanick:2019oxb,Restrepo:2019ilz,Mandal:2019oth,Avila:2019hhv,Escribano:2020iqq,Nomura:2020dzw,Beniwal:2020hjc,DeRomeri:2021yjo,De:2021crr,Kang:2021jmi,Sarazin:2021nwo,Nagao:2022oin,Ahriche:2022bpx,Cepedello:2022xgb,Chun:2023vbh,Escribano:2023hxj,Borah:2023hqw,Garbrecht:2024bbo,Nomura:2024zca,Cardenas:2024ojd,Singh:2025jtn,CarcamoHernandez:2025eyt,Escribano:2025kys,AbuSiam:2025voc},
based on other kinds of global discrete symmetries~\cite{Ma:2008ym,Adulpravitchai:2009re,Ma:2012ez,Bhattacharya:2013mpa,Kajiyama:2013rla,Ma:2013xqa,Ma:2014eka,Okada:2015bxa,Baek:2017qos,Ahriche:2020pwq,Chen:2020ark,deAnda:2021jzc,ChuliaCentelles:2022ogm,Barreiros:2022aqu,Bonilla:2023pna,Kumar:2023moh,Ganguly:2023jml,Arora:2024mhg,Kumar:2024zfb,Kim:2024cwp,delaVega:2024tuu,CarcamoHernandez:2024ycd,Luong:2025pjj},
and other embedding varieties, including those based on continuous-symmetry cases~\cite{Kubo:2006rm,Chang:2011kv,Kajiyama:2013lja,Ma:2013yga,AristizabalSierra:2014irc,Hatanaka:2014tba,Nishiwaki:2015iqa,Okada:2015vwh,Kanemura:2015bli,Yu:2016lof,Nomura:2016seu,Wang:2017mcy,Nomura:2017vzp,Nomura:2017tzj,Ma:2017zyb,Nomura:2017psk,Hagedorn:2018spx,Han:2018zcn,Calle:2018ovc,Carvajal:2018ohk,CentellesChulia:2019gic,Ma:2019yfo,Kang:2019sab,Nomura:2019jxj,Jana:2019mez,Ma:2019iwj,Nomura:2019yft,Fuentes-Martin:2019bue,Okada:2019xqk,Bonilla:2019ipe,Han:2019diw,Nomura:2019lnr,Leite:2019grf,Jana:2019mgj,Wang:2019byi,delaVega:2020jcp,Okada:2020dmb,Kim:2020aua,Wong:2020obo,Behera:2020lpd,Bernal:2021ezl,Okada:2021nwo,Nomura:2021aep,Escribano:2021ymx,Dasgupta:2021ggp,Berbig:2022nre,Portillo-Sanchez:2023kbz,deBoer:2023phz,Nomura:2023vmh,Leite:2023gzl,Nomura:2024jxc,VanLoi:2024ptt,Nomura:2024vzw,CentellesChulia:2024iom,Garnica:2024wur,Pathak:2024sei,Agudelo:2024luc,Babu:2024jdw,Gola:2024tfx,Batra:2025gzy,Borboruah:2025bwx,Dorsner:2025rzj,Kumar:2025cte,Nomura:2025raf,Ma:2025ymy,Leite:2025yuq}.\footnote{
Scenarios with bound-state dark matter are discussed in~\cite{Reig:2018mdk,Reig:2018ztc}.
Another fascinating class (without DM candidates) is loop-induced neutrino masses under extra coloured scalars; see e.g.,~\cite{AristizabalSierra:2007nf,FileviezPerez:2009ud,Kohda:2012sr} as early works.
{
A recent related development is neutrino mass generations from generalised symmetry breaking and under non-invertible selection rules, see e.g.,~\cite{Cordova:2022fhg,Kobayashi:2025cwx,Nomura:2025yoa} (also refer to~\cite{Suzuki:2025bxg}).
Another recent direction is to investigate precise calculations of how to probe a scotogenic model through flavours and electroweak precision measurements, see e.g.,~\cite{Darricau:2025vcs}.
\cred{See also~\cite{Cacciapaglia:2020psm} as a scotogenic scenario in the so-called composite Higgs framework, alleviating the electroweak hierarchy problem.}
}
}

However, it should be noted here that the ordinary Weakly Interacting Massive Particle~(WIMP) DM paradigm based on the thermal freeze-out scenario has recently reached a crossroads due to significant progress with null observations in the LUX-ZEPLIN~(LZ) direct observation experiment~\cite{LZ:2022lsv,LZ:2024zvo}.
Several ideas to evade the tight bounds in the freeze-out DM scenario \cred{include}, e.g.,
fermionic DM with pseudoscalar mediator~\cite{Freytsis:2010ne,Ipek:2014gua,Arcadi:2017wqi,Bell:2018zra,Abe:2018emu,Abe:2019wjw}, and pseudo-Nambu-Goldstone boson~(pNGB) DM~\cite{Barger:2010yn,Barducci:2016fue,Gross:2017dan,Balkin:2017aep,Ishiwata:2018sdi} (see also~\cite{Abe:2021byq,Okada:2021qmi,Chiang:2023omu} for embeddings to grand-unified scenarios).
{We can find other scenarios, for example, two-component dark matter (see e.g.,~\cite{Bhattacharya:2024ohh}) and freeze-in production (see e.g.,~\cite{Wang:2024qhe}).}

Another simple scenario is to rely on the enhancement of the DM pair annihilation near a resonance of a mediator in an $s$-channel,
where we can keep the coupling between DM and its mediator sizably small, and thereby, it is not difficult to suppress the (spin-independent) cross section in DM direct detections.
A favoured region of the DM-mass range would be a TeV region since more severe bounds have been imposed in the sub-TeV domain~\cite{LZ:2022lsv,LZ:2024zvo}.
We propose a new Scotogenic scenario based on the global $Z_4$ symmetry, where the vacuum expectation value~(VEV) of an additional singlet scalar breaks it to $Z_2$. The residual $Z_2$ symmetry ensures the stability of DM, and a large VEV of the singlet scalar naturally realises a TeV-scale Majorana fermionic DM.

This paper is organised as follows.
In Section~\ref{sec:Model}, we introduce our Scotogenic model based on a $Z_4$ global symmetry and derive the form of the one-loop induced neutrino mass terms.
In Section~\ref{sec:Non-DM}, we provide constraints on our scenario, namely, the lepton flavour violation, the oblique correction and the stability of the scalar potential. 
In Section~\ref{sec:DM}, we summarise our formulas for calculating the dark matter relic abundance and direct detection.
In Section~\ref{sec:Analysis}, we provide our results of numerical calculations and show that we can reproduce the observed neutrino texture at one-loop level and the observed dark matter relic abundance consistently.
In Section~\ref{sec:trilinear-Higgs}, we discuss the current constraint and future prospects of the constraint on our scenario via the measurement of the Higgs trilinear self-coupling at the {Large Hadron Collider~(LHC)}.
In Section~\ref{sec:domain-wall}, we comment on possible prescriptions to evade the domain-wall problem in our scenario.
In Section~\ref{sec:Summary}, we summarise our results and provide some comments.
In Appendix~\ref{sec:decay-s}, the decay width of the $SU(2)_\tx{L}$-singlet scalar is provided.
In Appendix~\ref{sec:DM-result-with-no-coannihilation}, the results of the numerical scans without the coannihilation effect are shown.

\section{A $Z_4$-based Scotogenic Model
\label{sec:Model}}

\subsection{Model Setup \label{sec:Model-Setup}}

\begin{table}[t]
\centering
\begin{tabular}{| c || c | c | c | c | c |}
\hline
New Particle	& $SU(3)_\tx{C}$ & $SU(2)_\tx{L}$ & $U(1)_\tx{Y}$ & $Z_4$ & Spin \\
\hline\hline
$S$            &   $1$ & $1$ & $0$ & $-1$ & $0$ (Real) \\ \hline
$\eta_A$    &   $1$ & $2$ & $1$ & $+i$ & $0$ (Complex) \\ \hline
$\eta_B$    &   $1$ & $2$ & $1$ & $-i$ & $0$ (Complex) \\ \hline
$N_{A_i R}\,(i=1,2,3)$  &   $1$ & $1$ & $0$ & $+i$ & $1/2$ (Right-handed) \\ \hline
$N_{B_i R}\,(i=1,2,3)$  &   $1$ & $1$ & $0$ & $-i$ & $1/2$ (Right-handed) \\ \hline
\end{tabular}
\caption{
Matter contents added to this model are summarised.
$SU(3)_\tx{C}$, $SU(2)_\tx{L}$ and $U(1)_\tx{Y}$ represent the degrees of freedom of the colour, weak-isospin and hypercharge, respectively.
All Standard Model fields take $+1$ of the $Z_4$ parity.
The index $i$ takes one to three and discriminates three generations.
}
\label{tab:matter-contents}
\end{table}

Here, we introduce our model, which is a Scotogenic-type model based on a discrete $Z_4$ symmetry,
where no extension of the gauge groups from those of the Standard Model.
Table~\ref{tab:matter-contents} shows a summary of the matter contents we introduced,
where we introduce one real $SU(2)_\tx{L}$-singlet scalar $S$, two complex $SU(2)_\tx{L}$-doublet scalars $\eta_A$ and $\eta_B$,
and six (two kinds of three-generation) $SU(2)_\tx{L}$-singlet right-handed neutrinos $N_{A_i R}$ and $N_{B_i R}$ $(i=1,2,3)$.
Note that for the scalar doublets and right-handed fermions, type-$A$ and type-$B$ particles take opposite $Z_4$ parities as $+i$ and $-i$, respectively. The trivial $Z_4$ parity ($+1$) is assigned for all of the SM particles.

In our scenario, the possible leptonic Yukawa terms are as follows,
\al{
-{\cal L}_\tx{Yukawa}
	&=
		Y^{i\alpha}_{A} \ol{N'_{A_i R}} \, \eta_A^{\tx{T}} \paren{i\sigma_2} L_{\alpha L} +
		Y^{i\alpha}_{B} \ol{N'_{B_i R}} \, \eta_B^{\tx{T}} \paren{i\sigma_2} L_{\alpha L} \notag \\
	&\quad
		+ {m_{N_{ij}} \ov 2} \ol{\paren{N'_{A_i R}}^\tx{c}} N'_{B_j R}
		+ {m_{N_{ij}} \ov 2} \ol{\paren{N'_{B_i R}}^\tx{c}} N'_{A_j R}
		+ {y_{A_{ij}} \ov 2} S \ol{\paren{N'_{A_i R}}^\tx{c}} N'_{A_j R}
		+ {y_{B_{ij}} \ov 2} S \ol{\paren{N'_{B_i R}}^\tx{c}} N'_{B_j R} + \tx{h.c.},
	\label{eq:Yukawa-terms}
}
and the possible scalar potential terms are written down as
\al{
{\cal V}
	&=	+ m_H^2 \paren{H^\dagger H}
		+ m_{\eta_A}^2 \paren{\eta_A^\dg \eta_A} + m_{\eta_B}^2 \paren{\eta_B^\dg \eta_B} + {m_S^2 \ov 2} S^2
		+ {\lambda_{HS} \ov 2} S^2 \paren{H^\dagger H} \notag \\
	&\quad
		+ {\lambda_{H} \ov 2} \paren{H^\dagger H}^2 + {\lambda_S \ov 4} S^4
		+ {\lambda_{2a} \ov 2} \paren{\eta_A^\dagger \eta_A}^2
		+ {\lambda_{2b} \ov 2} \paren{\eta_B^\dagger \eta_B}^2
		+ {\lambda_{2c} \ov 2} \paren{\eta_A^\dagger \eta_A}\paren{\eta_B^\dagger \eta_B} \notag \\
	&\quad
		+ {\lambda_{2d} \ov 2} \paren{\eta_A^\dagger \eta_B}\paren{\eta_B^\dagger \eta_A}
		+ \lambda_{3A} \paren{H^\dagger H} \paren{\eta_A^\dagger \eta_A}
		+ \lambda_{3B} \paren{H^\dagger H} \paren{\eta_B^\dagger \eta_B} \notag \\
	&\quad
		+ \lambda_{4A} \paren{H^\dagger \eta_A} \paren{\eta_A^\dagger H}
		+ \lambda_{4B} \paren{H^\dagger \eta_B} \paren{\eta_B^\dagger H}
		+ {1\ov 2} \sqbr{\lambda_5 \paren{H^\dagger \eta_A}\paren{H^\dagger \eta_B} 
			+ \lambda_6 \paren{\eta_A^\dagger \eta_B}^2 + \tx{h.c.} } \notag \\
	&\quad
		+ \mu_S \, S \paren{\eta^\dagger_A \eta_B} +\mu_S^\ast\, S \paren{\eta^\dagger_B \eta_A}
		+ {\lambda_{S\eta_A} \ov 2} S^2 \paren{\eta^\dagger_A \eta_A}
		+ {\lambda_{S\eta_B} \ov 2} S^2 \paren{\eta^\dagger_B \eta_B},
	\label{eq:potential}
}
where $H$ denotes the $SU(2)_\tx{L}$ Higgs doublet to trigger the spontaneous electroweak symmetry breaking;
$i$ and $\alpha$ run from one to three, representing the three generations of the right-handed neutrinos and left-handed Lepton doublets $L_{\alpha L}$, respectively.
{$\sigma_2$ is the 2nd Pauli matrix.}

In the following analysis, we will assume that  $\mu_S$ is a real parameter and 
$m_{N_{ij}}$, $y_{A_{ij}}$ and $y_{B_{ij}}$ are real and diagonal for simplicity.
Also, we will take $\lambda_6 = 0$ {to remove unnecessary complexity from the discussion without compromising the essence of the model, as the term does not contribute to mass terms under the inert condition for $\eta_A$ and $\eta_B$.}
The fermion with the superscript $\paren{N_R}^\tx{c}$ denotes the charge conjugation of $N_R$.

The $SU(2)_\tx{L}$ doublets are parametrised as follows,\footnote{
Note that
\als{
\ol{\paren{N'_{A_i R}}^\tx{c}} N'_{B_j R} 
	&=
		\ol{\paren{N'_{B_j R}}^\tx{c}} N'_{A_i R}. \\
Y^{i\alpha}_{A} \ol{N'_{A_i R}} \, \eta_A^{\tx{T}} \paren{i\sigma_2} L_{\alpha L} + \tx{h.c.}
	&=
	Y^{i\alpha}_{A} \ol{N'_{A_i R}} \sqbr{ \nu'_{\alpha L} {\eta'}^0_A - {l'}^-_{\alpha L} {\eta'}^+_A } +
	\paren{Y^{i\alpha}_{A}}^\ast  \sqbr{ \ol{\nu'_{\alpha L}} \paren{{\eta'}^0_A}^\ast - \ol{{l'}^-_{\alpha L}} {\eta'}^-_A } {N'_{A_i R}}.
}
}
\al{
L_{\alpha L}
	&=
		\bp \nu'_\alpha \\ {l'}^-_\alpha \ep_L,&
H
	&=
		\bp 0 \\ {1 \ov \sqrt{2}} \paren{v_H + h'} \ep,&
\eta_{A,B}
	&=
		\bp \eta'^+_{A,B} \\ \eta'^0_{A,B} \ep
		=
		\bp \eta'^+_{A,B} \\ {1 \ov \sqrt{2}} \paren{\eta'_{R_{A,B}} + i \eta'_{I_{A,B}} } \ep,&
	\label{eq:multiplet-components}
}
and the singlet scalar,
\al{
	S =\paren{v_S + s'},
	\label{eq:singlet-scalar-components}
}
where we took the unitary gauge and the prime symbol for components, representing that they are gauge eigenstates.
We focus on the following configuration of the VEVs of the scalars,
\al{
\braket{H}
	&=
		\bp 0 \\  {v_H/\sqrt{2}} \ep,&
\braket{S}
	&=
		v_S,&
	\label{}
\braket{\eta_{A}}
	&=
		\bp 0 \\ 0 \ep,&
\braket{\eta_{B}}
	&=
		\bp 0 \\ 0 \ep,&
	\label{eq:scalar-VEV-configuration}
}
where the $Z_4$ symmetry is spontaneously broken down to $Z_2$ {due to the nonzero VEV of $S$ (around the electroweak scale)},
and therefore, the lightest particle with $+i$ or $-i$ $Z_4$ parity is protected by a residual symmetry and becomes a dark matter candidate.
The inert condition for $\eta_A$ and $\eta_B$ is necessary to keep the residual $Z_2$ symmetry.
{Note that the situation in Eq.~\eqref{eq:scalar-VEV-configuration} with nonzero VEVs $v_H$ and $v_S$ are valid after the spontaneous electroweak symmetry breaking.}

The tadpole conditions,
\al{
{\pal {\cal V} \ov \pal h}\Big|_{h=s=0} = 0 = {\pal {\cal V} \ov \pal s}\Big|_{h=s=0},
}
lead to the forms
\al{
{\lambda_H \ov 2} v_H^2 + {\lambda_{HS} \ov 2} v_S^2 &= - m_H^2,&
{\lambda_{HS}\ov 2} v_H^2 + {\lambda_{S}} v_S^2 &= - m_S^2,&
	\label{eq:tadpole-conditions}
}
which are solved as
\al{
v_H^2
	&=
		{ 2 \sqbr{ 2\paren{-m_H^2}\lambda_S - \paren{-m_S^2} \lambda_{HS} } \ov {2\lambda_H\lambda_S - \lambda_{HS}^2} },&
v_S^2
	&=
		{ 2 \sqbr{ \paren{-m_S^2}\lambda_H - \paren{-m_H^2} \lambda_{HS} } \ov {2\lambda_H\lambda_S - \lambda_{HS}^2} },&
}
where the Higgs VEV {$v_H$} should take $\simeq 246\,\tx{GeV}$.
After the application of the tadpole conditions in Eq.~\eqref{eq:tadpole-conditions},
the mass matrix squared for $h'$ and $s'$ in this order reads
\al{
{\cal M}^2\!\sqbr{h',s'}
	=
		\bb  \lambda_H v_H^2 & \lambda_{HS} v_H v_S \\ \lambda_{HS} v_H v_S & 2\lambda_S v_S^2 \eb
		=
		\bb c_\alpha &- s_\alpha \\ s_\alpha & c_\alpha \eb
		\bb  M_{h}^2 & 0 \\ 0 & M_{s}^2 \eb
		\bb c_\alpha & s_\alpha \\ -s_\alpha & c_\alpha \eb,
}
where $h$ is the SM-like Higgs boson with $M_{h} = 125\,\tx{GeV}$ and $s$ an additional $\tx{CP}$-even scalar boson.
The mixing angle $\alpha$ is determined as
\al{\label{eq:higgsmix}
\sin\fn{2\alpha}
	=
		{-}{2 \lambda_{HS} v_H v_S \ov M_{h}^2 - M_{s}^2},
}
where the gauge eigenstates and the mass eigenstates (denoted without the prime symbol) are related as
\al{
\bb h' \\ s' \eb
	=
		\bb c_\alpha & -s_\alpha \\ s_\alpha & c_\alpha \eb
		\bb {h} \\ {s} \eb.
}

Next, we see the mass structure of the other scalars.
Under our assumption that $\mu_S$ is real,
the mass matrices squared for $\eta'_{R_A}$ and $\eta'_{R_B}$, and $\eta'_{I_A}$ and $\eta'_{I_B}$ yield
\al{
{\cal M}^2 \!\sqbr{\eta'_{R_A}, \eta'_{R_B}}
	&=
		\bb
		m_{R_{AA}}^2 & m_{R_{AB}}^2 \\
		m_{R_{AB}}^2 & m_{R_{BB}}^2
		\eb
		=
		U_{R}
		\bb M_{R_1}^2 & 0 \\ 0 & M_{R_2}^2  \eb
		U_{R}^\dagger, \\
{\cal M}^2 \!\sqbr{\eta'_{I_A}, \eta'_{I_B}}
	&=
		\bb
		m_{I_{AA}}^2 & m_{I_{AB}}^2 \\
		m_{I_{AB}}^2 & m_{I_{BB}}^2
		\eb
		=
		U_{I}
		\bb M_{I_1}^2 & 0 \\ 0 & M_{I_2}^2  \eb
		U_{I}^\dagger,
}
where the components are defined as
\al{\label{eq:neutralmass}
m_{R_{AA}}^2
	&=
		m_{I_{AA}}^2
		=
		m_{\eta_A}^2 + {\lambda_{3A} \ov 2} v_H^2 + {\lambda_{4A} \ov 2} v_H^2 + {\lambda_{S\eta_A} \ov 2} v_S^2,  \\
m_{R_{BB}}^2
	&=
		m_{I_{BB}}^2
		=
		m_{\eta_B}^2 + {\lambda_{3B} \ov 2} v_H^2 + {\lambda_{4B} \ov 2} v_H^2 + {\lambda_{S\eta_B} \ov 2} v_S^2, \\
\label{eq:mR_AB}
m_{R_{AB}}^2
	&=
		\mu_S v_S + {\lambda_5 \ov 2} v_H^2, \\
\label{eq:mI_AB}
m_{I_{AB}}^2
	&=
		\mu_S v_S - {\lambda_5 \ov 2} v_H^2.
}
The matrices diagonalise the mass matrices,
\al{
U_R
	&=
		\bb c_{\theta_R} & -s_{\theta_R} \\ s_{\theta_R} & c_{\theta_R} \eb,&
U_I
	&=
		\bb c_{\theta_I} & -s_{\theta_I} \\ s_{\theta_I} & c_{\theta_I} \eb;&
\bb \eta'_{R_A} \\ \eta'_{R_B} \eb
&=
U_R
\bb {\eta}_{R_1} \\ {\eta}_{R_2} \eb,&
\bb \eta'_{I_A} \\ \eta'_{I_B} \eb
&=
U_I
\bb {\eta}_{I_1} \\ {\eta}_{I_2} \eb,& \\
\tan\fn{2\theta_R}
	&=
		{ 2 m_{R_{AB}}^2 \ov m_{R_{AA}}^2 - m_{R_{BB}}^2 },&
\tan\fn{2\theta_I}
	&=
		{ 2 m_{I_{AB}}^2 \ov m_{I_{AA}}^2 - m_{I_{BB}}^2 },&
}
where their mass eigenvalues are given as
\al{
M_{R_1}^2
	&=
		{1\ov 2} \sqbr{ m_{R_{AA}}^2 + m_{R_{BB}}^2 + \sqrt{ 4 m_{R_{AB}}^4 + \paren{ m_{R_{AA}}^2 - m_{R_{BB}}^2 }^2 } },
		\label{eq:Metaneutral-first} \\
M_{R_2}^2
	&=
		{1\ov 2} \sqbr{ m_{R_{AA}}^2 + m_{R_{BB}}^2 - \sqrt{ 4 m_{R_{AB}}^4 + \paren{ m_{R_{AA}}^2 - m_{R_{BB}}^2 }^2 } }, \\
M_{I_1}^2
	&=
		{1\ov 2} \sqbr{ m_{I_{AA}}^2 + m_{I_{BB}}^2 + \sqrt{ 4 m_{I_{AB}}^4 + \paren{ m_{I_{AA}}^2 - m_{I_{BB}}^2 }^2 } }, \\
M_{I_2}^2
	&=
		{1\ov 2} \sqbr{ m_{I_{AA}}^2 + m_{I_{BB}}^2 - \sqrt{ 4 m_{I_{AB}}^4 + \paren{ m_{I_{AA}}^2 - m_{I_{BB}}^2 }^2 } }.
		\label{eq:Metaneutral-last}
}
{Note that the expressions from Eq.~\eqref{eq:Metaneutral-first} to Eq.~\eqref{eq:Metaneutral-last} hold at the tree-level;
This also applies to the following Eqs.~\eqref{eq:Metacharged-1}, \eqref{eq:Metacharged-2}, \eqref{eq:MNAi}, and \eqref{eq:MNBi}.}
Similarly, the mass matrix for the charged scalars yields
\al{
\label{eq:CSmass}
{\cal M}^2 \!\sqbr{ \eta'^+_A, \eta'^+_B }
	=
		\bb
		m_{\eta_A^+}^2  & m_{\eta_{AB}^+}^2  \\
		m_{\eta_{AB}^+}^2  & m_{\eta_B^+}^2 
		\eb,
}
with
\al{
m_{\eta_A^+}^2 
	&= 
		m_{\eta_A}^2 + {\lambda_{3A} \ov 2} v_H^2 + {\lambda_{S\eta_A} \ov 2} v_S^2, \\
m_{\eta_B^+}^2 
	&= 
		m_{\eta_B}^2 + {\lambda_{3B} \ov 2} v_H^2 + {\lambda_{S\eta_B} \ov 2} v_S^2, \\
m_{\eta_{AB}^+}^2 
	&= 
		{\mu_S v_S},
}
\al{
 {\cal M}^2 \!\sqbr{ \eta'^+_A, \eta'^+_B }
&=
{U_\tx{cs}}
\bb 
M_{\eta^+_1}^2 & 0 \\ 0 &  M_{\eta^+_2}^2
\eb
{U^\dagger_\tx{cs}},&
\bb \eta'^+_A \\ \eta'^+_B \eb
&=
U_\tx{cs}
\bb {\eta}^+_1 \\ {\eta}^+_2 \eb,&
U_\tx{cs}
&=
\bb
c_{\theta_\tx{cs}} & -s_{\theta_\tx{cs}} \\
s_{\theta_\tx{cs}} & c_{\theta_\tx{cs}}
\eb.&
}
Deriving the formulas for mass eigenvalues and $\tan\fn{2 \theta_\tx{cs}}$ is straightforward as
\al{
M_{\eta^+_1}^2
	&=
		{1\ov 2} \sqbr{m_{\eta_A^+}^2 + m_{\eta_B^+}^2 + \sqrt{ 4 m_{\eta_{AB}^+}^2 + \paren{ m_{\eta_A^+}^2 - m_{\eta_B^+}^2 }^2 } },
		\label{eq:Metacharged-1} \\
M_{\eta^+_2}^2
	&=
		{1\ov 2} \sqbr{m_{\eta_A^+}^2 + m_{\eta_B^+}^2 - \sqrt{ 4 m_{\eta_{AB}^+}^2 + \paren{ m_{\eta_A^+}^2 - m_{\eta_B^+}^2 }^2 } },
		\label{eq:Metacharged-2} \\
\tan\fn{2\theta_\tx{cs}}
	&=
		{ 2 \, {m_{\eta^+_{AB}}}^2 \ov {m_{\eta^+_{A}}}^2 - {m_{\eta^+_{B}}}^2 }.
}

{For the mass terms of the heavy fermions, to eliminate unnecessary complexity from the phenomenological discussion of neutrinos and dark matter---our subject of interest---and to make the essence of the model more transparently understandable, we manually constrain the degrees of freedom of the coupling constants under the following assumptions:
all of $m_{N_{ij}}$, $y_{A_{ij}}$ and $y_{B_{ij}}$ ($i,j = 1,2,3$) are real and diagonal.
Also, since physics involving CP violation (beyond the SM) is not the subject of this paper's discussion, physical complex coupling constants are not necessarily required.
Note that, as we will see in Eq.~\eqref{eq:CI_parametrisation}, the Yukawa couplings with the two inert doublets ($Y_A^{i\alpha}$ and $Y_B^{i\alpha}$) should be complex to generate a nonzero CP-violating phase of the PMNS matrix.

The mass terms of the right-handed neutrinos \cred{decompose} into each generation as follows,}
\al{
- {1\ov 2}
\sum_{i=1}^{3}
\bb \ol{\paren{N'_{A_i R}}^\tx{c}}, &  \ol{\paren{N'_{B_i R}}^\tx{c}}  \eb
{\cal M}_{N_{ii}}
\bb N'_{A_i R} \\ N'_{B_i R} \eb
+ \tx{h.c.},
	\label{eq:diagonal-Yukawa-texture}
}
where a (complex) orthogonal matrix diagonalises the symmetric Majorana mass terms as
\al{
{\cal M}_{N_{ii}}
&=
\bb
y_{A_{ii}} v_S & m_{N_{ii}} \\
m_{N_{ii}} & y_{B_{ii}} v_S
\eb
=
U_{N_i}
\bb
M_{N_{A_i}} & 0 \\ 0 & M_{N_{B_i}} 
\eb
\paren{U_{N_i}}^\tx{T},&
U_{N_i}
	=
		\bb
		c_{\theta_{N_i}} & - s_{\theta_{N_i}} \\
		s_{\theta_{N_i}} & c_{\theta_{N_i}}
		\eb,
}
with
\al{
M_{N_{A_i}}
	&=
		{1\ov 2} \sqbr{ y_{A_{ii}} v_S + y_{B_{ii}} v_S + \sqrt{ 4 m_{N_{ii}}^2 + \paren{ y_{A_{ii}} v_S - y_{B_{ii}} v_S }^2 } },
		\label{eq:MNAi} \\
M_{N_{B_i}}
	&=
		{1\ov 2} \sqbr{ y_{A_{ii}} v_S + y_{B_{ii}} v_S - \sqrt{ 4 m_{N_{ii}}^2 + \paren{ y_{A_{ii}} v_S - y_{B_{ii}} v_S }^2 } }.
		\label{eq:MNBi}
}
The diagonalised basis (without the prime symbol) of the Majorana mass terms is defined as
\al{
\bb N'_{A_i R} \\ N'_{B_i R} \eb
=
U_{N_i}
\bb {N}_{A_i R} \\ {N}_{B_i R} \eb.
}
We introduce the following compact notation for clarity in the mass eigenbases:
\al{
\label{eq:Yukawa_compact}
{\cal L}_\tx{Yukawa}
	&\supset
		- \sum_{s^0 = \eta_{R_1}, \eta_{R_2}, \eta_{I_1}, \eta_{I_2}}
		\sqbr{
			{\cal Y}^{\paren{s^0}}_{A_i \alpha} \, \ol{{N}_{A_i R}} \, \nu_{\alpha L} {s^0} +
			{\cal Y}^{\paren{s^0}}_{B_i \alpha} \, \ol{{N}_{B_i R}} \, \nu_{\alpha L} {s^0}
		} \notag \\
	&\quad \, \,
		{+} \sum_{s^+ = \eta_{1}^+, \eta_{2}^+}
		\sqbr{
			{\cal Y}^{\paren{s^+}}_{A_i \alpha} \, \ol{{N}_{A_i R}} \, \ell^-_{\alpha L} {s^+} +
			{\cal Y}^{\paren{s^+}}_{B_i \alpha} \, \ol{{N}_{B_i R}} \, \ell^-_{\alpha L} {s^+}
		} 
  + \tx{h.c.},
}
with
\al{
{\cal Y}^{\paren{\eta_{R_1}}}_{A_i \alpha} 
	&=
		+{Y^{i\alpha}_A \ov \sqrt{2}} c_{N_i} c_{\theta_R} + {Y^{i\alpha}_B \ov \sqrt{2}} s_{N_i} s_{\theta_R},&
{\cal Y}^{\paren{\eta_{R_1}}}_{B_i \alpha} 
	&=
		-{Y^{i\alpha}_A \ov \sqrt{2}} s_{N_i} c_{\theta_R} + {Y^{i\alpha}_B \ov \sqrt{2}} c_{N_i} s_{\theta_R},& \notag \\
{\cal Y}^{\paren{\eta_{R_2}}}_{A_i \alpha} 
	&=
		-{Y^{i\alpha}_A \ov \sqrt{2}} c_{N_i} s_{\theta_R} + {Y^{i\alpha}_B \ov \sqrt{2}} s_{N_i} c_{\theta_R},&
{\cal Y}^{\paren{\eta_{R_2}}}_{B_i \alpha} 
	&=
		+{Y^{i\alpha}_A \ov \sqrt{2}} s_{N_i} s_{\theta_R} + {Y^{i\alpha}_B \ov \sqrt{2}} c_{N_i} c_{\theta_R},& \notag  \\
{\cal Y}^{\paren{\eta_{I_1}}}_{A_i \alpha} 
	&=
		+i {Y^{i\alpha}_A \ov \sqrt{2}} c_{N_i} c_{\theta_I} + i {Y^{i\alpha}_B \ov \sqrt{2}} s_{N_i} s_{\theta_I},&
{\cal Y}^{\paren{\eta_{I_1}}}_{B_i \alpha} 
	&=
		-i {Y^{i\alpha}_A \ov \sqrt{2}} s_{N_i} c_{\theta_I} + i {Y^{i\alpha}_B \ov \sqrt{2}} c_{N_i} s_{\theta_I},& \notag  \\
{\cal Y}^{\paren{\eta_{I_2}}}_{A_i \alpha} 
	&=
		-i {Y^{i\alpha}_A \ov \sqrt{2}} c_{N_i} s_{\theta_I} + i {Y^{i\alpha}_B \ov \sqrt{2}} s_{N_i} c_{\theta_I},&
{\cal Y}^{\paren{\eta_{I_2}}}_{B_i \alpha} 
	&=
		+i {Y^{i\alpha}_A \ov \sqrt{2}} s_{N_i} s_{\theta_I} + i {Y^{i\alpha}_B \ov \sqrt{2}} c_{N_i} c_{\theta_I},& \notag  \\
{\cal Y}^{\paren{\eta_{1}^+}}_{A_i \alpha} 
	&=
		+{Y^{i\alpha}_A } c_{N_i} c_{\theta_\tx{cs}} + {Y^{i\alpha}_B } s_{N_i} s_{\theta_\tx{cs}},&
{\cal Y}^{\paren{\eta_{1}^+}}_{B_i \alpha} 
	&=
		-{Y^{i\alpha}_A} s_{N_i} c_{\theta_\tx{cs}} + {Y^{i\alpha}_B } c_{N_i} s_{\theta_\tx{cs}},& \notag \\
{\cal Y}^{\paren{\eta_{2}^+}}_{A_i \alpha} 
	&=
		-{Y^{i\alpha}_A} c_{N_i} s_{\theta_\tx{cs}} + {Y^{i\alpha}_B } s_{N_i} c_{\theta_\tx{cs}},&
{\cal Y}^{\paren{\eta_{2}^+}}_{B_i \alpha} 
	&=
		+{Y^{i\alpha}_A } s_{N_i} s_{\theta_\tx{cs}} + {Y^{i\alpha}_B } c_{N_i} c_{\theta_\tx{cs}},&
	\label{eq:Yukawa-in-mass-basis}
}
where the relative sign originates from the $i\sigma_2$ matrix in Eq.~\eqref{eq:Yukawa-terms}.

\subsection{Loop-Induced Neutrino Mass Terms} {\label{sec:massterm}}

\begin{figure}[t]
\centering
\includegraphics[width=0.45\linewidth]{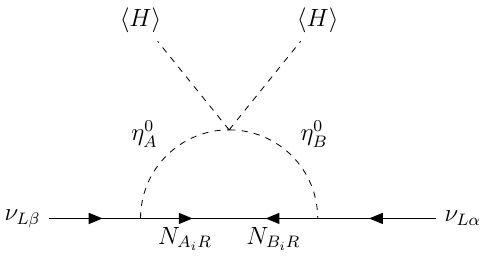}
\caption{
The topology of loop-induced mass generation is depicted in gauge eigenstates.
{Note that all of the Feynman diagrams in this manuscript were drawn by {\tt Tikz-Feynman}~\cite{Ellis:2016jkw}.}
}
\label{fig:neutrino-mass}
\end{figure}

In our scenario, as in the original Scotogenic model, the left-handed active neutrino mass terms are radiatively generated at a one-loop level,
where we adopt the parametrisation
\al{
{\cal L}_\tx{this model}
	\supset
		- {1 \ov 2} {\cal M}^\nu_{\alpha\beta} \ol{\paren{\nu'_{\alpha L}}^\tx{c}} \nu'_{\beta L} + \tx{h.c.}.
}
The calculation of the one-loop diagrams in Fig.~\ref{fig:neutrino-mass} with the dimensional regularisation brings us the result,
\al{
\label{eq:mass_formula}
{\cal M}^\nu_{\alpha\beta}
	&=
		- \sum_{s^0 = \eta_{R_1}, \eta_{R_2}, \eta_{I_1}, \eta_{I_2}} \sum_{i=1}^{3}
		{M_{N_{A_i}}    {\cal Y}^{\paren{s^0}}_{A_i \alpha}  {\cal Y}^{\paren{s^0}}_{A_i \beta}    \ov \paren{4\pi}^2}
		\br{
			{1\ov \vep} -\gamma + \log\fn{4\pi} - \int_0^1 dx \log\sqbr{ x M_{s^0}^2 + (1-x) M_{N_{A_i}}^2 - i \epsilon }
		} \notag \\
	&\quad \,
		- \sum_{s^0 = \eta_{R_1}, \eta_{R_2}, \eta_{I_1}, \eta_{I_2}} \sum_{i=1}^{3}
		 {M_{N_{B_i}}   {\cal Y}^{\paren{s^0}}_{B_i \alpha} {\cal Y}^{\paren{s^0}}_{B_i \beta}    \ov \paren{4\pi}^2}
		\br{
			{1\ov \vep} -\gamma + \log\fn{4\pi} - \int_0^1 dx \log\sqbr{ x M_{s^0}^2 + (1-x) M_{N_{B_i}}^2 - i \epsilon }
		} \notag \\
	&=
		- \sum_{s^0 = \eta_{R_1}, \eta_{R_2}, \eta_{I_1}, \eta_{I_2}} \sum_{i=1}^{3}
		 {M_{N_{A_i}}   {\cal Y}^{\paren{s^0}}_{A_i \alpha} {\cal Y}^{\paren{s^0}}_{A_i \beta}    \ov \paren{4\pi}^2}
		\br{
			- \int_0^1 dx \log\sqbr{ x M_{s^0}^2 + (1-x) M_{N_{A_i}}^2 - i \epsilon }
		} \notag \\
	&\quad \,
		- \sum_{s^0 = \eta_{R_1}, \eta_{R_2}, \eta_{I_1}, \eta_{I_2}} \sum_{i=1}^{3}
		 {M_{N_{B_i}}   {\cal Y}^{\paren{s^0}}_{B_i \alpha} {\cal Y}^{\paren{s^0}}_{B_i \beta}    \ov \paren{4\pi}^2}
		\br{
			- \int_0^1 dx \log\sqbr{ x M_{s^0}^2 + (1-x) M_{N_{B_i}}^2 - i \epsilon }
		},
}
where $1/\vep$ and $\gamma$ represent a divergent pole and the Euler-Mascheroni constant.
As expected from the nature of higher-dimensional operators, the divergences in the active-neutrino mass terms are
cancelled out in each of the $A$ and $B$ parts by use of Eq.~\eqref{eq:Yukawa-in-mass-basis}.
Note that this matrix is non-diagonal due to the degrees of freedom of $\alpha$ and $\beta$ ($\alpha, \beta = 1,2,3$).
Ignoring the $i\epsilon$ term in the loop integral of {heavy particles} Eq.~\eqref{eq:mass_formula} we get, 
\al{
\int_0^1 dx \log\sqbr{ x M_{s^0}^2 + (1-x) M_{N_{A_i}}^2 } = -1 +  {M_{N_{A_i}}^2 \log\paren{M_{N_{A_i}}^2} - M_{s^0}^2\log\paren{M_{s^0}^2} \ov M_{N_{A_i}}^2 - M_{s^0}^2 } .
}

As is apparent from the summation of the neutral scalars in  Eq.~\eqref{eq:mass_formula},  
${\cal M}^\nu_{\alpha\beta}$ is dependent on the mass difference between the real and imaginary scalars,
since their contributions are opposite to each other.
If $M_{R_1} = M_{I_1}$ and  $M_{R_2} = M_{I_2}$ all mass
contribution to ${\cal M}^\nu_{\alpha\beta}$ vanishes. The difference in their masses is
parameterised by the coupling $\lambda_5$ as seen in Eqs.~\eqref{eq:mR_AB} and~\eqref{eq:mI_AB}.
$v_S$ must also be nonzero for a finite mass contribution.

Hereafter, we restrict the Yukawa parameter under the condition for further simplicity,
\al{
Y^{i\alpha}_{A} \longrightarrow Y^{i\alpha} \longleftarrow Y^{i\alpha}_{B}, 
}
where so ${\cal M}^\nu_{\alpha\beta}$ becomes
\al{
{\cal M}^\nu_{\alpha\beta}
	&\to
		\sum_{i=1}^{3} Y^{i\alpha} {{\cal X}_i} Y^{i\beta},
	\label{eq:active-neutrino-mass-form-1}
}
where the derivation of ${\cal X}_i$ is straightforward as
\al{
{\cal X}_i
	&:=
		+ \sum_{s^0 = \eta_{R_1}, \eta_{R_2}, \eta_{I_1}, \eta_{I_2}}
		 {M_{N_{A_i}}   \paren{ {\cal Y}^{\paren{s^0}}_{A_i \alpha}/Y^{i\alpha} }   \paren{ {\cal Y}^{\paren{s^0}}_{A_i \beta}/Y^{i\beta} }    \ov \paren{4\pi}^2}
			\int_0^1 dx \log\sqbr{ x M_{s^0}^2 + (1-x) M_{N_{A_i}}^2  }
		 \notag \\
	&\quad \,\,
		+ \sum_{s^0 = \eta_{R_1}, \eta_{R_2}, \eta_{I_1}, \eta_{I_2}}
		 {M_{N_{B_i}}   \paren{ {\cal Y}^{\paren{s^0}}_{B_i \alpha}/Y^{i\alpha} }   \paren{ {\cal Y}^{\paren{s^0}}_{B_i \beta}/Y^{i\beta} }    \ov \paren{4\pi}^2}
			\int_0^1 dx \log\sqbr{ x M_{s^0}^2 + (1-x) M_{N_{B_i}}^2 },
	\label{eq:active-neutrino-mass-form-2}
}
and this form follows the Casas-Ibarra parametrisation~\cite{Casas:2001sr},\footnote{
A systematic generalisation of this parametrisation is found in~\cite{Cordero-Carrion:2019qtu}.
}
\al{
Y^{i\alpha}
	=
		\sqrt{ {\cal X}^{-1} } R_\nu \sqrt{M_{\wh{m}_{\nu}}} U^\dagger_\tx{PMNS},
		\label{eq:CI_parametrisation}
}
where the following three-by-three matrices are introduced
\al{
{\cal X}
	&:=
		\bb
			{\cal X}_1 & 0 & 0 \\
			0 & {\cal X}_2 & 0 \\
			0 & 0 & {\cal X}_3
		\eb,&
\sqrt{ {\cal X}^{-1} }
	&=
		\bb
			\paren{ {\cal X}_1 }^{-1/2} & 0 & 0 \\
			0 & \paren{ {\cal X}_2 }^{-1/2} & 0 \\
			0 & 0 & \paren{ {\cal X}_3 }^{-1/2}
		\eb,&
\sqrt{M_{\wh{m}_{\nu}}}
	&=
		\bb
			\sqrt{\wh{m}_{\nu_1}} & 0 & 0 \\
			0 & \sqrt{\wh{m}_{\nu_2}} & 0 \\
			0 & 0 & \sqrt{\wh{m}_{\nu_3}}
		\eb.&
}
$\wh{m}_{\nu_i} \, (i=1,2,3)$ are the mass eigenvalues of the three active neutrinos.
Also, $R_\nu$ and $U^\dagger_\tx{PMNS}$ are an arbitrary three-by-three orthogonal matrix and the {PMNS} matrix,
respectively.
As the only condition on $R_\nu$ is that it is orthogonal, it provides the physical degrees of freedom in the Yukawa couplings after realising the observed masses and mixings of the active neutrinos.

While any orthogonal matrix $R_{\nu}$ may be used in the Casas-Ibarra parameterisation, we scanned for an instance that produced a favourable Yukawa texture, i.e., high Yukawa for the 3rd generation Majorana neutrinos (DM and co-annihilating candidate).
In the following analysis, we focus on the choice of $R_{\nu}$,\footnote{
{The $R_\nu$ form in Eq.~\eqref{eq:Rnu-matrix} was adopted in our analysis as a preferred example based on our empirical understanding from numerical scans, which showed that a form where all components are highly mixed more readily satisfies all phenomenological requirements in our model.}
}
\al{
R_{\nu}
=
\begin{bmatrix}
 0.11641348 & 0.78579266 & 0.60743542 \\
-0.74517440 & -0.33523731 & 0.57648163 \\
 0.65663005 & -0.51975555 & 0.54652643
\end{bmatrix}.
	\label{eq:Rnu-matrix}
}
Also, we assume the normal ordering of the three neutrino masses.
Based on the data~\cite{ParticleDataGroup:2024cfk}, we \cred{determine}
$\wh{m}_{\nu 1} = 0.01\,\tx{eV}$,
$\wh{m}_{\nu 2} = \wh{m}_{\nu 1} + \sqrt{\Delta m_{21}^2} = \wh{m}_{\nu 1} + 8.6 \times 10^{-3} \,\tx{eV} =  0.0186 \,\tx{eV}$, and
$\wh{m}_{\nu 3} = \wh{m}_{\nu 1} + \sqrt{\Delta m_{32}^2 +\Delta m_{21}^2} =\wh{m}_{\nu 1} + 0.05 \,\tx{eV} = 0.060 \,\tx{eV}$.
For the mixing angles, we follow the parametrisation,
\als{
U_\tx{PMNS}
	&=
		\begin{pmatrix}
		c_{12} c_{13} & s_{12} c_{13} & s_{13} e^{-i \delta_\tx{CP}} \\
		-s_{12} c_{23} - c_{12} s_{13} s_{23} e^{i \delta_\tx{CP}} & c_{12} c_{23} - s_{12} s_{13} s_{23} e^{i \delta_\tx{CP}} &
			c_{13} s_{23} \\
		s_{12} s_{23} - c_{12} s_{13} c_{23} e^{i \delta_\tx{CP}} & -c_{12} s_{23} - s_{12} s_{13} c_{23} e^{i \delta_\tx{CP}} &
			c_{13} c_{23}
		\end{pmatrix},
}
where $c_{ij} = \cos\theta_{ij}$, $s_{ij} = \sin\theta_{ij}$, and $\delta_\tx{CP}$ the CP-violating phase, and we will adopt the best-fit result for the normal mass ordering in Table~1 of~\cite{Esteban:2024eli};
$\theta_{12} = 33.68^{\circ}$,
$\theta_{23} = 43.3^{\circ}$,
$\theta_{13} = 8.56^{\circ}$, and
$\delta_\tx{CP} = 212^{\circ}$;
{we set the two possible Majorana phases as zeros because we do not address the physics related to CP violation in this manuscript.}

\section{Non-DM Constraints on the Model
\label{sec:Non-DM}}

This section will discuss constraints other than DM physics on our $Z_4$-based Scotogenic scenario.

\subsection{Lepton Flavour Violation }{\label{sec:LFV}}

\begin{figure}[t]
    \centering
     \includegraphics[width=0.8\linewidth]{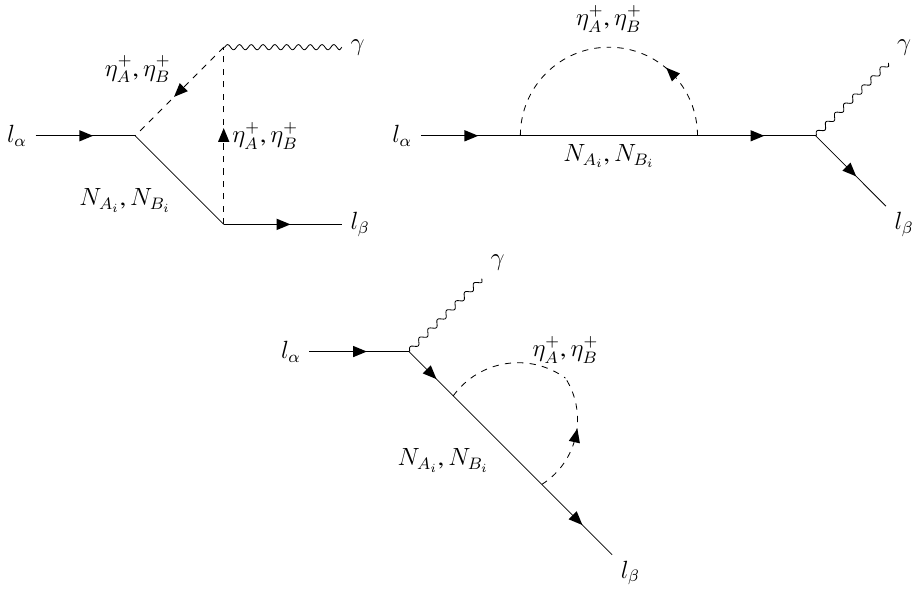}
    \caption{The Feynman diagrams contributing to $l_\alpha \to l_\beta + \gamma$ are shown in mass eigenstates.}
    \label{fig:LFV}
\end{figure}

We present the \cred{formulas} involved in calculating the branching ratios of 
$l_\alpha \rightarrow l_\beta + \gamma$ ($l_{1} = e$, $l_{2} = \mu$, $l_{3} = \tau$), as shown in Fig.~\ref{fig:LFV}.
The general form of the decay amplitudes squared is given by 
\al{
\ab{{\cal M}_{\alpha\beta}}^2 = {e^2 m_{l_\alpha}^6 \ov 4\paren{4\pi}^4}
        \ab{\sum_{i=1}^3 \sum_{s^+=\eta_1^+,\eta_2^+}
        \sqbr{
        {\paren{{\cal Y}^{\paren{s^+}}_{A_i \beta}}^* {\cal Y}^{\paren{s^+}}_{A_i \alpha} \ov M_{s^+}^2 }
        F_2\paren{M_{N_{A_i}}^2 \ov M_{s^+}^2} +
        {\paren{{\cal Y}^{\paren{s^+}}_{B_i \beta}}^* {\cal Y}^{\paren{s^+}}_{B_i \alpha}\ov M_{s^+}^2 }
        F_2\paren{M_{N_{B_i}}^2 \ov M_{s^+}^2} 
        }
        }^2,
        \label{eq:formula-LFV}
}
where $\alpha$ and $\beta$ represent the initial and final lepton external states, assuming $m_{l_\alpha}>m_{l_\beta}$ and the function
$F_2\fn{x}$ is defined as
\al{
F_2\paren{x} = {1 - 6x + 3x^2 + 2x^3 - 6x^2 \ln{x} \ov 6\paren{1-x^4}}.
}
The decay width can be calculated using \cite{Thomson:2013zua},
\al{
\Gamma_{l_\alpha \rightarrow l_\beta + \gamma}
	&=
		{p^* \ov 8\pi m_{l_\alpha} ^2}  \ab{\mathcal{M}_{\alpha\beta}}^2,&
p^*
	&=
		{1 \ov 2m_{l_\alpha}}\paren{m_{l_\alpha}^2 - m_{l_\beta}^2}.
}
Using the total decay width $\Gamma_\mu$, the branching ratio of $l_\alpha \rightarrow l_\beta + \gamma$ is estimated as
\al{\tx{Br}\fn{ l_\alpha \rightarrow l_\beta + \gamma }
	=
		{\Gamma_{l_\alpha \rightarrow l_\beta + \gamma} \ov \Gamma_\alpha}.
}
We will use the lifetimes $\tau_{\mu} = 2.1969811 \times 10^{-6}\,\tx{s}$ and $\tau_\tau = 2.903 \times 10^{-13}\,\tx{s}$~\cite{ParticleDataGroup:2024cfk} and the experimental constraints on the three processes, summarised in Table~\ref{tab:LFVconstraints}.

\begin{table}[t]
\centering
        \begin{tabular}{ |c|c| } 
            \hline
             Channel    & Observational upper limits \\
            \hline
            $\tx{Br}\fn{\mu^- \rightarrow e^- + \gamma}$ &  $< 3.1 \times10^{-13}$ \cite{MEGII:2023ltw} \\ 
            \hline
            $\tx{Br}\fn{\tau^- \rightarrow \mu^- + \gamma}$ & $<4.4\times 10^{-8}$ \cite{BaBar:2009hkt} \\ 
            \hline
            $\tx{Br}\fn{\tau^- \rightarrow e^- + \gamma}$ & $<3.3\times 10^{-8}$ \cite{BaBar:2009hkt} \\ 
            \hline
        \end{tabular}
        \caption{Lepton flavour violation observational bounds at 90\% {confidence level~(CL)}.}
	\label{tab:LFVconstraints}
\end{table}

\subsection{Oblique Constraint}


Following the prescription in~\cite{Grimus:2007if},
we define two matrices $U$ and $V$ having dimensions $n_d \times n$ and $n_d \times m$.
Here, $n_d \, (=3)$ is the number of $SU(2)_\tx{L}$ doublet scalars.
In terms of the numbers of $n_c \, (= 0)$ and $n_n \, (=1)$, which are the numbers of complex $SU(2)_\tx{L}$-singlet and
real $SU(2)_\tx{L}$-singlet scalars with zero hypercharge,
$n$ and $m$ are defined as $n := n_d + n_c \, (= 3)$ and $m := n_n + 2n_d \, (=7)$,
which represent the singly charged and neutral components' numbers (including the would-be Nambu-Goldstone~(NG) degrees of freedom), respectively.
$U$ and $V$ must satisfy,
\al{
\bb G^+ \\ {\eta'}_A^+ \\ {\eta'}_B^+ \eb
&=
U
\bb G^+ \\ {\eta}_{1}^+ \\ {\eta}_{2}^+ \eb,&
\bb h' + iG^{0}  \\ \eta'_{R_A} + i\eta'_{I_A} \\ \eta'_{R_B} +  i\eta'_{I_B} \eb
&=
V
\bb G^{0} \\ {h} \\ {\eta}_{R_1} \\ {\eta}_{I_1} \\ {\eta}_{R_2} \\ {\eta}_{I_2} \\ {s} \eb,
}
where $G^+$ and $G^0$ are the would-be NG bosons of the SM Higgs doublet, corresponding to the longitudinal degrees of freedom, respectively. The structures of $U$ and $V$ are identified as
\al{
\label{eq:Umatrix}
U &= \begin{pmatrix}
1 & 0 & 0 \\
0 & c_{\theta_\tx{cs}} & -s_{\theta_\tx{cs}} \\
0 & s_{\theta_\tx{cs}} & c_{\theta_\tx{cs}}
\end{pmatrix}, \\
\label{eq:Vmatrix}
V &= \begin{pmatrix}
i & c_\alpha & 0 & 0 & 0 & 0 & -s_\alpha \\
0 & 0 & c_{\theta_R} & ic_{\theta_I} & -s_{\theta_R} & -is_{\theta_I} & 0 \\
0 & 0 & s_{\theta_R} & is_{\theta_I} & c_{\theta_R} & ic_{\theta_I} & 0
\end{pmatrix}.
}

From this setup, we can calculate the deviation of the $\rho$ parameter $\Delta\rho$ using the general formula in~\cite{Grimus:2007if},
\al{
\label{eq:del_rho}
\Delta\rho =& {g^2 \ov 64 \pi^2 M_W^2} \Bigg[ \sum_{a=2}^{n} \sum_{b=2}^{m}\ab{ \paren{U^\dagger V}_{ab} }^2 F\paren{m_a^2, m_b^2} \notag\\  
            &- \sum_{b=2}^{m-1} \sum_{b'=b+1}^{m}\sqbr{\text{Im}\paren{V^\dagger V}_{bb'}}^2 F\paren{m_b^2, m_{b'}^2}   \notag\\
            &-2\sum_{a=2}^{n-1} \sum_{a'=a+1}^{n}\sqbr{\paren{U^\dagger U}_{aa'}}^2 F\paren{m_a^2, m_{a'}^2}   \notag\\
            &+ 3 \sum_{b=2}^{m} \sqbr{\text{Im}\paren{V^\dagger V}_{1b}}^2 \sqbr{F\paren{M_Z^2, m_b^2}  - F\paren{M_W^2, m_b^2}  } \notag\\
            &- 3 \sqbr{ F\paren{M_Z^2, M_{{h}}^2}  - F\paren{M_W^2, M_{{h}}^2}} \Bigg],       
}
where $g$ is the $SU(2)_\tx{L}$ gauge coupling, and $M_W$ and $M_Z$ are the $W$-boson and $Z$-boson masses.
$m_a$ (also $m_{a'}$) and $m_b$ (also $m_{b'}$) denote the masses of the charged scalars and neutral scalars in the mass eigenstates, respectively.
The two states identified by $a=1$ or $b=1$ are unphysical; therefore, they are not considered in Eq.~\eqref{eq:del_rho}.
Equation~\eqref{eq:del_rho} is simplified in the models with no $SU(2)_\tx{L}$-singlet
charged scalars, as in our model, where the third term proportional to $U^\dagger U$ vanishes
as $U^\dagger U$ in unity, and since the summation is over off-diagonal elements, the term has no contributions.
The function $F\fn{x,y}$ in Eq.~\eqref{eq:del_rho} is defined as
\al{
F(x,y) = \begin{cases}
    0 & \text{if } x = y, \\
    \displaystyle {x + y \ov 2} - {x y \ln(x/y) \ov x - y} & \text{otherwise}.
\end{cases}
}
To compare with observational results \cite{particle2022review}, we use the parameter $T$,
\al{
T={\Delta\rho \ov \alpha\fn{M_Z}}.
}
We consider the case when $U=0$, and we use the central value of 
$\alpha\paren{M_Z}=\paren{127.951 \pm 0.009}^{-1}$, which is the
quantum-electrodynamical fine structure constant defined in the $\ol{\tx{MS}}$ scheme with \cred{five} quark flavours
\cite{particle2022review}.
We will target the following value via statistical analysis~\cite{ParticleDataGroup:2024cfk}:
\al{
\label{eq:Tbound}
T = 0.01 \pm 0.12.
}

\subsection{Vacuum Stability}

\subsubsection{Boundness from below}

For the constraints via bounded from below, we need to look at the quartic part of the potential ${\cal V}^{(4)}$ defined as
\al{
\label{eq:potential4}
{\cal V}^{(4)}(H,\eta_A,\eta_B, S)
	&= {\lambda_{HS} \ov 2} S^2 \paren{H^\dagger H} + {\lambda_{S\eta_A} \ov 2} S^2 \paren{\eta^\dagger_A \eta_A}
		+ {\lambda_{S\eta_B} \ov 2} S^2 \paren{\eta^\dagger_B \eta_B} \notag \\
	&\quad
		+ {\lambda_{H} \ov 2} \paren{H^\dagger H}^2 + {\lambda_S \ov 4} S^4
		+ {\lambda_{2a} \ov 2} \paren{\eta_A^\dagger \eta_A}^2
		+ {\lambda_{2b} \ov 2} \paren{\eta_B^\dagger \eta_B}^2
		+ {\lambda_{2c} \ov 2} \paren{\eta_A^\dagger \eta_A}\paren{\eta_B^\dagger \eta_B} \notag \\
	&\quad
		+ {\lambda_{2d} \ov 2} \paren{\eta_A^\dagger \eta_B}\paren{\eta_B^\dagger \eta_A}
		+ \lambda_{3A} \paren{H^\dagger H} \paren{\eta_A^\dagger \eta_A}
		+ \lambda_{3B} \paren{H^\dagger H} \paren{\eta_B^\dagger \eta_B} \notag \\
	&\quad
		+ \lambda_{4A} \paren{H^\dagger \eta_A} \paren{\eta_A^\dagger H}
		+ \lambda_{4B} \paren{H^\dagger \eta_B} \paren{\eta_B^\dagger H}
		+ {1\ov 2} \sqbr{\lambda_5 \paren{H^\dagger \eta_A}\paren{H^\dagger \eta_B} + \tx{h.c.} },
}
{where we took $\lambda_6 = 0$ [refer to Eq.~\eqref{eq:potential}].}
Assuming fields are real, we can rewrite the potential ${\cal V}^{(4)}$ as described in~\cite{Kannike:2016fmd},
\al{
{\cal V}^{(4)} = x^\tx{T} {\Lambda} \, x,
}
with
\al{
x=\begin{pmatrix}
    H^2\\\eta_A^2\\ \eta_A \eta_B\\ \eta_B^2\\ S^2 
\end{pmatrix},
}
and 
\al{
{\Lambda}
	=
\begin{pmatrix}
\frac{\lambda_H}{2} & \frac{\lambda_{3A} + \lambda_{4A}}{2} & \frac{\lambda_5}{2} & \frac{\lambda_{3B} + \lambda_{4B}}{2} & \frac{\lambda_{HS}}{4} \\
\frac{\lambda_{3A} + \lambda_{4A}}{2} & \frac{\lambda_{2a}}{2} & 0 & \frac{(1 - c)}{2} \left(\frac{\lambda_{2c}}{2} + \frac{\lambda_{2d}}{2}\right) & \frac{\lambda_{S\eta A}}{4} \\
\frac{\lambda_5}{2} & 0 & c \left(\frac{\lambda_{2c}}{2} + \frac{\lambda_{2d}}{2}\right) & 0 & 0 \\
\frac{\lambda_{3B} + \lambda_{4B}}{2} & \frac{(1 - c)}{2} \left(\frac{\lambda_{2c}}{2} + \frac{\lambda_{2d}}{2}\right) & 0 & \frac{\lambda_{2b}}{2} & \frac{\lambda_{S\eta B}}{4} \\
\frac{\lambda_{HS}}{4} & \frac{\lambda_{S\eta A}}{4} & 0 & \frac{\lambda_{S\eta B}}{4} & \frac{\lambda_S}{4}
\end{pmatrix}.
}
$c$ is an arbitrary constant due to the ambiguity in $(\eta_A\eta_B)^2$ terms.


The requirement of stability means demanding that the quartic part of the potential ${\cal V}^{(4)} > 0$ as the fields $\phi_i \rightarrow \infty$.
To find the conditions for ${\cal V}^{(4)} > 0$ we can impose \cred{the} positivity condition on {${\Lambda}$}~\cite{Kannike:2016fmd}.
We use the Sylvester criterion for the positivity of {${\Lambda}$}, assuming $\phi_i\phi_j \in \mathcal{R}$. We must apply the Sylvester 
criterion to all permutations of the basis vector $x$.
Sylvester's criterion states that a $n\times n$ Hermitian matrix {${\Lambda}$} is positive-definite if and only if all the following matrices have a positive determinant: the upper left 1-by-1 corner of {${\Lambda}$},
the upper left 2-by-2 corner of {${\Lambda}$}, and \cred{so} on.
This sequence continues till {${\Lambda}$} itself.

We can then write down the general conditions,
\al{
\lambda_{ii}>0,
}
\al{
\lambda_{ii}\lambda_{jj}- \lambda_{ij} ^2 > 0 \quad \tx{for } i\neq j,
}
\al{
&\begin{vmatrix}
    \lambda_{ii} & \lambda_{ij} & \lambda_{ik} \\
    \lambda_{ji} & \lambda_{jj} & \lambda_{jk} \\
    \lambda_{ki} & \lambda_{kj} & \lambda_{kk} 
\end{vmatrix}
>0 \quad \tx{for } i\neq j \neq k,
&\begin{vmatrix}
    \lambda_{ii} & \lambda_{ij} & \lambda_{ik} & \lambda_{il}\\
    \lambda_{ji} & \lambda_{jj} & \lambda_{jk} & \lambda_{jl}\\
    \lambda_{ki} & \lambda_{kj} & \lambda_{kk} & \lambda_{kl}\\
    \lambda_{li} & \lambda_{lj} & \lambda_{lk} & \lambda_{ll}
\end{vmatrix}
>0  \quad \tx{for }  i\neq j \neq k \neq l,
}
\al{\tx{Det}[\Lambda]>0.}
$\lambda_{ij}$'s are the elements of matrix $\Lambda$  where $i, j, k = 1, 2, 3, 4, 5$ \cred{represent} the element position of the matrix.\\


We get 5 constraints from the $1\times 1$ matrix,  
\al{
&\lambda_H > 0,&&\lambda_{2a} > 0, &&\lambda_{2b} > 0,&&c\left(\lambda_{2c} + \lambda_{2d}\right) > 0,&&\lambda_S > 0,
\label{eq:Det-condition-start}
}
{We get the following independent constraints from the $2\times 2$ matrices,}
\al{
&\lambda_H\lambda_{2a} - (\lambda_{3A} + \lambda_{4A})^2 >0, 
&&\lambda_H\lambda_{2b} - (\lambda_{3B} + \lambda_{4B})^2 >0,\\
&2\lambda_{2a} \lambda_S - \lambda_{S\eta_A}^2 >0,
&&2\lambda_{2b} \lambda_S - \lambda_{S\eta_B}^2 >0,\\
&2\lambda_H\lambda_S - \lambda_{HS}^2>0, && c\lambda_H \left(\lambda_{2c} + \lambda_{2d}\right)-\lambda_5 ^2 >0,\label{eq:c1}\\
&4\lambda_{2a}\lambda_{2b} - (1 - c)^2 \left(\lambda_{2c}+ \lambda_{2d}\right)^2>0,&& 
\label{eq:c2} 
}
{where 10 constraints appear from them if we do not remove redundancies.}
{We get the following independent constraints from the $3\times 3$ matrices,}
\al{
&\begin{vmatrix}
     \frac{\lambda_H}{2} & \frac{\lambda_{3A} + \lambda_{4A}}{2} &  \frac{\lambda_5}{2}\\
     \frac{\lambda_{3A} + \lambda_{4A}}{2} & \frac{\lambda_{2a}}{2} & 0\\
     \frac{\lambda_5}{2} & 0 & c \left(\frac{\lambda_{2c}}{2} + \frac{\lambda_{2d}}{2}\right)  \\ 
\end{vmatrix}>0,
&&\begin{vmatrix}
     \frac{\lambda_H}{2} & \frac{\lambda_{3B} + \lambda_{4B}}{2} &  \frac{\lambda_5}{2}\\
     \frac{\lambda_{3B} + \lambda_{4B}}{2} & \frac{\lambda_{2b}}{2} & 0\\
     \frac{\lambda_5}{2} & 0 & c \left(\frac{\lambda_{2c}}{2} + \frac{\lambda_{2d}}{2}\right)  \\ 
\end{vmatrix}>0,\\ \notag\\
&\begin{vmatrix}
     \frac{\lambda_H}{2} & \frac{\lambda_{HS}}{4} &  \frac{\lambda_{3A} + \lambda_{4A}}{2}\\
     \frac{\lambda_{HS}}{4} & \frac{\lambda_S}{4}& \frac{\lambda_{S\eta A}}{4}\\
     \frac{\lambda_{3A} + \lambda_{4A}}{2} & \frac{\lambda_{S\eta A}}{4} & \frac{\lambda_{2a}}{2}   \\ 
\end{vmatrix}>0, 
&&\begin{vmatrix}
     \frac{\lambda_H}{2} & \frac{\lambda_{HS}}{4} &  \frac{\lambda_{3B} + \lambda_{4B}}{2}\\
     \frac{\lambda_{HS}}{4} & \frac{\lambda_S}{4}& \frac{\lambda_{S\eta B}}{4}\\
     \frac{\lambda_{3B} + \lambda_{4B}}{2} & \frac{\lambda_{S\eta B}}{4} & \frac{\lambda_{2b}}{2}   \\ 
\end{vmatrix}>0, \\ \notag\\
&\begin{vmatrix}
     \frac{\lambda_H}{2} & \frac{\lambda_{3A} + \lambda_{4A}}{2} &  \frac{\lambda_{3B} + \lambda_{4B}}{2}\\
     \frac{\lambda_{3A} + \lambda_{4A}}{2} & \frac{\lambda_{2a}}{2} & \frac{(1 - c)}{2} \left(\frac{\lambda_{2c}}{2}+\frac{\lambda_{2d}}{2}\right)\\
     \frac{\lambda_{3B} + \lambda_{4B}}{2} & \frac{(1 - c)}{2} \left(\frac{\lambda_{2c}}{2}+\frac{\lambda_{2d}}{2}\right) & \frac{\lambda_{2b}}{2} 
\end{vmatrix}>0, 
&&\begin{vmatrix}
     \frac{\lambda_S}{4} & \frac{\lambda_{S\eta A}}{4} &  \frac{\lambda_{S\eta B}}{4}\\
     \frac{\lambda_{S\eta A}}{4} & \frac{\lambda_{2a}}{2} & \frac{(1 - c)}{2} \left(\frac{\lambda_{2c}}{2}+\frac{\lambda_{2d}}{2}\right)\\
     \frac{\lambda_{S\eta B}}{4} & \frac{(1 - c)}{2} \left(\frac{\lambda_{2c}}{2}+\frac{\lambda_{2d}}{2}\right) & \frac{\lambda_{2b}}{2} 
\end{vmatrix}>0,
}
\al{
&\begin{vmatrix}
     \frac{\lambda_H}{2} & \frac{\lambda_{HS}}{4} &  \frac{\lambda_5}{2}\\
     \frac{\lambda_{HS}}{4} & \frac{\lambda_S}{4}& 0\\
     \frac{\lambda_5}{2} & 0 & \frac{\lambda_{2b}}{2}   \\ 
\end{vmatrix}>0,
}
{where 10 constraints appear from them if we do not remove redundancies.}
{We get the following independent constraints from the $4\times 4$ matrices,}
\al{\label{eq:c_ineq}
&\begin{vmatrix}
\frac{\lambda_H}{2} & \frac{\lambda_{3A} + \lambda_{4A}}{2} & \frac{\lambda_5}{2} & \frac{\lambda_{3B} + \lambda_{4B}}{2}  \\
\frac{\lambda_{3A} + \lambda_{4A}}{2} & \frac{\lambda_{2a}}{2} & 0 & \frac{(1 - c)}{2} \left(\frac{\lambda_{2c}}{2} + \frac{\lambda_{2d}}{2}\right)  \\
\frac{\lambda_5}{2} & 0 & c \left(\frac{\lambda_{2c}}{2} + \frac{\lambda_{2d}}{2}\right) & 0  \\
\frac{\lambda_{3B} + \lambda_{4B}}{2} & \frac{(1 - c)}{2} \left(\frac{\lambda_{2c}}{2} + \frac{\lambda_{2d}}{2}\right) & 0 & \frac{\lambda_{2b}}{2} 
\end{vmatrix}>0,
\\ \notag\\
&\begin{vmatrix}
\frac{\lambda_H}{2} & \frac{\lambda_{3A} + \lambda_{4A}}{2} & \frac{\lambda_5}{2}  & \frac{\lambda_{HS}}{4} \\
\frac{\lambda_{3A} + \lambda_{4A}}{2} & \frac{\lambda_{2a}}{2} & 0 &  \frac{\lambda_{S\eta A}}{4} \\
\frac{\lambda_5}{2} & 0 & c \left(\frac{\lambda_{2c}}{2} + \frac{\lambda_{2d}}{2}\right)  & 0 \\
\frac{\lambda_{HS}}{4} & \frac{\lambda_{S\eta A}}{4} & 0  & \frac{\lambda_S}{4}
\end{vmatrix}>0,\\ \notag\\
&\begin{vmatrix}
\frac{\lambda_H}{2} & \frac{\lambda_{3B} + \lambda_{4B}}{2} & \frac{\lambda_5}{2}  & \frac{\lambda_{HS}}{4} \\
\frac{\lambda_{3B} + \lambda_{4B}}{2} & \frac{\lambda_{2b}}{2} & 0 &  \frac{\lambda_{S\eta B}}{4} \\
\frac{\lambda_5}{2} & 0 & c \left(\frac{\lambda_{2c}}{2} + \frac{\lambda_{2d}}{2}\right)  & 0 \\
\frac{\lambda_{HS}}{4} & \frac{\lambda_{S\eta B}}{4} & 0  & \frac{\lambda_S}{4}
\end{vmatrix}>0,\\ \notag\\
&\begin{vmatrix}
\frac{\lambda_H}{2} & \frac{\lambda_{3A} + \lambda_{4A}}{2}  & \frac{\lambda_{3B} + \lambda_{4B}}{2} & \frac{\lambda_{HS}}{4} \\
\frac{\lambda_{3A} + \lambda_{4A}}{2} & \frac{\lambda_{2a}}{2} &  \frac{(1 - c)}{2} \left(\frac{\lambda_{2c}}{2} + \frac{\lambda_{2d}}{2}\right) & \frac{\lambda_{S\eta A}}{4} \\
\frac{\lambda_{3B} + \lambda_{4B}}{2} & \frac{(1 - c)}{2} \left(\frac{\lambda_{2c}}{2} + \frac{\lambda_{2d}}{2}\right)  & \frac{\lambda_{2b}}{2} & \frac{\lambda_{S\eta B}}{4} \\
\frac{\lambda_{HS}}{4} & \frac{\lambda_{S\eta A}}{4} &\frac{\lambda_{S\eta B}}{4} & \frac{\lambda_S}{4}
\end{vmatrix}>0,
\label{eq:Det-condition-end}
}
{where 5 constraints appear from them if we do not remove redundancies.}

We must also constrain the value of $c$ as described in~\cite{Kannike:2016fmd}, from \cred{the} second constraint of Eq.~\eqref{eq:c1} requires $|c|\geq |c_0|$,
\al{
c_0 = {1 \ov \paren{\lambda_{2c} + \lambda_{2d}} }\paren{\lambda_5 ^2 \ov \lambda_H}
}
The extrema solutions of the left-hand side of the inequality Eq.~\eqref{eq:c_ineq} with respect to $c$ are $c_+$, $c_-$. We check which solution gives the maximum value for the inequality of Eq.~\eqref{eq:c_ineq} by substituting for $c$ in its second derivative.

Finally, we have constraints,
\al{
c = \begin{cases}
    \tx{max}\paren{c_0,c_+} & \text{if } \paren{\lambda_{2c} + \lambda_{2d}}>0, \\
    \tx{max}\paren{c_0,c_-} & \text{if } \paren{\lambda_{2c} + \lambda_{2d}}<0,\\
    0 & \text{if } \paren{\lambda_{2c} + \lambda_{2d}}=0.
\end{cases}
}
We can compare and check our results with \cite{Keus:2013hya} for the $Z_4$ three doublet model if we turn off coupling with the $S$ field and take its mass to zero. Our result is a match.

\subsubsection{Inert condition via Hessian}

The potential in Eq.~\eqref{eq:potential} should not develop a VEV in the inert direction.
We find these constraints by constructing the \cred{Hessian} matrix $H$ described in~\cite{Keus:2013hya}.
\al{
(H)_{ij} = {\partial^2 {\cal V}\ov \partial \phi_i \partial\phi_j}\Big|_{ \phi_i= <\phi_i>, \, \phi_j=<\phi_j>},
}
where $\phi_{i},\phi_j = \{H,\eta_A,\, \eta_B, S \}$ assuming $\phi_i \in \mathcal{R}$.
We consider the minimum case
$\{v_H,0,0,v_S\}$.

\al{
H=\Bigg(\begin{smallmatrix}
2 m_H^2 + 3 v_H^2 \lambda_H + v_S^2 \lambda_{HS} & 0 & 0 &  { \sqrt{2} v_H v_S \lambda_{HS} }  \\
0 & 2 m_{\eta_A}^2 +  v_H^2 (\lambda_{3A} + \lambda_{4A}) + v_S^2 \lambda_{S\eta_A} & {\lambda_5 v_h^2 \ov 4} & 0 \\
0 & {\lambda_5 v_h^2 \ov 4} & m_{\eta_B}^2 + v_H^2 (\lambda_{3B} + \lambda_{4B}) + v_S^2 \lambda_{S\eta_B} & 0 \\
{ \sqrt{2} v_H v_S \lambda_{HS} } & 0 & 0 & m_S^2 + {v_H^2\ov 2} \lambda_{HS} + 3 v_S^2 \lambda_S
\end{smallmatrix}\Bigg)
}
The Hessian must be positive, therefore, we impose the Sylvestor criterion to get the following constraints,
\al{
&2 m_H^2 + 3 v_H^2 \lambda_H + v_S^2 \lambda_{HS}>0, && 2 m_{\eta_A}^2 +  v_H^2 (\lambda_{3A} + \lambda_{4A}) + v_S^2 \lambda_{S\eta_A}>0, \notag\\
&2 m_{\eta_B}^2 +  v_H^2 (\lambda_{3B} + \lambda_{4B}) + v_S^2 \lambda_{S\eta_B}>0, 
&& m_S^2 + {v_H^2 \ov 2} \lambda_{HS} + 3 v_S^2 \lambda_S>0.
	\label{eq:Inert-Condition-1}
}
\vspace{-6mm}
\al{
&\left(2 m_{\eta_A}^2 +  v_H^2 (\lambda_{3A} +
\lambda_{4A}) + v_S^2 \lambda_{S\eta_A}\right)
\left(m_{\eta_B}^2 +  v_H^2 (\lambda_{3B} + 
\lambda_{4B}) + v_S^2 \lambda_{S\eta_B}\right) -
\left({\lambda_5 v_h^2 \ov 4}\right)^2 >0, \notag\\
&\left(2 m_H^2 + 3 v_H^2 \lambda_H + \lambda_{HS}v_S^2\right)\left(m_S^2 + {v_H^2\ov2} \lambda_{HS} + 3 v_S^2 \lambda_S\right)- 2\left( v_H v_S \lambda_{HS} \right)^2 >0.
	\label{eq:Inert-Condition-2}
}
Here, we have simplified and removed redundant constraints of the criteria.

\section{Dark Matter Phenomenology
\label{sec:DM}}

As we discussed in Section~\ref{sec:Model-Setup},
the lightest fermioin that possesses the $Z_4$ parity $+i$ or $-i$ is stabilised by the residual $Z_2$ parity,
and it can be a dark matter candidate $\chi$. 
In this paper, we focus on the scenario that one of the heavy Majorana neutrinos $N_{A_i}$ or $N_{B_i}$ $(i=1,2,3)$ is a dark matter candidate.
Which one is appropriate will be debated later.

We will consider the ordinary thermal freeze-out scenario in our Scotogenic model.
First, we will briefly summarise the scenario of dark matter.
We follow the procedure in~\cite{Edsjo:1997bg} in our implementation.

\subsection{Relic Density through Thermal Freeze-out}

\subsubsection{Thermal averaging}

We first calculate the unpolarized annihilation rate, 
\al{
W_{ij\rightarrow kl}
    	&= {p_{kl} \ov 16\pi^2 g_i g_j S_{kl} \sqrt{s}}\int \ol{ \ab{ {\cal M}_{ij \to kl}}^2 } d\Omega,&
\ol{ \ab{ {\cal M}_{ij \to kl}}^2 }
	&:=
		\sum_{\tx{internal d.o.f.}} \ab{ {\cal M}_{ij \to kl}}^2,
    \label{eq:annihilationrate}
}
where $\sqrt{s}$ is the total energy of the centre-of-the-mass system and $d\Omega$ is the solid angle for the final state.
$\sum_{\tx{internal d.o.f.}} \ab{ {\cal M}_{ij \to kl} }^2$ is the square amplitude of the process $ij \to kl$ averaged
over initial internal degrees of freedoms (d.o.f.) and summing over final d.o.f.~\cite{Srednicki:1988ce}.
The subscripts $i,j$ in Eq.~\eqref{eq:annihilationrate} represent different species of initial dark matter particles and $k,l$ are final SM particle 
species.
The indices $k$ and $l$ discriminate particles in final states, and $p_{kl}$ is the final centre of mass momentum, which is defined as,
\al{
	p_{kl}=\frac{\sqrt{s-(m_k+m_l)^2} \sqrt{s-(m_k-m_l)^2} }{2\sqrt{s}},
}
where $g_i$ is the degrees of freedom for the dark matter particles ($g_1=2$ for  Majorana-fermion dark matter), 
and $ S_{kl}$ is the symmetry factor $S_{kl}=2$ for identical final particles and $S_{kl}=1$ otherwise.
If we focus on a single DM candidate with no co-annihilation, $i$ and $j$ take only one, the effective unpolarized annihilation rate is,
\al{
	W_{\tx{eff}}=\sum_{k,l}  W_{11\rightarrow kl}.
}
$W_{\tx{eff}}$ is the annihilation rate summed over all final state SM particles ($k,l$).


At equilibrium, we approximate the momentum distribution of the dark matter candidate to follow the Maxwell-Boltzmann distribution.
We now  integrate the annihilation rate over the momentum distribution of the dark matter population,
\al{
\label{eq:momentum_int}
    {\cal A} =\frac{g^2 T}{4\pi^4}\int_0 ^\infty dp_{\tx{eff}} \,p_{\tx{eff}}^2 W_{\tx{eff}} K_1\paren{\sqrt{s}\ov T},
}
\al{
    p_{\tx{eff}} = p_{11} &=\frac{1}{2}\sqrt{s-4M_{\chi}^2},
}
where $T$ describes the temperature,
$p_\tx{eff}$ is the incoming momentum of the DM particle (in the centre of mass frame) and
$K_1(x)$ is a modified Bessel function of the second kind of order one.
%
Equilibrium number density of the dark matter $n_{\tx{eq}}$ is
\al{
\label{eq:neq}
    n_{\tx{eq}} = {T \ov 2\pi^2} g_\chi M_{\chi}^2 K_2\fn{{ {M_\chi} \ov T}},
}
where $K_2(x)$ is a modified Bessel function of the second kind of order two.
%
The thermalised cross-section can then be put together,
\al{
\label{eq:thermalcross}
    \left< \sigma_{\tx{eff}}v\right> = \frac{{\cal A}}{n_\tx{eq}^2}.
}

In the case of co-annihilating dark matter, there is an additional normalisation for the case of co-annihilation.
We calculate the effective unpolarised annihilation rate,
\al{
\label{eq:Weff_co}
    W_{\tx{eff}} =  \sum_{i,j} \sum_{k,l} {p_{ij}\ov p_{11}} {g_ig_j\ov g_1^2} W_{ij\rightarrow kl},
}
$p_{11}$ is the initial centre of mass momentum when the two initial-state particles are the dark matter ($i=j=1$),
while $p_{ij}$ is that when the initial-state particles are the $i$-th and $j$-th dark matter candidates, defined as
\al{
p_{ij}=&\frac{\sqrt{s-(M_{\chi_i}+M_{\chi_j})^2} \sqrt{s-(M_{\chi_i}-M_{\chi_j})^2} }{2\sqrt{s}}.
}
Also, the equilibrium number density in Eq.~\eqref{eq:neq} is modified as
\al{
    n_{\tx{eq}} = {T \ov 2\pi^2} \sum_i g_i M_{\chi_i}^2 K_2\paren{{M_{\chi_i}}\ov{T}},
    \label{eq:neq_co}
}
where the \cred{summation} over $i$ in Eq.~\eqref{eq:neq_co} is taken over the dark-matter candidates relevant in the co-annihilation.\footnote{
Furthermore, we should modify the variable $Y$ [defined in Eq.~\eqref{eq:definition-Y}] in the thermal equilibrium, appearing in the Boltzmann equation to describe the thermal \cred{freeze-out} DM scenario in Eq.~\eqref{eq:boltz} for the coannihilation case as~\cite{Edsjo:1997bg}
\als{
    Y_\tx{eq} ={45 x^2 \ov 4\pi^4 h_{\tx{eff}}}\sum_i g_i \paren{{M_{\chi_i}}\ov{M_{\chi_1}}}^2 K_2\paren{x {M_{\chi_i}\ov M_{\chi_1}}}.
}
We make the same scaling modification to the rescaled variable [defined in Eq.~\eqref{eq:scaled_boltz}] as described earlier, the new yield at equilibrium is
\als{
y_{\tx{eq}}= {M_\tx{PL} \sqrt{g^\ast} x^2\ov \sqrt{90} h_{\tx{eff}} \pi^3}\sum_i g_i \paren{{M_{\chi_i}}\ov{M_{\chi_1}}}^2 K_2\paren{x {M_{\chi_i}\ov M_{\chi_1}}}.
}
}

To calculate the thermal cross section in Eq.~\eqref{eq:thermalcross}, we approximate $T={T_f}_i$ and $x=x_f$ in the equilibrium number density as the different particles decouple at different temperatures.
\al{
    {T_f}_i = {M_{\chi_i} \ov x_f}.
} 
In the momentum integration of Eq.~\eqref{eq:momentum_int}, we replace $T$ with ${T_f}_1$ as it is integrated over dark matter temperature.

\subsubsection{Boltzmann equation}

Let us consider the quantity $Y$; it is the ratio of dark matter number density 
$n$ to entropy density of the universe $s$,
\begin{equation}
    Y=\frac{n}{s}.
    \label{eq:definition-Y}
\end{equation}
$Y$ is the measure of dark matter number density; solving the following Boltzmann equation allows us to compare 
$Y_\infty$ after the freeze-out mechanism as measured by Planck~\cite{Planck:2018vyg} $\Omega h^2 = 0.120\pm0.001$,

\al{\label{eq:boltz}
    \frac{dY}{dx} = -\sqrt{\frac{\pi}{45 G}} { \sqrt{g^\ast} M_{\chi} \ov x^2}\left< \sigma_{\tx{eff}}v\right>\left(Y^2-Y_\tx{eq}^2\right),
}
and the distribution of $Y$ in the thermal equilibrium is represented as
\al{
    Y_{\tx{eq}} =& \frac{n_\tx{eq}}{s}= \frac{45 x^2}{4\pi^4 h_{\tx{eff}}}g_1 K_2(x),
}
where we introduced the dimensionless inverse temperature $x$ as
\al{
x := {m_1 \ov T}.
}
$g^\ast$ is the energy density degrees of freedom,
$h_\tx{eff}$ is the entropy density degrees of freedom of the \cred{Universe}, 
$G$ is the gravitational constant.
We use the scaling trick suggested by \cite{tanedo2011defense} to rescale the Boltzmann equation Eq.~\eqref{eq:boltz}
by multiplying it with a constant ($M_\tx{PL}(\frac{8\pi^2 g^\ast}{45})^{1/2}$), \cred{and} we make the required changes to the Boltzmann equation. 
\al{
\label{eq:scaled_boltz}
\frac{dy}{dx} 
    &= 
    	-\left<\sigma_{\tx{eff}}v\right>M_{\chi}\frac{\left(y^2 - y_{\tx{eq}}^2\right)}{x^2},&
y_{\paren{\tx{eq}}}
	&:=
		M_\tx{PL} \paren{\frac{8\pi^2 g^\ast}{45} }^{1/2} Y_{\paren{\tx{eq}}}, & \\
	&&
y_{\tx{eq}}
	&= 
		{M_\tx{PL} \sqrt{g^\ast} x^2\ov \sqrt{90} h_{\tx{eff}} \pi^3}g_1 K_2(x),&
}
where $y_{\paren{\tx{eq}}}$ is the new rescaled abundance and
$M_\tx{PL}$ is \cred{the} reduced Planck mass defined by $M_\tx{PL} = \paren{8\pi G}^{{-1/2}}$.
This modification prevents numerical instabilities, which are common while solving the Boltzmann equation.

To solve Eq.~\eqref{eq:scaled_boltz}, we approximate $T=T_f$ and $x=x_f$ when we calculate Eq.~\eqref{eq:thermalcross} which makes 
$\left<\sigma v\right>_{\tx{eff}}$
a constant within the region of integration, where $T_f$ and $x_f$ represent the temperature 
at which dark matter decouples or freezes out of the thermal bath $\paren{x_f = {M_{\chi}/T_f}}$.
$x_f$ typically ranges from 20 to 25, so we use {$x_f = 20$} for our numerical analysis.
Above $100\,\tx{GeV}$, the degrees of freedom $g^\ast \simeq h_\tx{eff} \approx 100 $ and significantly decreases only at QCD phase transition ($150-214 \,\tx{MeV}$)~\cite{Husdal:2016haj}. 
Since we are considering TeV-scale dark matter mass, we can use the typical value of  $g^* \simeq h_\tx{eff} \approx 100$ as a constant throughout the integration range of the Boltzmann equation because decoupling will occur before a significant change in $g^\ast$ and $h_\tx{eff}$.

To obtain relic density analytically, we need to integrate the Boltzmann equation from $x=0$ to $x_{\infty} := M_{\chi}/{T_0}$
where $T_0$ is the present photon temperature of the \cred{Universe} ($T_0 \sim 10^{-4}\,\tx{eV}$).
We also have a boundary condition that is imposed $Y_\tx{eq}=Y(x=0)$, but practically,
this is outside of the range of validity of the non-relativistic expressions used.
In our analysis we take $Y_\tx{eq} = Y\paren{x=1}$ and integrate from $x=1$ to $x = 10^3$.
We have verified there is no strong dependence on $x$ for small $x$ for the lower limit of the integration. 
We also check that $x=10^3$ is sufficiently large, beyond which the changes in relic density \cred{are} insignificant.
\\
At $x=10^3$, we obtain the final abundance of dark matter, which can be plugged into the following formula
to get the present dark matter relic density~\cite{Edsjo:1997bg},
\al{
\label{eq:relic}
    \Omega h^2 = 2.755\times10^8 \times  {M_{\chi} \ov \tx{GeV}}  \times Y_\infty,
}
where $Y_\infty$ is defined as $Y\fn{x = 10^3}$.

\subsubsection{Lepton-portal contributions}

In our scenario, where one of the heavy Majorana neutrinos is a thermal dark matter particle $\chi$,
through the Yukawa interaction in Eq.~\eqref{eq:Yukawa_compact},
the pair-annihilation channels $\chi\chi\rightarrow l_\alpha {\ol{l_\beta}}$ and $\chi\chi\rightarrow \nu_\alpha {\ol{\nu_\beta}}$
$(\alpha, \beta = 1,2,3)$ contribute.
We get the following scattering amplitudes contributing to the relic abundance in the freeze-out scenario;
${\cal M}_t$ and ${\cal M}_u$ are the total $t$-channel and $u$-channel amplitudes, respectively.
\al{
\label{eq:t_channel}
{4 \ol{\ab{{\cal M}^{\paren{\tx{L}}}_t}^2}}
	=&\sum_{\alpha,\beta=e,\mu,\tau} 
    \sqbr{\paren{t-M_\chi^2-m_{l_\alpha}^2 }\paren{t-M_\chi^2-m_{l_\beta}^2 }
    }
    \ab{\sum_{s^+=\eta_{1}^+, \eta_{2}^+} {\paren{{\cal Y}^{\paren{s^+}} _{\chi \alpha}}^{*} {\cal Y}^{\paren{s^+}} _{\chi \beta} \ov t-M_{s^+}^2}}^2
    \notag \\
 +&\sum_{\alpha,\beta=e,\mu,\tau} 
    \sqbr{\paren{t-M_\chi^2-m_{\nu_\alpha}^2 }\paren{t-M_\chi^2-m_{\nu_\beta}^2 } }
    \ab{\sum_{s^0 = \eta_{R_1}, \eta_{R_2}, \eta_{I_1}, \eta_{I_2}} 
    {\paren{{\cal Y}^{\paren{s^0}} _{\chi \alpha}}^{*} {\cal Y}^{\paren{s^0}} _{\chi \beta} \ov t-M_{s^0}^2}}^2,   
}
\al{
\label{eq:u_channel}
{4 \ol{{\ab{{\cal M}^{\paren{\tx{L}}}_u}^2}}}
	=&\sum_{\alpha,\beta=e,\mu,\tau} 
    \sqbr{\paren{u-M_\chi^2-m_{l_\alpha}^2 }\paren{u-M_\chi^2-m_{l_\beta}^2 }
    }
    \ab{\sum_{s^+=\eta_{1}^+, \eta_{2}^+} {\paren{{\cal Y}^{\paren{s^+}}_{\chi \alpha}}^* {\cal Y}^{\paren{s^+}}_{\chi \beta} \ov u-M_{s^+}^2}}^2 \notag \\
 +&\sum_{\alpha,\beta=e,\mu,\tau} 
    \sqbr{\paren{u-M_\chi^2-m_{\nu_\alpha}^2 }\paren{u-M_\chi^2-m_{\nu_\beta}^2 } }
    \ab{\sum_{s^0 = \eta_{R_1}, \eta_{R_2}, \eta_{I_1}, \eta_{I_2}} 
    {\paren{{\cal Y}^{\paren{s^0}}_{\chi \alpha}}^* {\cal Y}^{\paren{s^0}}_{\chi \beta} \ov u-M_{s^0}^2}}^2,    
}
\al{
\label{eq:tu_channel}
&{
	4 \paren{\ol{ {\cal M}^{\paren{\tx{L}}\dagger}_{t} {\cal M}^{\paren{\tx{L}}}_{u}  + {\cal M}^{\paren{\tx{L}}\dagger}_{u}{\cal M}^{\paren{\tx{L}}}_{t} }}
	  }
	= \notag \\
    -&\sum_{\alpha,\beta=e,\mu,\tau} 2
    \paren{s-m_{l_\alpha}^2-m_{l_\beta}^2 }M_\chi^2 \, \tx{Re}
    \Bigg[ \sum_{s_1^+=\eta_{1}^+, \eta_{2}^+} 
    \paren{{{\cal Y}^{\paren{s_1^+}}_{\chi \alpha} \paren{{\cal Y}^{\paren{s_1^+}}_{\chi \beta}}^* \ov t-M_{s_1^+}^2}}\sum_{s_2^+=\eta_{1}^+, \eta_{2}^+}
    \paren{{\paren{{\cal Y}^{\paren{s_2^+}}_{\chi \alpha}}^* {\cal Y}^{\paren{s_2^+}}_{\chi \beta} \ov u-M_{s_2^+}^2}} \Bigg] \notag \\
 -&\sum_{\alpha,\beta=e,\mu,\tau} 2
    \paren{s-m_{\nu_\alpha}^2-m_{\nu_\beta}^2 }M_\chi^2 \, \tx{Re}
   \Bigg[\sum_{s_1^0 = \eta_{R_1}, \eta_{R_2}, \atop \eta_{I_1}, \eta_{I_2}}
    \paren{{{\cal Y}^{\paren{s_1^0}}_{\chi \alpha} \paren{{\cal Y}^{\paren{s_1^0}}_{\chi \beta}}^* \ov t-M_{s_1^0}^2}}
   \sum_{s_2^0 = \eta_{R_1}, \eta_{R_2}, \atop \eta_{I_1}, \eta_{I_2}}
    \paren{{\paren{{\cal Y}^{\paren{s_2^0}}_{\chi \alpha}}^* {\cal Y}^{\paren{s_2^0}}_{\chi \beta} \ov u-M_{s_2^0}^2}
    }\Bigg],
}
where {$\tx{Re}[X]$} represents the real part of $X$, and we introduced the notation about the suffix for the Yukawa couplings in Eq.~\eqref{eq:Yukawa_compact},
\al{
	{\cal Y}^{\paren{s}}_{\chi \alpha}
	&:=
	\tx{the corresponding one from }
	{\cal Y}^{\paren{s}}_{A_1 \alpha}, \, {\cal Y}^{\paren{s}}_{A_2 \alpha}, \, {\cal Y}^{\paren{s}}_{A_3 \alpha}, \,
	{\cal Y}^{\paren{s}}_{B_1 \alpha}, \, {\cal Y}^{\paren{s}}_{B_2 \alpha}, \, {\cal Y}^{\paren{s}}_{B_3 \alpha}.
	\label{eq:Yukawa-convention-for-lepton-portal}
}

In the case of coannihilation, there are additional diagrams associated with the co-annihilating particle; we use the convention $\chi_1$ for the DM candidate and $\chi_2$ for the co-annihilating particle.
In addition to the $t$-channel diagrams for $\chi_2\chi_2\rightarrow l_\alpha \ol{l_\beta}$ that is added to dark matter annihilation, 
we will also have $\chi_1\chi_2\rightarrow l_\alpha \ol{l_\beta}$ as described above.
The $u$-channel and mixed interaction only contribute to
$\chi_i\chi_i\rightarrow l_\alpha \ol{l_\beta}$ since $\chi_1$ and $\chi_2$ are not identical particles.
The corresponding amplitudes squared for the coannihilation are written down as
\al{
\label{eq:t_channel_co}
{ 4 \paren{ \ol{\ab{{\cal M}^{\paren{\tx{L,co}}}_t}^2}}_{ij}}
	=&\sum_{\alpha,\beta=e,\mu,\tau} 
    \sqbr{\paren{t-M_{\chi_i}^2-m_{l_\alpha}^2 }\paren{t-M_{\chi_j}^2-m_{l_\beta}^2 }
    }
    \ab{\sum_{s^+=\eta_{1}^+, \eta_{2}^+} {\paren{{\cal Y}^{\paren{s^+}} _{\chi_i \alpha}}^{*} {\cal Y}^{\paren{s^+}} _{\chi_j \beta} \ov t-M_{s^+}^2}}^2
    \notag \\
 +&\sum_{\alpha,\beta=e,\mu,\tau} 
    \sqbr{\paren{t-M_{\chi_i}^2-m_{\nu_\alpha}^2 }\paren{t-M_{\chi_j}^2-m_{\nu_\beta}^2 } }
    \ab{\sum_{s^0 = \eta_{R_1}, \eta_{R_2}, \eta_{I_1}, \eta_{I_2}} 
    {\paren{{\cal Y}^{\paren{s^0}} _{\chi_i\alpha}}^{*} {\cal Y}^{\paren{s^0}} _{\chi_j \beta} \ov t-M_{s^0}^2}}^2,   
}
\al{
\label{eq:u_channel_co}
{ 4 \paren{\ol{\ab{{\cal M}^{\paren{\tx{L,co}}}_u}^2}}_{ii}}
	=&\sum_{\alpha,\beta=e,\mu,\tau} 
    \sqbr{\paren{u-M_{\chi_i}^2-m_{l_\alpha}^2 }\paren{u-M_{\chi_i}^2-m_{l_\beta}^2 }
    }
    \ab{\sum_{s^+=\eta_{1}^+, \eta_{2}^+} {\paren{{\cal Y}^{\paren{s^+}}_{{\chi_i} \alpha}}^* {\cal Y}^{\paren{s^+}}_{{\chi_i} \beta} \ov u-M_{s^+}^2}}^2 \notag \\
 +&\sum_{\alpha,\beta=e,\mu,\tau} 
    \sqbr{\paren{u-M_{\chi_i}^2-m_{\nu_\alpha}^2 }\paren{u-M_{\chi_i}^2-m_{\nu_\beta}^2 } }
    \ab{\sum_{s^0 = \eta_{R_1}, \eta_{R_2}, \eta_{I_1}, \eta_{I_2}} 
    {\paren{{\cal Y}^{\paren{s^0}}_{{\chi_i} \alpha}}^* {\cal Y}^{\paren{s^0}}_{{\chi_i} \beta} \ov u-M_{s^0}^2}}^2,    
}
\al{
\label{eq:tu_channel_co}
& {4 \paren{ \ol{{\cal M}^{\paren{\tx{L,co}}\dagger}_{t} {\cal M}^{\paren{\tx{L,co}}}_{u} + {\cal M}^{\paren{\tx{L,co}}\dagger}_{u}{\cal} M^{\paren{\tx{L,co}}}_{t}}}_{ii}}
	= \notag \\
    -&\sum_{\alpha,\beta=e,\mu,\tau} 2
    \paren{s-m_{l_\alpha}^2-m_{l_\beta}^2 }M_{\chi_i}^2 \, \tx{Re}
    \Bigg[ \sum_{s_1^+=\eta_{1}^+, \eta_{2}^+} 
    \paren{{{\cal Y}^{\paren{s_1^+}}_{\chi_i \alpha} \paren{{\cal Y}^{\paren{s_1^+}}_{\chi_i \beta}}^* \ov t-M_{s_1^+}^2}}
    \sum_{s_2^+=\eta_{1}^+, \eta_{2}^+}
    \paren{{\paren{{\cal Y}^{\paren{s_2^+}}_{\chi_i \alpha}}^* {\cal Y}^{\paren{s_2^+}}_{\chi_i \beta} \ov u-M_{s_2^+}^2}}\Bigg] \notag \\
 -&\sum_{\alpha,\beta=e,\mu,\tau} 2
    \paren{s-m_{\nu_\alpha}^2-m_{\nu_\beta}^2 }M_{\chi_i}^2 \, \tx{Re}
    \Bigg[\sum_{s_1^0 = \eta_{R_1}, \eta_{R_2}, \atop \eta_{I_1}, \eta_{I_2}}
    \paren{{{\cal Y}^{\paren{s_1^0}}_{\chi_i \alpha} \paren{{\cal Y}^{\paren{s_1^0}}_{\chi_i \beta}}^* \ov t-M_{s_1^0}^2}}
    \sum_{s_2^0 = \eta_{R_1}, \eta_{R_2}, \atop \eta_{I_1}, \eta_{I_2}}
    \paren{{\paren{{\cal Y}^{\paren{s_2^0}}_{\chi_i \alpha}}^* {\cal Y}^{\paren{s_2^0}}_{\chi_i \beta} \ov u-M_{s_2^0}^2}}\Bigg].
}

\subsubsection{Higgs-portal contributions}

In our scenario, the six heavy right-handed neutrinos $N_{A_i}$ and $N_{B_i}$ couple with the singlet scalar $S$, and
after the spontaneous symmetry breaking with a nonzero VEV for $S$, the residual $Z_2$ symmetry remains, and
the quantum fluctuation of $S$ ($s'$) mixes with the physical Higgs field $h'$ [refer to Eqs.~\eqref{eq:multiplet-components}, \eqref{eq:singlet-scalar-components} and \eqref{eq:scalar-VEV-configuration}].
Therefore, $\chi \chi \to \sqbr{\tx{a pair of SM particles}}$ is also possible as a Higgs portal.

We introduce the following compact notation for clarity to the relevant part of our calculation:
\al{\label{eq:higgs_yukawa}
-{\cal L}_\tx{Higgs Yukawa}
	\supset
		\sum_{i = 1}^{3} \sum_{h^0 = h, s}
		\Bigg[
			&{\cal Y}^{\paren{h^0}}_{A_i } \, \ol{\paren{{N}_{A_i R}}^\tx{c}} \, {N}_{A_i R} {h^0} +
			{\cal Y}^{\paren{h^0}}_{B_i } \, \ol{\paren{{N}_{B_i R}}^\tx{c}} \, {N}_{B_i R} {h^0} \notag \\
            & +
            {\cal Y}^{\paren{h^0}}_{AB_i } \, \ol{\paren{{N}_{B_i R}}^\tx{c}} \, {N}_{A_i R} {h^0} +
			{\cal Y}^{\paren{h^0}}_{AB_i } \, \ol{\paren{{N}_{A_i R}}^\tx{c}} \, {N}_{B_i R} {h^0}\Bigg],
}
with 
\al{
{\cal Y}^{\paren{h}}_{A_i } 
	&=
		{y_{A_{ii}}\ov 2 } c_{N_i}^2 s_{\alpha} + {y_{B_{ii}}\ov 2} s_{N_i}^2 s_{\alpha},&
{\cal Y}^{\paren{h}}_{B_i } 
	&=
		{y_{A_{ii}}\ov 2 } s_{N_i}^2 s_{\alpha} + {y_{B_{ii}}\ov 2} c_{N_i}^2 s_{\alpha},& \notag \\
{\cal Y}^{\paren{s}}_{A_i } 
	&=
		{y_{A_{ii}}\ov 2 } c_{N_i}^2 c_{\alpha} + {y_{B_{ii}}\ov 2} s_{N_i}^2 c_{\alpha},&
{\cal Y}^{\paren{s}}_{B_i } 
	&=
		{y_{A_{ii}}\ov 2 } s_{N_i}^2 c_{\alpha} + {y_{B_{ii}}\ov 2} c_{N_i}^2 c_{\alpha},& \notag \\
{\cal Y}^{\paren{h}}_{AB_i } 
	&=
		{y_{A_{ii}}\ov 2 } s_{N_i}c_{N_i} s_{\alpha} - {y_{B_{ii}}\ov 2} s_{N_i}c_{N_i} s_{\alpha},&
{\cal Y}^{\paren{s}}_{AB_i } 
	&=
		{y_{A_{ii}}\ov 2 } s_{N_i}c_{N_i} c_{\alpha} + {y_{B_{ii}}\ov 2} s_{N_i}c_{N_i} c_{\alpha},&
	\label{eq:Higgs-portal-couplings-1}
}
where the six Yukawa couplings $y_{A_{ii}}$ and $y_{B_{ii}}$ {($i=1,2,3$)} do not appear in the active-neutrino mass formula and the lepton flavour violation,
as shown in Eqs.~\eqref{eq:active-neutrino-mass-form-1}, \eqref{eq:active-neutrino-mass-form-2} and \eqref{eq:formula-LFV}.
So, these couplings are not constrained through these phenomena and play an important role in the Higgs-portal freeze-out DM scenario.
{Note that, in our current parametrisation, the six effective couplings in Eq.~\eqref{eq:Higgs-portal-couplings-1} are real.}


The following part of the SM Higgs couplings is relevant for our study,
\al{
\label{eq:higgs_SM}
\L_\tx{Higgs, SM} 
	= 
		\sum_{h^0 = h, s}
                  {h^0}
                  \sqbr{ {\cal C}^{\paren{h^0}}_{W} W^+_\mu W^{-\mu} 
                        + {\cal C}^{\paren{h^0}}_{Z} {1\ov 2} Z_\mu Z^\mu }
                  - {\cal Y}^{\paren{h^0}}_{f} \sum_{h^0 = h, s} {h^0}\ol{f}f \,,
}
with 
\al{
{\cal C}^{\paren{h}}_{W}
	&=
		{2M_W^2 c_\alpha \ov v_H },&
{\cal C}^{\paren{h}}_{Z}
	&=
		{2M_Z^2 c_\alpha \ov v_H },& 
{\cal Y}^{\paren{h}}_{f}
	&=
		{ {M_f} c_\alpha \ov v_H },& \notag \\
{\cal C}^{\paren{s}}_{W}
	&=
		-{2M_W^2 s_\alpha \ov v_H },&
{\cal C}^{\paren{s}}_{Z}
	&=
		-{2M_Z^2 s_\alpha \ov v_H },&
{\cal Y}^{\paren{s}}_{f}
	&=
		-{ {M_f} s_\alpha \ov v_H },
	\label{eq:Higgs-portal-couplings-2}
}
where $M_f$ is the mass of a SM fermion $f$.\footnote{
{In the following analysis in \cred{Section}~\ref{sec:Analysis}, we will consider the top, bottom quarks (if kinematically accepted) as the final state $f$ in our numerical scans for the DM mass $\gtrsim\, 10^2\,\tx{GeV}$.}
}

Under the interaction in Eq.~\eqref{eq:higgs_yukawa} and Eq.~\eqref{eq:higgs_SM},
we get the $s$-channel scattering amplitudes as follows,
\al{
{ 4 \ol{\ab{ {\cal M}_s^{\paren{\tx{H}}} }^2}}
	&=
		\sum_f 16 \sqbr{ \paren{s-4 M_\chi^2} \paren{s - 4 {M_f^2}} }
		\left| \sum_{h^0 = h,s} {  {\cal Y}^{\paren{h^0}}_{\chi} {\cal Y}^{\paren{h^0}}_{f} \ov {s - M_{h^0}^2} 
		{+ i M_{h^0} \Gamma_{h^0}}   } \right|^2 \notag \\
	&\quad
		+8 \sqbr{ 
			\paren{s-4M_\chi^2}  {\paren{ 2 - { \paren{s-2M_W^2}^2 \ov 4M_W^4 } }} 	
			}
		\left| \sum_{h^0 = h,s} {  {\cal Y}^{\paren{h^0}}_{\chi} {\cal C}^{\paren{h^0}}_{W} \ov {s - M_{h^0}^2} {+ i M_{h^0} \Gamma_{h^0}}  } \right|^2 \notag \\
	&\quad
		+8 \sqbr{ \paren{s-4M_\chi^2} 
			{   
				\paren{ 2 - { \paren{s-2M_Z^2}^2 \ov 4M_Z^4 } } } 
			}
		\left| \sum_{h^0 = h,s} {  {\cal Y}^{\paren{h^0}}_{\chi} {\cal C}^{\paren{h^0}}_{Z} \ov {s - M_{h^0}^2} {+ i M_{h^0} \Gamma_{h^0}}  } \right|^2,
}
where $f$ is summed over all SM fermions (coupled with the Higgs boson) and $\Gamma_{h^0}$ represents the decay width of $h^0$.\footnote{
We will provide the calculation of $\Gamma_{s}$ in Appendix~\ref{sec:decay-s}.
}
The first term represents dark-matter pair annihilations into SM fermions, and the second and third terms represent those to SM massive gauge bosons $W$ and $Z$, respectively.

In addition to the $s$-channel diagrams, we include $t$-channel and $u$-channel diagrams for $\chi \chi \to hh$,
mediated via the Majorana fermions; this can be the dark matter particles and heavier ones.
The following forms will be used for our numerical calculations,
\al{
{4 \ol{\ab{{\cal M}_t ^{\paren{\tx{H}}}}^2}_{11}}
	&= 
		4\sum_{l,m=1,2}  \Bigg[ {1\ov2} \paren{M_{{\chi_1}}^2 +M_{{\chi_1}}M_{\chi_l} +M_{{\chi_1}}M_{\chi_m} } \paren{ s + u - t -4M_{{\chi_1}}^2} + {1\ov 2}M_{\chi_l}M_{\chi_m} \paren{ s  -4M_{{\chi_1}}^2} \notag \\
+&{1\ov 2} M_{{\chi_1}}^2 \paren{u-t} 
-{   M_{h}^4\ov 2} + {1\ov 2}\paren{M_{{\chi_1}}^2-t}\paren{M_{{\chi_1}}^2-u}  \Bigg]
{
	\paren{ {{\cal Y}^{\paren{h}}_{\chi_l {\chi_1} }  {\cal Y}^{\paren{h}}_{\chi_l {\chi_1} }\ov {t-M_{\chi_l}^2}} }
	\paren{ { {\cal Y}^{\paren{h}}_{ {\chi_1} \chi_m}  {\cal Y}^{\paren{h}}_{{\chi_1}\chi_m }\ov {t-M_{\chi_m}^2}}} 
}\,,\\
{4 \ol{\ab{{\cal M}_u ^{\paren{\tx{H}}}}^2}_{11}}
	&= 
		4\sum_{l,m=1,2} \Bigg[ {1\ov2} \paren{M_{{\chi_1}}^2 +M_{{\chi_1}}M_{\chi_l} +M_{{\chi_1}}M_{\chi_m} } \paren{ s + t - u -4M_{{\chi_1}}^2} + {1\ov 2}M_{\chi_l}M_{\chi_m} \paren{ s  -4M_{{\chi_1}}^2} \notag\\
+&{1\ov 2}M_{{\chi_1}}^2 \paren{ t-u}
-{   M_{h}^4\ov 2} + {1\ov 2}\paren{M_{{\chi_1}}^2-t}\paren{M_{{\chi_1}}^2-u}  \Bigg]
{
	\paren{ {{\cal Y}^{\paren{h}}_{\chi_l {\chi_1} }  {\cal Y}^{\paren{h}}_{\chi_l {\chi_1} }\ov {u-M_{\chi_l}^2}} } 
	\paren{{{\cal Y}^{\paren{h}}_{ {\chi_1} \chi_m}  {\cal Y}^{\paren{h}}_{ {\chi_1} \chi_m }\ov {u-M_{\chi_m}^2}}} 
}\,,
}
\al{
{4 \Bigg( \ol{\paren{{\cal M}_{t}^{\paren{\tx{H}}}}^\dagger{\cal M}_{u}^{\paren{\tx{H}}}}}
	&+ {\ol{\paren{{\cal M}_{u}^{\paren{\tx{H}}}}^\dagger{\cal M}_{t}^{\paren{\tx{H}}}} \Bigg)_{11}}
	= -4 \sum_{l,m=1,2} \Bigg[- M_h^4 +\paren{M_{{\chi_1}}^2-t}\paren{M_{{\chi_1}}^2-u}
\notag\\
 &- M_{{\chi_1}} \paren{M_{\chi_l}+M_{\chi_m}}\paren{u-t}-\paren{s-4M_{{\chi_1}}^2}\paren{M_{{\chi_1}}^2-M_{{\chi_1}} M_{\chi_l} + M_{{\chi_1}}M_{\chi_m} - M_{\chi_l}M_{\chi_m}}
 \Bigg]\notag\\
 &{\times} \sqbr{
 	\paren{ { {\cal Y}^{\paren{h}}_{\chi_l {\chi_1} }  {\cal Y}^{\paren{h}}_{\chi_l {\chi_1} }\ov {t-M_{\chi_l}^2}} } 
 	\paren{{ {\cal Y}^{\paren{h}}_{ {\chi_1} \chi_m}   {\cal Y}^{\paren{h}}_{ {\chi_1} \chi_m }\ov {u-M_{\chi_m}^2}}}  
	}\,,
}
where we put {the condition} that only the second-lightest Majorana fermion $\chi_{i=2} \, \paren{= \chi_2}$ and the lightest as a dark matter $\chi_{i=1} \,\paren{= \chi_1 = \chi}$ are taken into account.
Also, we introduced similar notations for the Yukawa couplings as for the lepton-portal channels in Eq.~\eqref{eq:Yukawa-convention-for-lepton-portal} [refer to Eqs.~\eqref{eq:Higgs-portal-couplings-1} and \eqref{eq:Higgs-portal-couplings-2}],
\al{
{\cal Y}^{\paren{h}}_{\chi_1}
	&:=
	\tx{the corresponding one from }
	{\cal Y}^{\paren{h}}_{A_1}, \, {\cal Y}^{\paren{h}}_{A_2}, \, {\cal Y}^{\paren{h}}_{A_3}, \,
	{\cal Y}^{\paren{h}}_{B_1}, \, {\cal Y}^{\paren{h}}_{B_2}, \, {\cal Y}^{\paren{h}}_{B_3}. 
	\label{eq:definition-Yhchione} \\
{\cal Y}^{\paren{h}}_{\chi_1\chi_2} {= {\cal Y}^{\paren{h}}_{\chi_2\chi_1}}
	&:=
	\tx{the corresponding one from }
	{\cal Y}^{\paren{h}}_{AB_1}, \, {\cal Y}^{\paren{h}}_{AB_2}, \, {\cal Y}^{\paren{h}}_{AB_3}.
	\label{eq:Yukawa-convention-for-Higgs-portal}
}


Similar to the \cred{lepton} portal, we ought to consider additional diagrams involved in the coannihilation processes in the Higgs-portal scenario, too;
$\chi_2 \chi_2 \rightarrow f \bar{f}$, $\chi_1 \chi_2 \rightarrow f \bar{f}$ as well as $\chi_2 \chi_2 \rightarrow b \bar{b}$, $\chi_1 \chi_2 \rightarrow b \bar{b}$ contribute to the relic in the co-annihilation scenario.
Here, $f$ and $b$ are the standard model \cred{fermions and bosons}, respectively. 
\al{
\label{eq:s_channel_co}
{4 \ol{\ab{{\cal M}^{(\tx{H,co})}_s}_{22}^2}}
	=&
	\sum_{f}  
    16\sqbr{\paren{s-4M_{\chi_2}^2 }\paren{s-4m_{f}^2 }
    }
    \ab{\sum_{h^0=h, s} {{\cal Y}^{\paren{h^0}}_{\chi_2 } {\cal Y}^{\paren{h^0}}_{f} \ov s-M_{h^0}^2 {+ i M_{h^0} \Gamma_{h^0}} }}^2
    \notag\\
 +& 
    8\sqbr{\paren{s-4M_{\chi_2}^2 } {\paren{2 - {{\paren{s-2M_W^2} }^2 \ov 4M_W^4 }}} }
    \ab{\sum_{h^0=h, s} 
    {{\cal Y}^{\paren{h^0}}_{\chi_2} {\cal C}^{\paren{h^0}}_{W} \ov s-M_{h^0}^2 {+ i M_{h^0} \Gamma_{h^0}} }}^2 \notag\\
 +& 
    8\sqbr{\paren{s-4M_{\chi_2}^2 } {\paren{2 - {{\paren{s-2M_Z^2} }^2 \ov 4M_Z^4 }}} }
    \ab{\sum_{h^0=h, s} 
    {{\cal Y}^{\paren{h^0}}_{\chi_2} {\cal C}^{\paren{h^0}}_{Z} \ov s-M_{h^0}^2 {+ i M_{h^0} \Gamma_{h^0}}}}^2 ,\\
{4 \ol{\ab{{\cal M}^{(\tx{H,co})}_s}_{12}^2}}
	=&
	\sum_{f}  
    16\sqbr{\paren{s-\paren{M_{\chi_1}+M_{\chi_2}}^2 }\paren{s-4m_{f}^2 }
    }
    \ab{\sum_{h^0=h, s} {{\cal Y}^{\paren{h^0}}_{\chi_{1} \chi_{2}} {\cal Y}^{\paren{h^0}}_{f} \ov s-M_{h^0}^2 {+ i M_{h^0} \Gamma_{h^0}}}}^2
    \notag\\
 +& 
    8\sqbr{\paren{s-\paren{M_{\chi_1}+M_{\chi_2}}^2 } {\paren{2 - {{\paren{s-2M_W^2} }^2 \ov 4M_W^4 }}} }
    \ab{\sum_{h^0=h, s} 
    {{\cal Y}^{\paren{h^0}}_{\chi_{1} \chi_{2}} {\cal C}^{\paren{h^0}}_{W} \ov s-M_{h^0}^2 {+ i M_{h^0} \Gamma_{h^0}} }}^2 \notag\\
 +& 
    8\sqbr{\paren{s-\paren{M_{\chi_1}+M_{\chi_2}}^2 } {\paren{2 - {{\paren{s-2M_Z^2} }^2 \ov 4M_Z^4 }}} }
    \ab{\sum_{h^0=h, s} 
    {{\cal Y}^{\paren{h^0}}_{\chi_{1} \chi_{2}} {\cal C}^{\paren{h^0}}_{Z} \ov s-M_{h^0}^2 {+ i M_{h^0} \Gamma_{h^0}} }}^2,
}
where ${\cal Y}^{\paren{h^0}}_{\chi_2}$ is defined as similar as ${\cal Y}^{\paren{h^0}}_{\chi_1}$ in Eq.~\eqref{eq:definition-Yhchione}.

In the co-annihilation scenario, we replace the mass difference between the two species with  $\delta M_{ij}=  \ab{M_{\chi_j}-M_{\chi_i}}$, where $i=1,2$ and $j=1,2$, and $\delta M_{ii}=0$. 
\al{
{4\overline{\ab{{\cal M}_t ^{\paren{\tx{H,co}}}}_{ij}^2}} 
	&= 2\sum_{l=1,2}\sum_{m=1,2}\Bigg[ -M_{h}^2\paren{s-\delta M_{ij}^2-\paren{2M_{\chi_i}+M_{\chi_l}+M_{\chi_m}}\delta M_{ij}}
\notag\\
& {+} \paren{s-M_{\chi_i}^2-M_{\chi_j}^2}\paren{
M_{\chi_i}^2 +M_{\chi_i}M_{\chi_l}+M_{\chi_i}M_{\chi_m}+M_{\chi_l}M_{\chi_m}
} \notag\\
&+\paren{2M_{\chi_i}+M_{\chi_l}+M_{\chi_m}}\paren{M_{\chi_j}\paren{M_{\chi_i}^{2}-t}-M_{\chi_i}\paren{M_{\chi_j}^{2}-u}}
\notag\\
&+ {1\ov 2} \paren{M_h^2 +M_{\chi_i}^2 -t}\paren{M_h^2 +M_{\chi_j}^2 -u}\Bigg]
{
\paren{ {\paren{{\cal Y}^{\paren{h}}_{\chi_l \chi_i } } {\cal Y}^{\paren{h}}_{\chi_l\chi_j }\ov {t-M_{\chi_l}^2}} }
\paren{{\paren{{\cal Y}^{\paren{h}}_{ \chi_i \chi_m} } {\cal Y}^{\paren{h}}_{\chi_j\chi_m }\ov {t-M_{\chi_m}^2}}},
} \\
{4\overline{\ab{{\cal M}_u ^{\paren{\tx{H,co}}}}_{22}^2}} 
	&= 4\sum_{l=1,2}\sum_{m=1,2}  \Bigg[ {1\ov2} \paren{M_{{\chi_2}}^2 +M_{{\chi_2}}M_{\chi_l} +M_{{\chi_2}}M_{\chi_m} } \paren{ s + t - u -4M_{{\chi_2}}^2} \notag\\
	+& {1\ov 2}M_{\chi_l}M_{\chi_m} \paren{ s  -4M_{{\chi_2}}^2} + {1\ov 2}M_{{\chi_2}}^2 \paren{t-u} - {M_{h}^4\ov 2} \notag\\ \hspace{-1cm}
	+& {1\ov 2}\paren{M_{{\chi_2}}^2-t}\paren{M_{{\chi_2}}^2-u}  \Bigg]
{
\paren{ {\paren{{\cal Y}^{\paren{h}}_{\chi_l {\chi_2} } } {\cal Y}^{\paren{h}}_{\chi_l{\chi_2} }\ov {u-M_{\chi_l}^2}} } 
\paren{{\paren{{\cal Y}^{\paren{h}}_{ {\chi_2} \chi_m} } {\cal Y}^{\paren{h}}_{{\chi_2}\chi_m }\ov {u-M_{\chi_m}^2}}},
} \\
{  4\Bigg(\overline{\paren{{\cal M}_{t}^{\paren{\tx{H,co}}}}^\dagger{\cal M}_{u}^{\paren{\tx{H,co}}}}  } &
	{  + \overline{\paren{{\cal M}_{u}^{\paren{\tx{H,co}}}}^\dagger{\cal M}_{t}^{\paren{\tx{H,co}}} }\Bigg)_{22}  }
	= -4\sum_{l=1,2}\sum_{m=1,2}\Bigg[- M_h^4 +\paren{M_{{\chi_2}}^2-t}\paren{M_{{\chi_2}}^2-u}
\notag\\
 &- M_{{\chi_2}} \paren{M_{\chi_l}+M_{\chi_m}}\paren{u-t}-\paren{s-4M_{{\chi_2}}^2}\paren{M_{{\chi_2}}^2-M_{{\chi_2}}M_{\chi_l}+M_{{\chi_2}}M_{\chi_m}-M_{\chi_l}M_{\chi_m}}
 \Bigg]\notag\\
&\times
{
\paren{ {\paren{{\cal Y}^{\paren{h}}_{\chi_l {\chi_2} } } {\cal Y}^{\paren{h}}_{\chi_l{\chi_2} }\ov {t-M_{\chi_l}^2}} } 
\paren{{\paren{{\cal Y}^{\paren{h}}_{ {\chi_2} \chi_m} } {\cal Y}^{\paren{h}}_{{\chi_2}\chi_m }\ov {u-M_{\chi_m}^2}}}.
}
}

Note that the other possible processes with $s$ in the final state, $\chi+\chi \rightarrow h + s$ and $\chi+\chi \rightarrow s+s$.
As we will discuss, we focus on the kinematic region in $2 M_\chi \lesssim M_s$, where the two processes are kinematically suppressed and may not contribute significantly. Therefore, we ignore them.

\subsection{Constraints via Direct Detection}

\subsubsection{Lepton-portal contributions}

We define $\A_S$ as the \cred{anapole} moment; it is the effective coupling between a photon and 
the dark matter candidate (Majorana fermion)~\cite{Kopp:2014tsa} via a charged \cred{scalar-SM} lepton loop.
\al{
\A_S &=   -{e \ov 96 \pi^2 M_\chi ^2} \sum_{s^+=\eta_{1}^+, \eta_{2}^+}
        \sum_{\alpha=\mu,\tau} \ab{{\cal Y}^{\paren{s^+}}_{\chi \alpha}}^2 \F\paren{{m_{l_\alpha}\ov M_\chi},{{M_{s^+}}\ov M_\chi}}\\
        &\quad -{e \ov 32 \pi^2 M_\chi ^2} \sum_{s^+=\eta_{1}^+, \eta_{2}^+}
         \ab{{\cal Y}^{\paren{s^+}}_{\chi e}}^2 \F_e\paren{{\sqrt{|q|^2}\ov M_{\chi}},{{M_{s^+}}\ov M_\chi}}\, .\label{eq:anapole_moment_new}
}
This calculation assumes the mass of the electron to be negligible; hence $\alpha$ is only summed over $\mu$ and $\tau$, the second line in Eq.~\eqref{eq:anapole_moment_new} is the electron loop contribution.
The formula holds only when {momentum transfer from dark matter to the target nucleus $q^2$} is greater than the lepton mass {squared ($ q^2 > m^2_{l}$)}, as in our case.
We use shorthand notations, $\mu := {{M_{s^+}}/M_\chi} $ and $\epsilon := {m_{l_\alpha}/M_\chi}$ below,
\al{
\F\paren{\mu, \epsilon}
	&= 
		{3 \ov 2} \log\paren{\mu^2 \ov \epsilon^2} 
		- \frac{1 + 3\mu^2 - 3\epsilon^2}{\sqrt{\left(\mu^2 - 1 - \epsilon^2\right)^2 - 4\epsilon^2}} 
		\operatorname{Arctanh}\left(\frac{\sqrt{\left(\mu^2 - 1 - \epsilon^2\right)^2 - 4\epsilon^2}}{\mu^2 - 1 + \epsilon^2}\right), \\
\F_e\paren{{\sqrt{|q|^2}\ov M_{\chi}}, \mu} 
	&= 
		\frac{-10 + 12 \log\left({\sqrt{|q|^2}\ov M_{\chi}}\right) - \left(3 + 9\mu^2\right) \log\left(\mu^2 - 1\right) 
		- \left(3 - 9\mu^2\right) \log\left(\mu^2\right)}{9 \left(\mu^2 - 1\right)},
}
We quote (the relevant part of) the spin-independent differential \cred{cross section} of dark matter scattering off a nucleus having atomic
number $Z$ formula from~\cite{Kopp:2014tsa},\footnote{
Note that the full form of Eq.~\eqref{eq:anapolediffcross_new} is as follows,
\als{
{d\sigma^\tx{L} _{\chi } \ov dE_R} 
	=
		4\A_S ^2\sqbr{\alpha_{\tx{EM}} Z^2\paren{2m_T - \paren{1+{m_T \ov M_\chi}}^2{E_R \ov v^2}}F_Z ^2 \paren{q^2} 
		+ {d^2_A} \paren{J+1 \ov 3J} {2E_R m_T ^2 \ov \pi v^2}F_S ^2(q^2)},
}
where the second term in the cross section describes the scattering with the magnetic dipole moment of the nucleus.
This is subdominant in the \cred{xenon} detectors, which is relevant to our paper.
Hence, it is dropped going forward.
$d_A$, $J$ and $F_S\fn{q^2}$ are the nuclear dipole moment, the spin form factor, and the spin of the target nucleus.
}
\al{
\label{eq:anapolediffcross_new}
{d\sigma^\tx{L} _{\chi } \ov dE_R} 
	=
		4\A_S ^2\sqbr{\alpha_{\tx{EM}} Z^2\paren{2m_T - \paren{1+{m_T \ov M_\chi}}^2{E_R \ov v^2}}F_Z ^2 \paren{q^2} },
}
where $E_R$ is the recoil energy, $\alpha_\tx{EM}$ the fine-structure constant in Electromagnetism, $Z$ is the nuclear charge of the detector atom,  $m_T$ is the detector nucleus mass, and $v$ is the velocity of the dark matter particle relative to that of the nucleus.  
$F_Z(q)$ is the nuclear charge form factor~\cite{Kopp:2014tsa},
\al{
F_Z(q) =3 \, e^{-\frac{q^2 s^2}{2}} {\paren{ \sin(q r) - q r \cos(q r) } \ov (q r)^3} \, ,
}
where $s = 1 \, \tx{fm}$, $r = \sqrt{R^2-5s^2}$, $R = 1.2 {A}^{1/3} \, \tx{fm}$ (with the nuclear mass number $A$).
The (squared) momentum transfer from dark matter to the target nucleus $q^2$ takes
\al{
q^2 = 2m_T E_R.
}

To compare with the latest dark matter direct detection experiments~\cite{LZ:2024zvo},
we must write down the total DM-nucleon spin-independent cross section.
We can express the DM velocity relative to the nucleus in terms of nucleus recoil energy~\cite{Bai:2014osa},
\al{
    v=\sqbr{{m_T E_R \ov \mu_{\chi T}^2(1-\cos\theta)}}^{1/2},
}
where ${\mu_{\chi T}}$ is the reduced mass of the dark matter-nucleus system.
As described in \cite{Bai:2014osa}, we integrate Eq.~\eqref{eq:anapolediffcross_new} over the solid angle and the recoil energy the detector is sensitive to. In the case of the LZ detector, we integrate from $5.5$ to $55$ KeV recoil energies, in which region the detector efficiency is $>50\%$~\cite{LZ:2024zvo}

We can now compare the latest results of the LZ cross section per nucleon~\cite{LZ:2024zvo} with our calculated cross section 
\al{
\label{eq:lepton_cross_nucleon}
\sigma_{\chi p}^\tx{L} 
	= 
		{\sigma_{\chi}^\tx{L}\ov A} ,
}
where $A=129$ for Xenon.
The detector uses a mixture of isotopes of which  $\tx{Xe}^{129}$
has the largest contribution.
In Eq.~\eqref{eq:lepton_cross_nucleon}, we have used a conservative, simple way to convert our calculated total \cred{cross section} 
into \cred{cross section} per nucleon.

\subsubsection{Higgs-portal contributions}

We quote the spin-independent cross-section of dark matter scattering off a nucleon, having formula from~\cite{Arcadi:2021mag},
\al{
y_{\chi H} 
	&= 
		\begin{cases}
			\displaystyle {y_{A_{ii}}\ov 2 } c_{N_i}^2 + {y_{B_{ii}}\ov 2} s_{N_i}^2 & \text{if $\chi$ is } N_{A_i} \, (i=1,2,3), \\
			\displaystyle {y_{A_{ii}}\ov 2 } s_{N_i}^2 + {y_{B_{ii}}\ov 2} c_{N_i}^2 & \text{if $\chi$ is } N_{B_i} \, (i=1,2,3), 
		\end{cases}
	\label{eq:higgs_coupling} \\
\sigma^\tx{H}_{\chi p} 
	&= 
		{\mu^2_{\chi p} \ov \pi } {{y_{\chi H}}^2 s_\alpha^2 c_\alpha^2 m_p ^2 \ov v_H ^2}
		\F_H\paren{M_\chi\,, M_{h}\,,M_{s}\,, v}f_p^2.
	\label{eq:higgsdetection_new}
}
We define \cred{coupling} $y_{\chi H}$ between the neutral singlet $S$ (gauge basis) and DM 
candidate $\chi$ (mass basis) as is described in \cite{Arcadi:2021mag},

\al{
\F_H\paren{M_\chi, M_{H_1}\,,M_{H_2}, v} 
	&= 
		{1 \ov 4M_\chi^2 v^2} \Bigg[\sum_{i=1,2}
        		\paren{{1\ov M_{H_i}^2} - {1 \ov 4M_\chi^2 v^2 + M_{H_i}^2} }\notag \\
	&\quad
		-{2 \ov \paren{M_{H_2}^2 -M_{H_1}^2}}\sum_{i=1}^2
    		    (-1)^{i-1}\log\paren{1+{4M_\chi^2v^2 \ov M_{H_i}^2}}\Bigg], \\
\mu_{\chi p } &= {m_p M_\chi \ov m_p + M_\chi},
}
where we took $M_{H_1}=M_{h}$ and $M_{H_2}=M_{s}$.
Here, proton mass $m_p$ is assumed to be nucleon mass, $\mu_{\chi p}$ is dark matter-nucleon reduced mass. $v$ is dark matter velocity in the laboratory frame and $f_p \approx 0.326$ is \cred{the} nucleon form factor.
To compare with the results of \cred{the LZ Collaboration}~\cite{LZ:2024zvo},
we input the dark matter velocity $v$ in Eq.~\eqref{eq:higgsdetection_new} for the spin-independent cross section per nucleon.
The mean velocity of the dark matter halo is $v=220\,\tx{km/s} = 73.5\times 10^{-5}\approx 10^{-3}$ in natural units is used in our analysis.

\section{Numerical Analysis
\label{sec:Analysis}}

\begin{figure}[t]
    \centering
    \begin{minipage}{0.48\textwidth}
        \centering
        \includegraphics[width=\linewidth]{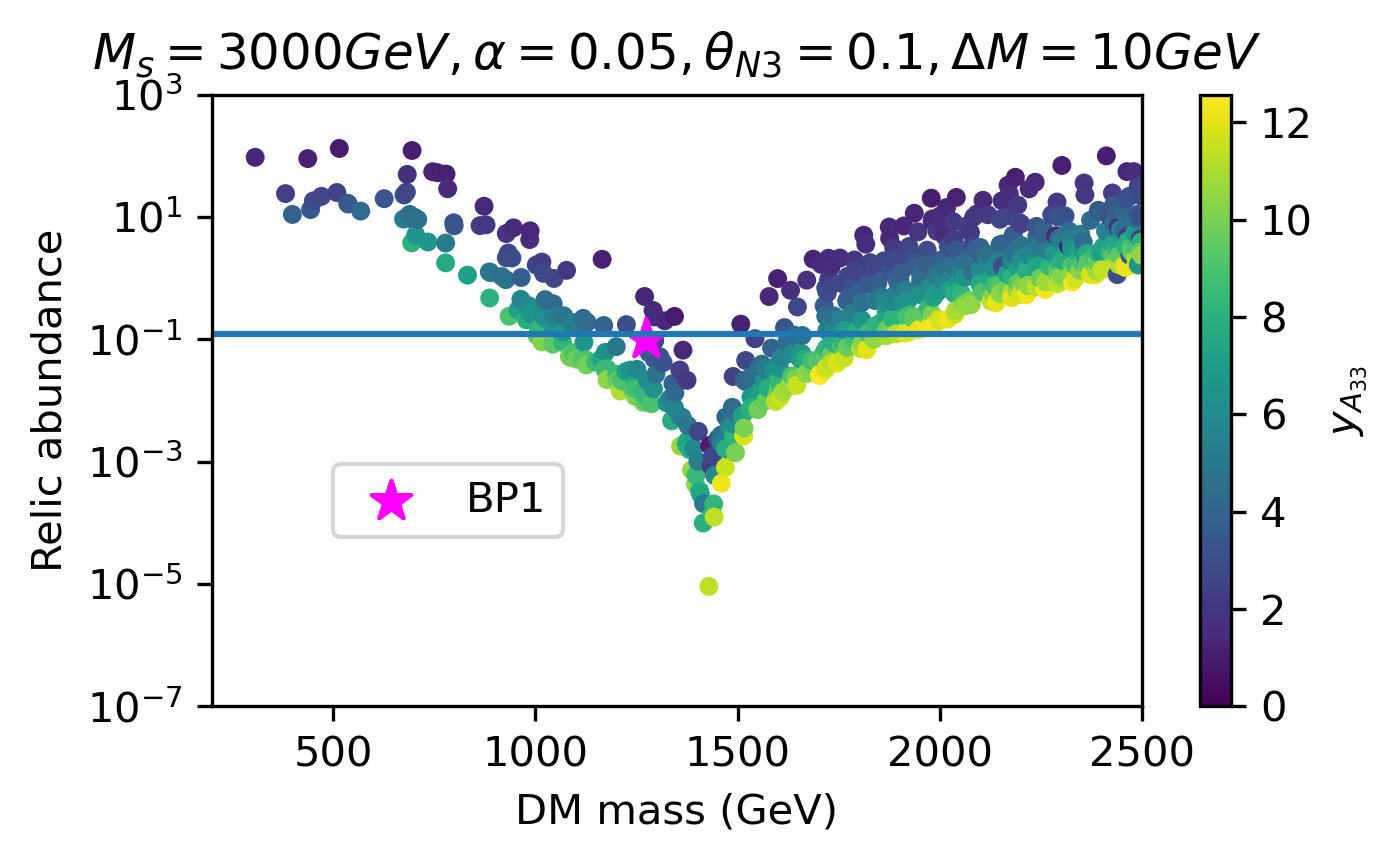} 
    \end{minipage}
    \hfill
    \begin{minipage}{0.48\textwidth}
        \centering
        \includegraphics[width=\linewidth]{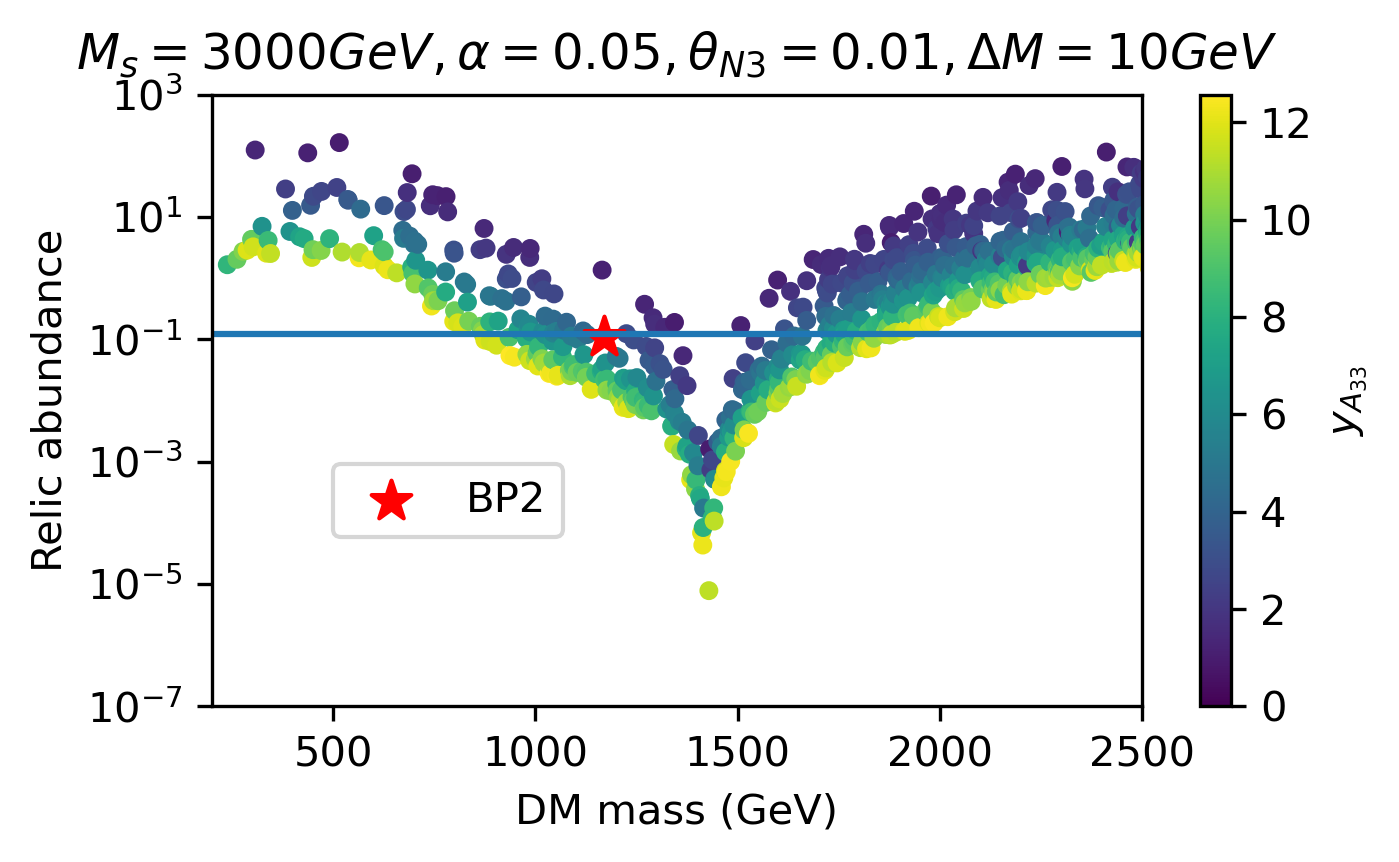} 
    \end{minipage}
    \caption{
Results of the relic abundance calculations under $M_s = 3\,\tx{TeV}$, $M_{\chi_2} - M_{\chi_1} \,(=\Delta M) = 10\,\tx{GeV}$, $\alpha = 0.05$; $\theta_{N_3} = 0.1$ (Left panel) and $\theta_{N_3} = 0.01$ (Right panel), taking account of coannihilation.
The horizontal blue line depicts the experimental observation of the DM relic abundance {($\Omega h^2 = 0.120 \pm 0.001$)~\cite{Planck:2018vyg}}.
Note that each point of the panels is consistent with {the observed neutrino profiles}, the constraints via the lepton flavour violation ($l_\alpha \to l_\beta + \gamma$) and  $T$ parameter.
The colours of the points indicate the magnitudes of $y_{A_{33}}$, while we take $y_{B_{33}} = 0$.
The details of the two benchmark points are available in Table~\ref{tab:lepton_Higgs_portal_BP-updated}.
}
    \label{fig:DM-result-1-updated}
\end{figure}

\begin{figure}[t]
    \centering
    \begin{minipage}{0.48\textwidth}
        \centering
        \includegraphics[width=\linewidth]{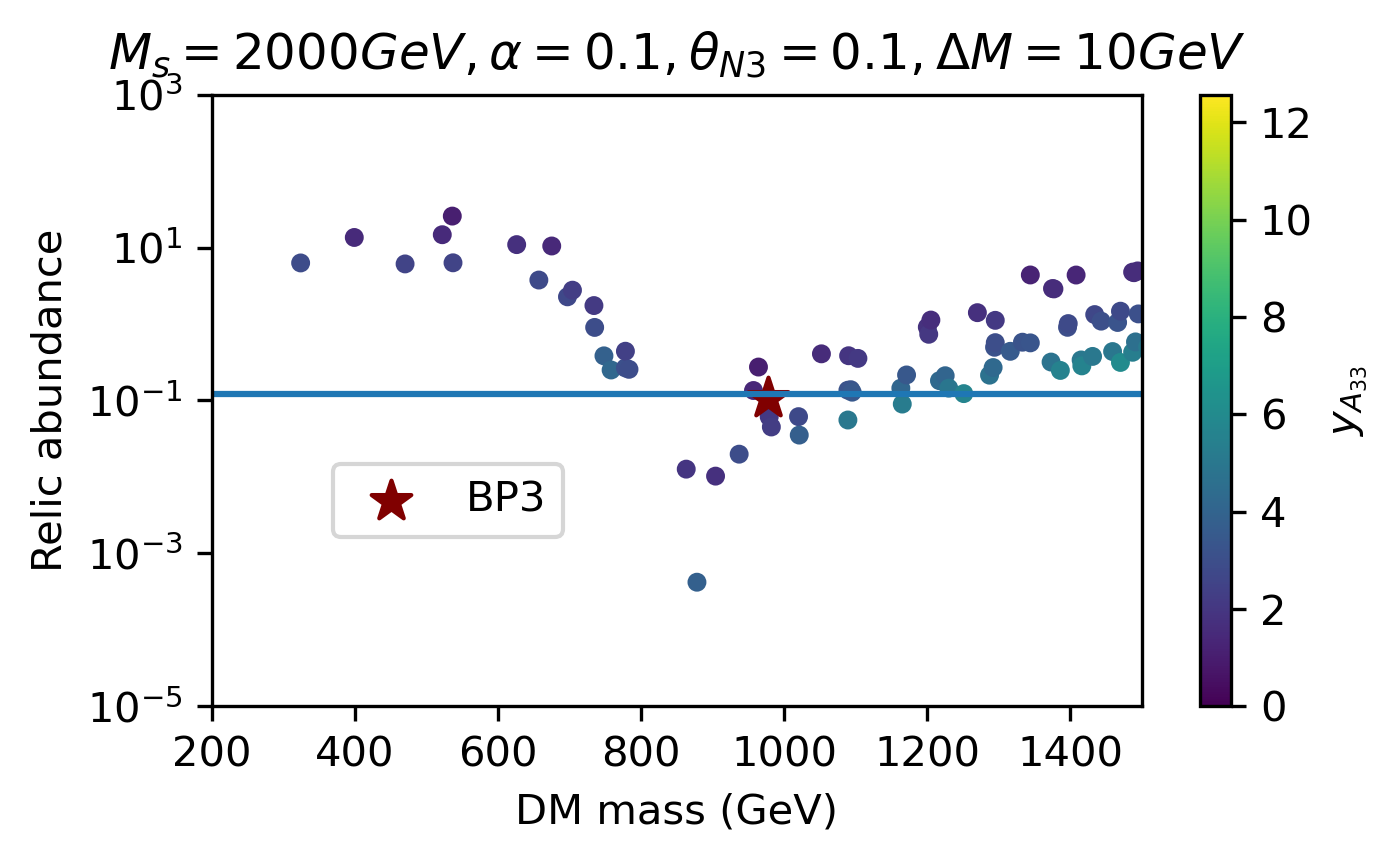} 
    \end{minipage}
    \hfill
    \begin{minipage}{0.48\textwidth}
        \centering
        \includegraphics[width=\linewidth]{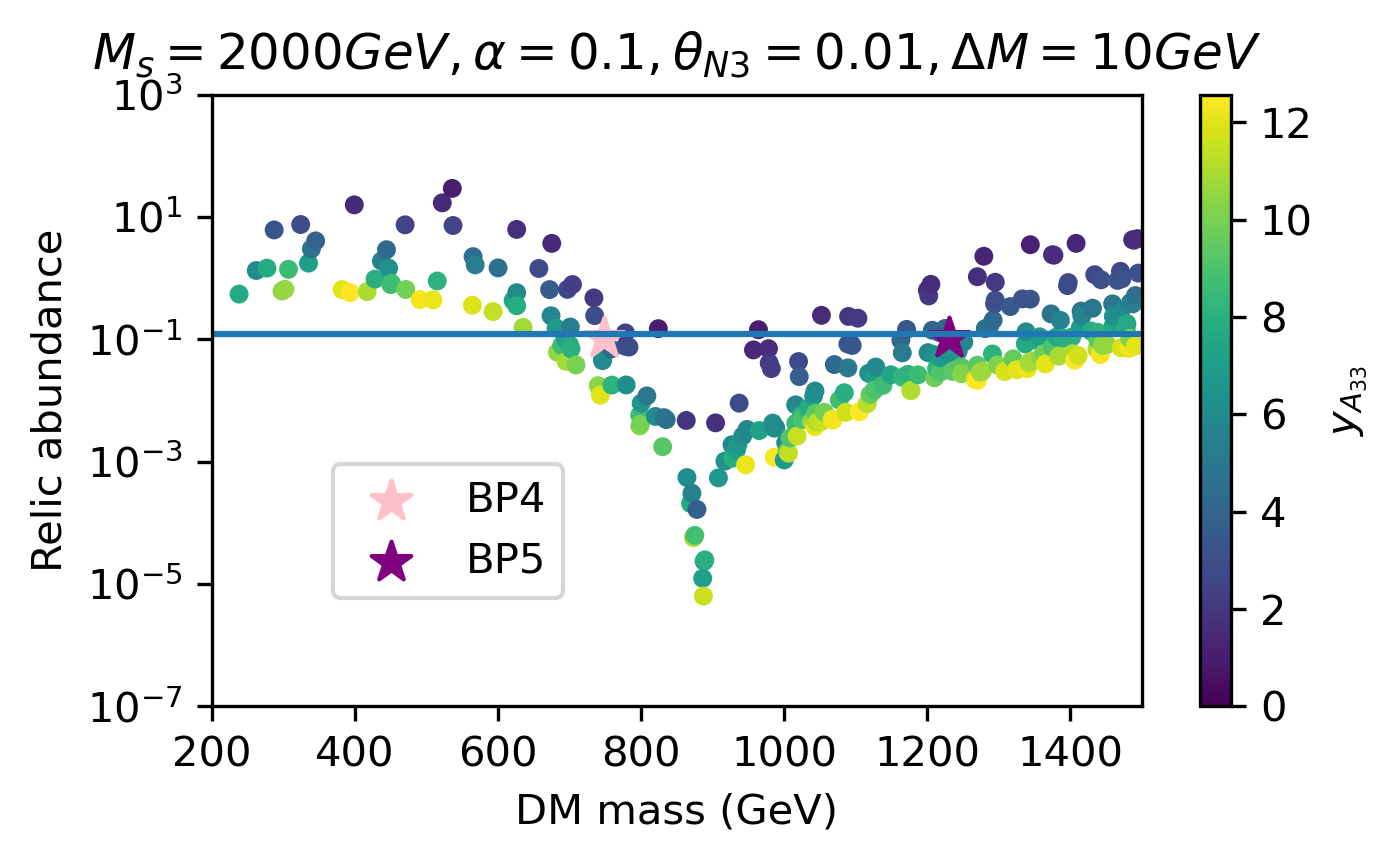} 
    \end{minipage} \\
        \begin{minipage}{0.48\textwidth}
        \centering
        \includegraphics[width=\linewidth]{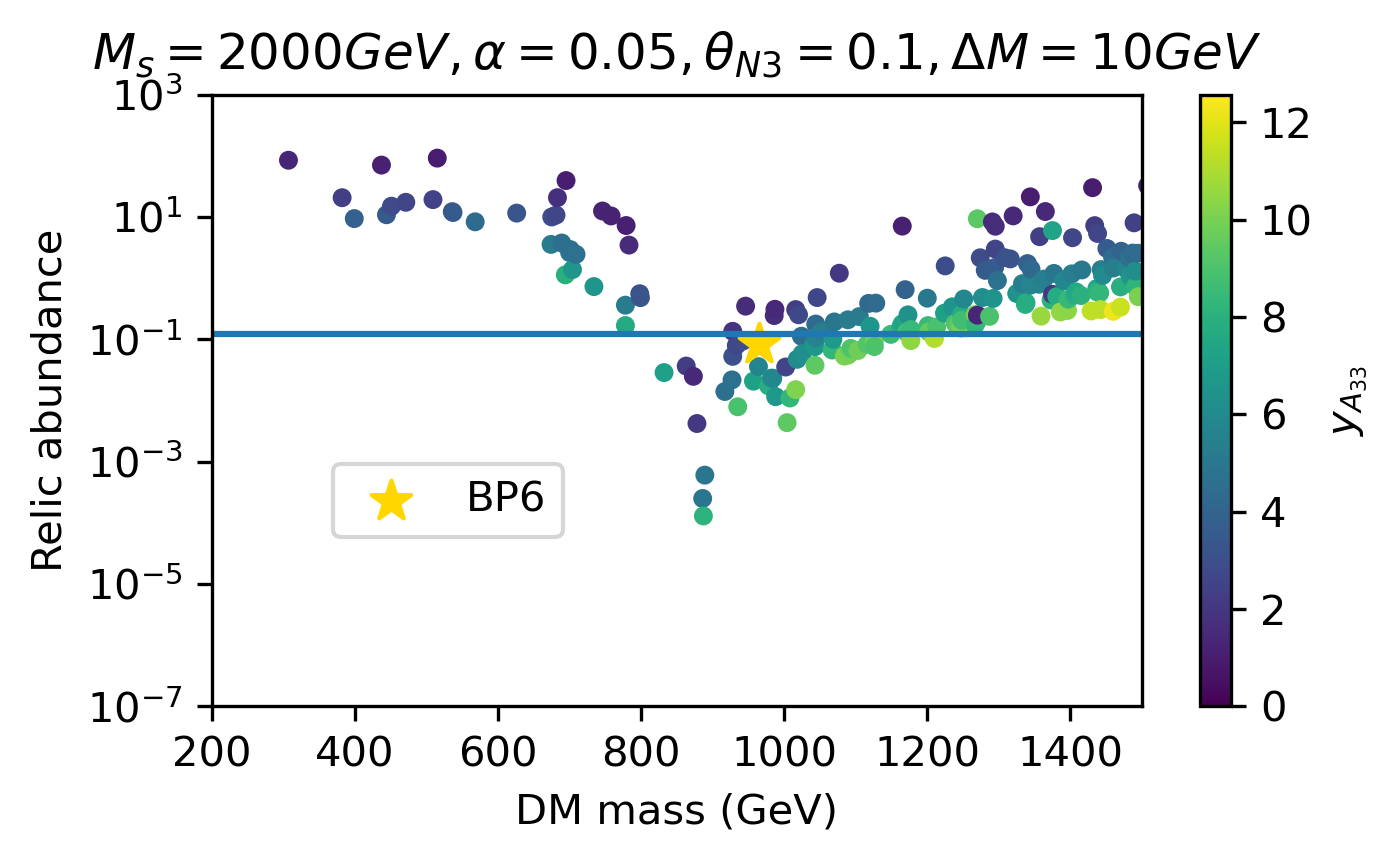} 
    \end{minipage}
    \hfill
    \begin{minipage}{0.48\textwidth}
        \centering
        \includegraphics[width=\linewidth]{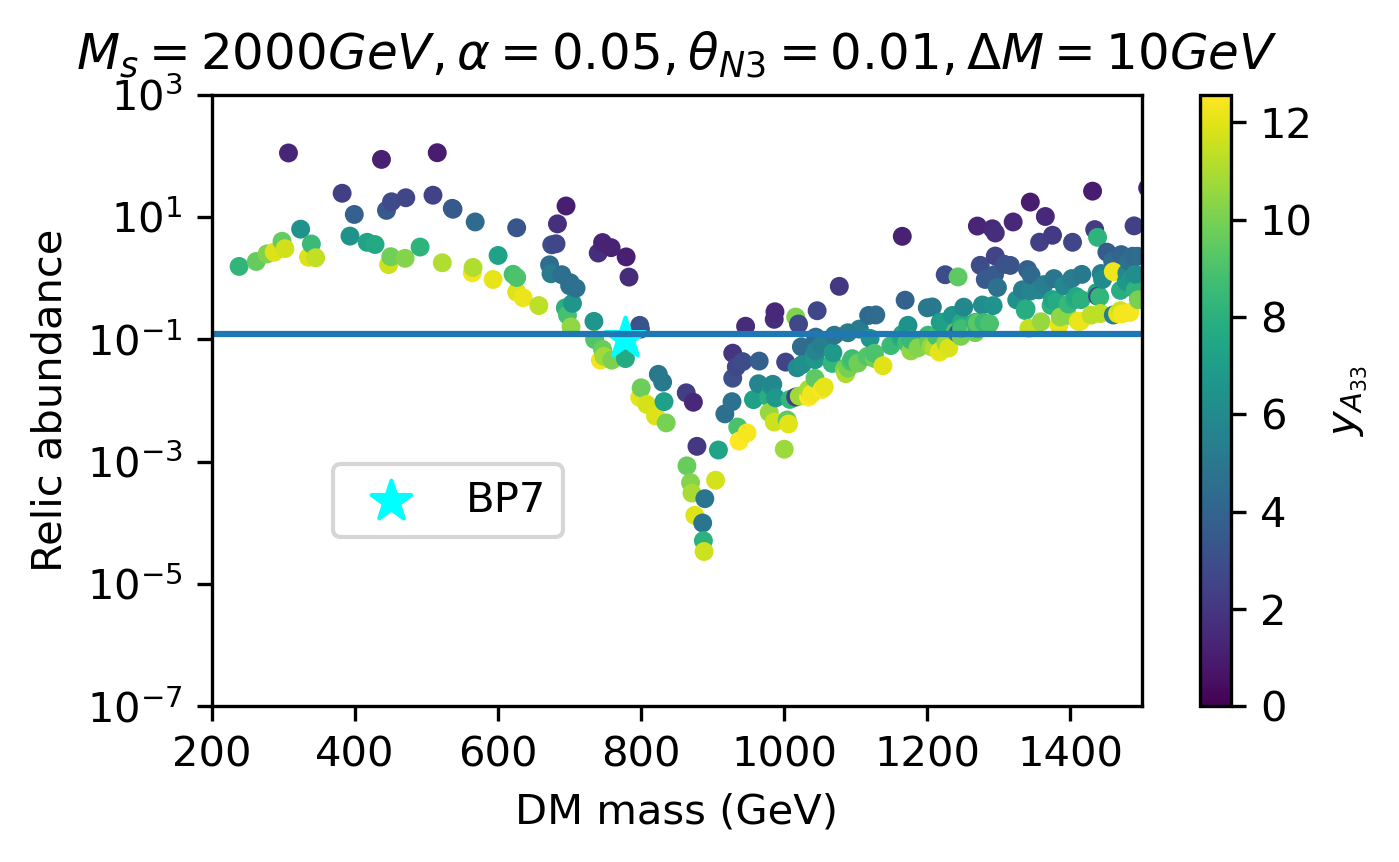} 
    \end{minipage}
    \caption{
Results of the relic abundance calculations under $M_s = 2\,\tx{TeV}$, $M_{\chi_2} - M_{\chi_1} \,(=\Delta M) = 10\,\tx{GeV}$;
$\paren{\alpha,\theta_{N_3}} =$
	$\paren{0.1,0.1}$ (Left upper),
	$\paren{0.1,0.01}$ (Right upper),
	$\paren{0.05,0.1}$ (Left lower),
	$\paren{0.05,0.01}$ (Right lower), respectively, taking account of coannihilation.
The other parameter choices and conventions are the same as those of Fig.~\ref{fig:DM-result-1-updated}.
The details of the {five} benchmark points are available in Table~\ref{tab:lepton_Higgs_portal_BP-updated}.
}
    \label{fig:DM-result-2-updated}
\end{figure}

\begin{table}[t]
    \centering
\begin{tabular}{|l||c|c||c|c|c|c|c|c|}
\hline
\textbf{Parameter} & \textbf{BP1} & \textbf{BP2} & \textbf{BP3} & \textbf{BP4} & \textbf{BP5} & \textbf{BP6} & \textbf{BP7} \\
\hline
\hline
$\Omega h^2$ & 0.1019 & 0.1119 & {0.1075} & 0.1042 & 0.1039 & 0.0830 & 0.1041 \\
\hline \hline
$M_s$ (TeV) & 3.0 & 3.0 & 2.0 & 2.0 & 2.0 & 2.0 & 2.0 \\
\hline
$M_{\chi_1} (= M_{N_{B_3}})$ (GeV) & 1274 & 1169 & 978 & 748 & 1230 & 965 & 778 \\
\hline
$M_{\chi_2} (= M_{N_{A_3}})$ (GeV) & 1284 & 1179 & 988 & 758 & 1240 & 975 & 788 \\
\hline
$M_{\eta_1^+}$ (GeV) & 2199.48 & 2795.66 & 2015.27 & 1299.27 & 2138.41 & 2157.74 & 1393.35 \\
\hline
$M_{\eta_2^+}$ (GeV) & 2006.75 & 2405.21 & 1173.00 & 897.00 & 1814.11 & 1706.14 & 933.00 \\
\hline
$M_{R_1}$ (GeV) & 2235.39 & 2808.58 & 2056.27 & 1305.33 & 2182.50 & 2206.11 & 1408.59 \\
\hline
$M_{R_2}$ (GeV) & 2019.67 & 2414.53 & 1190.37 & 857.16 & 1819.99 & 1716.45 & 962.39 \\
\hline
$M_{N_{A_1}}$ (GeV) & 1534 & 1408 & 1179 & 903 & 1482 & 1164 & 939 \\
\hline
$M_{N_{A_2}}$ (GeV) & 1533 & 1407 & 1178 & 902 & 1481 & 1163 & 938 \\
\hline
$M_{N_{B_1}}$ (GeV) & 1535 & 1409 & 1180 & 904 & 1483 & 1165 & 940 \\
\hline
$M_{N_{B_2}}$ (GeV) & 1532 & 1406 & 1177 & 901 & 1480 & 1162 & 937 \\
\hline
$\theta_R$ & 0.5859 & 0.5484 & 0.6322 & 0.6864 & 0.5482 & 0.6958 & 0.7500 \\
\hline
$\theta_{N_1} (= \theta_{N_2}), \theta_{N_3}$ & 0.7, 0.1 & 0.7, 0.01 & 0.7, 0.1 & 0.7, 0.01 & 0.7, 0.01 & 0.7, 0.1 & 0.7, 0.01 \\
\hline
$\theta_\tx{cs}$ & 0.6 & 0.6 & 0.6 & 0.6 & 0.6 & 0.6 & 0.6 \\
\hline
$\alpha$ & 0.05 & 0.05 & 0.1 & 0.1 & 0.1 & 0.05 & 0.05 \\
\hline
$\log_{10}(\lambda_5)$ & -6.0307 & -6.1558 & -6.8970 & -6.7433 & -6.1206 & -6.4266 & -6.6757 \\
\hline
$\lambda_6$ & 0 & 0 & 0 & 0 & 0 & 0 & 0 \\
\hline
$y_{A_{33}}$ & 2.8509 & 3.8996 & 1.5849 & 3.8393 & 4.9502 & 3.7451 & 5.2354 \\
\hline
$y_{B_{33}}$ & 0 & 0 & 0 & 0 & 0 & 0 & 0 \\
\hline \hline
$M_{I_1}$ (GeV) & 2235.39 & 2808.58 & 2056.27 & 1305.33 & 2182.50 & 2206.11 & 1409.59 \\
\hline
$M_{I_2}$ (GeV) & 2019.67 & 2414.53 & 1190.37 & 857.16 & 1819.99 & 1716.45 & 962.39 \\
\hline
\end{tabular}
\caption{
The values of the relevant parameters for the seven benchmark points in Figs.~\ref{fig:DM-result-1-updated} and \ref{fig:DM-result-2-updated} and realised relic abundances ($\Omega h^2$) are shown.
Note that they are consistent with {the observed neutrino profiles}, the lepton flavour violation ($l_\alpha \to l_\beta + \gamma$), $T$ parameter, and DM direct detection.
}
    \label{tab:lepton_Higgs_portal_BP-updated}
\end{table}

\begin{table}[t]
\centering
\begin{tabular}{|l||c|c||c|c|c|c|c|}
\hline
\textbf{Parameter} & \textbf{BP1} & \textbf{BP2} & \textbf{BP3} & \textbf{BP4} & \textbf{BP5} & \textbf{BP6} & \textbf{BP7} \\
\hline
\hline
$\lambda_{S\eta_A}$ & 0.7105 & 0.9196 & 0.5929 & 0.4385 & 0.6829 & 0.6594 & 0.4845 \\
\hline
$\lambda_{S\eta_B}$ & 0.6551 & 0.8175 & 0.4982 & 0.3963 & 0.6030 & 0.6052 & 0.3879 \\
\hline
$\lambda_{2a}$ & 2.8 & 3.7 & 4 & 3.5 & 4 & 3.5 & 3.5 \\
\hline
$\lambda_{2b}$ & 3 & 4 & 4 & 3.5 & 4 & 3.8 & 3.6 \\
\hline
$\lambda_{2c}$ & 1 & 1 & 3 & 0.1 & 4 & 4.8 & 2.5 \\
\hline
$\lambda_{2d}$ & 1 & 1 & 3 & 0.1 & 4.4 & 5 & 2.5 \\
\hline
$\lambda_{3A}$ & -3.5287 & -4.3015 & -0.3693 & 4.0333 & -5.7771 & 1.7173 & -8.6457\\
\hline
$\lambda_{3B}$ & -1.4539 & 2.4240 & -4.5045 & -0.2452 & 0.7748 & -7.8595 & 7.3934 \\
\hline
$\lambda_{4A}$ & 4.5287 & 5.3014 & 1.3693 & -3.0333 & 6.7771 & -0.7173 & 9.6456 \\
\hline
$\lambda_{4B}$ & 2.4539 & -1.4240 & 5.5045 & 1.2452 & 0.2252 & 8.8595 & -6.3934 \\
\hline
$m_{\eta_A}$ (GeV) & 2048.17\,$i$ & 2048.69\,$i$ & 2048.30\,$i$ & 2048.54\,$i$ & 2048.72\,$i$ & 2048.13\,$i$ & 2048.18\,$i$ \\
\hline
$m_{\eta_B}$ (GeV) & 1964.41\,$i$ & 1964.48\,$i$ & 1964.47\,$i$ & 1964.56\,$i$ & 1965.08\,$i$ & 1964.62\,$i$ & 1964.71\,$i$ \\
\hline
$\lambda_{HS}$ & 0.3646 & 0.36461 & 0.3218 & 0.3218 & 0.3218 & 0.1617 & 0.1617 \\
\hline
$\lambda_H$ & 0.6290 & 0.629044 & 0.9144 & 0.9144 & 0.9144 & 0.4227 & 0.4227 \\
\hline
$\lambda_S$ & 0.1796 & 0.179551 & 0.0792 & 0.0792 & 0.0792 & 0.0798 & 0.0798 \\
\hline
\end{tabular}
\caption{
For each of the seven benchmark points (BP1 to BP7) of Figs.~\ref{fig:DM-result-1-updated} and \ref{fig:DM-result-2-updated}, we provide a set of valid parameters describing the potential (also not being relevant for the phenomena considered in the numerical scans).
In any one of the sets, all of the conditions for 
the {bounded} from below [from Eq.~\eqref{eq:Det-condition-start} to Eq.~\eqref{eq:Det-condition-end}]
and the inert condition [from Eqs.~\eqref{eq:Inert-Condition-1} and \eqref{eq:Inert-Condition-2}] are satisfied.
}
    \label{tab:VstabilityCon-updated}
\end{table}

We will discuss the dark matter phenomenology of the current scenario, taking into account the observed active neutrino texture,
and constraints via the lepton flavour violation ($l_\alpha \to l_\beta + \gamma$), $T$ parameter oblique constraint and the stability of the scalar potential. For simplicity, we focus on the normal ordering in the active neutrino mass matrix below.\footnote{
{All numerical calculations except for Fig.~\ref{fig:kappa3} in this paper were performed using {\tt Python\,3}~\cite{10.5555/1593511} with the following packages:
{\tt Math}~\cite{van1995python}, {\tt NumPy}~\cite{harris2020array}, {\tt Pandas}~\cite{reback2020pandas}, {\tt Random}~\cite{van1995python} and {\tt SciPy}~\cite{2020SciPy-NMeth};
also {\tt Matplotlib}~\cite{Hunter:2007} for plotting.
For Fig.~\ref{fig:kappa3}, {\tt Mathematica} was used.}
}

A key parameter of our dark matter scenario is the mixing angle $\alpha$ between the Higgs doublet and the singlet,
which determines the magnitudes of the Higgs-portal channels in the relic abundance and constraints via the direct detection.
Note that $\alpha = 0$ corresponds to the pure lepton-portal scenario.
We choose the Majorana fermion $N_{B_3} \, (= \chi_1 = \chi)$ as a dark matter candidate, where under the texture in {Eqs.~\eqref{eq:Yukawa_compact} and \eqref{eq:higgs_yukawa}.}
$N_{A_3} \, (= \chi_2)$ is an efficient partner of coannihilation.
As shown in Eqs.~\eqref{eq:Higgs-portal-couplings-1} and \eqref{eq:higgs_coupling}, the relevant Yukawa couplings to the gauge singlet among them, namely $y_{A_{33}}$ and $y_{B_{33}}$, play a primary role in relic abundance and direct detection.
{We take the other $y_{A_{ii}}$ and $y_{B_{ii}}$ as zero ($y_{A_{11}} = y_{A_{22}} = y_{B_{11}} = y_{B_{22}} = 0$).}
Another important parameter is the mixing angle for the `third'-generation Majorana fermions $\theta_{N_3}$, which modulates the interactions for relic abundance and direct detection.
Finally, we should point out that the dark matter candidate can hit the pole of the (mostly-singlet) scalar $s$ if the mass is around half of it. The resonance enhancement in the annihilation (without large couplings) helps dark matter phenomenology.

As a benchmark, we choose $M_s = 3\,\tx{TeV}$, $\Delta M := M_{\chi_2} - M_{\chi_1} = 10\,\tx{GeV}$, $\alpha = 0.05$,
where the Higgs-portal channel is switched on and the coannihilation mechanism may work.
Figure~\ref{fig:DM-result-1-updated} provides the results of our numerical scans for $\theta_{N_3} = 0.1$ (Left panel) and $\theta_{N_3} = 0.01$ (Right panel), where other relevant parameters are scanned.

{For each scan, we randomly selected the parameters $M_{\chi_1} (= M_{N_{B_3}})$, $M_{N_{A_1}}$, $M_{N_{B_1}}$, $M_{N_{A_2}}$, $M_{N_{B_2}}$, $M_{\eta^+_1}$,
$M_{\eta^+_2}$, $M_{R_1}$, $M_{R_2}$, $\theta_R$, $\lambda_5$, $y_{A_{33}}$, while the following other independent relevant parameters are fixed;
$M_s$, $M_{\chi_2} (= M_{N_{A_3}})$, $\theta_{N_1}$, $\theta_{N_2}$, $\theta_{N_3}$, $\theta_\tx{cs}$, $\alpha$, $\lambda_6 \, (=0)$, and $y_{B_{33}} \, (=0)$.
Note that the values of $M_{I_1}$ and $M_{I_2}$ are determined by the parameters above, but we show the corresponding digits for convenience.
Also, the two differences between $M_{R_i}$ and $M_{I_i}$ $(i=1,2)$ are not observed within the shown digits in Table~\ref{tab:lepton_Higgs_portal_BP-updated}, but they should be nonzero to produce the minuscule mass differences of the active neutrinos (see Section~\ref{sec:massterm}).
{Due to the same reason, $\theta_I$ (not shown in Table~\ref{tab:lepton_Higgs_portal_BP-updated}) becomes extremely close to $\theta_R$.}
See Table~\ref{tab:lepton_Higgs_portal_BP-updated} for concrete information.}\footnote{
{
As mentioned above, the vital parameters for the residual amount of dark matter have been identified through phenomenological discussions, namely $M_{\chi_1}$, $M_{\chi_2}$, $M_s$, $\alpha$ and $\theta_{N_3}$, other than couplings such as $y_{A_{33}}$.
We determined that fixing these parameters for comparison would clarify the characteristics of the models.
In addition, several parameters that are not expected to affect the overall phenomenon significantly, such as $\theta_{N_1}$ and $\theta_{N_2}$, have been fixed appropriately.
}
}

Note that each point of the panels is consistent with the observed neutrino profiles, the constraints via the lepton flavour violation ($l_\alpha \to l_\beta + \gamma$) and the $T$ parameter.
The colours of the points indicate the magnitudes of $y_{A_{33}}$, while we take $y_{B_{33}} = 0$ to evade the direct-detection constraint.
These results show that if we introduce an order-one Yukawa $y_{A_{33}}$ (keeping $y_{B_{33}} = 0$),
our model can explain the TeV-scale dark matter remnant simultaneously with the preferred neutrino mass terms.
The effective area extends over a wide range, not only in the vicinity of the resonance around $1.5\,\tx{TeV}$.
It can also be read that $\theta_{N_3}$ affects the valid range of the parameters, as shown in Fig.~\ref{fig:DM-result-1-updated},
a smaller $\theta_{N_3}$ is slightly more significant for a small value of $\alpha$.
We provide the full sets of model parameters for the two benchmarks (BP1, BP2) in Table~\ref{tab:lepton_Higgs_portal_BP-updated}.
Note that for a sizable choice of $\alpha$, the Higgs-portal interaction increases rapidly, and the direct detection discards such a point.

We provide our investigations for a lighter $s$ of $M_s = 2\,\tx{TeV}$, keeping $\Delta M (= M_{\chi_2} - M_{\chi_1}) = 10\,\tx{GeV}$, taking into account coannihilation.
The results are available in Fig.~\ref{fig:DM-result-2-updated}, where we focus on the four choices of 
$\paren{\alpha,\theta_{N_3}} =$ $\paren{0.1,0.1}$ (Left upper), $\paren{0.1,0.01}$ (Right upper), $\paren{0.05,0.1}$ (Left lower), and $\paren{0.05,0.01}$ (Right lower), respectively.

Here, as a general trend, it becomes more difficult to avoid various restrictions because the masses of heavy particles tend to be smaller, compared with those in the scan for $M_s = 3\,\tx{TeV}$ (refer to Fig.~\ref{fig:DM-result-1-updated}), in effective parameter points.
As seen in the previous scans for $M_s = 3\,\tx{TeV}$ (refer to Fig.~\ref{fig:DM-result-1-updated}), the smaller $\theta_{N_3}$ tends to yield more effective configurations.
Furthermore, even with the same $\theta_{N_3}$, a smaller nonzero $\alpha$ tended to result in more effective configurations.
We provide the full sets of model parameters for the five benchmarks (BP3, BP4, BP5, BP6, BP7) in Table~\ref{tab:lepton_Higgs_portal_BP-updated}.\footnote{
{For the shift of the peak from the naive location $M_s/2 = 1\,\tx{TeV}$, see Chapter~6 of~\cite{Gondolo:1990dk} for details.}
}

The scan results in Figs.~\ref{fig:DM-result-1-updated} and \ref{fig:DM-result-2-updated} clearly show that this scenario can still correctly reproduce the coordination of active neutrinos and the relic abundance of thermal WIMP dark matter while avoiding the latest direct search limitations on dark matter and other theoretical limitations, where active-neutrino masses and mixings are induced at 1-loop level as shown in Fig.~\ref{fig:neutrino-mass}, and the Yukawa coupling $y_{A_{33}}$ tends to be ${\cal O}\fn{1}$.
We provide a few comments:
\begin{itemize}
\item
If we choose $M_{\chi_2}$ sizably above $M_{\chi_1}$, coannihilation is disactivated and most of the relevant parameter points addressing the DM relic abundance are gone, see Appendix~\ref{sec:DM-result-with-no-coannihilation} for checking this point.
\item
{If we switch off the Higgs portal coupling by $\alpha = 0$, only the lepton-portal coupling remains.
As far as we checked, no valid parameter point for addressing the neutrino masses/mixings and the DM relic abundance (even with coannihilation), also evading the latest bound on the DM direct detection~\cite{LZ:2024zvo} and other constraints.}\footnote{
{Note that we found parameter points with the correct amount of the DM relic abundance in the pure lepton-portal case ($\alpha=0$) with coannihilation, being consistent with the previous LZ constraint on the direct detection~\cite{LZ:2022lsv}, and the others.}
}
\item
Since, as shown in Table~\ref{tab:lepton_Higgs_portal_BP-updated}, also Figs.~\ref{fig:DM-result-1-updated} and \ref{fig:DM-result-2-updated}, the masses of the dark matter candidate and the other new fermions and scalars tend to be above (or around) $1\,\tx{TeV}$, and thus to investigate the current scenario at the {LHC} by focusing on events with large missing $E_\tx{T}$ is uneasy.
However, as we will debate in Section~\ref{sec:trilinear-Higgs}, we can investigate the current scenario with the help of the LHC by focusing on the measurement of the trilinear Higgs self-coupling.
\end{itemize}

In the previous numerical scans in Figs.~\ref{fig:DM-result-1-updated} and \ref{fig:DM-result-2-updated}, also in Table~\ref{tab:VstabilityCon-updated}, we did not include the constraints on the shape of the scalar potential;
the {bounded} from below [from Eq.~\eqref{eq:Det-condition-start} to Eq.~\eqref{eq:Det-condition-end}]
and the inert condition [from Eqs.~\eqref{eq:Inert-Condition-1} and \eqref{eq:Inert-Condition-2}].\footnote{
{In our model, the potential has many parameters, making it structurally impossible to uniquely determine the potential parameters
based solely on the information on the scalar's VEV, masses and mixings.
Due to this circumstance, we adopted the approach of separating the discussion of potential stability from other discussions.}
}
To confirm this point, in Table~\ref{tab:VstabilityCon-updated}, for each of the seven benchmark points (BP1 to BP7) of Figs.~\ref{fig:DM-result-1-updated} and \ref{fig:DM-result-2-updated}, we provide a set of valid parameters describing the potential (also not being relevant for the phenomena considered in the numerical scans), where, in any one of the sets, all of the conditions for 
the {bounded} from below [from Eq.~\eqref{eq:Det-condition-start} to Eq.~\eqref{eq:Det-condition-end}]
and the inert condition [from Eqs.~\eqref{eq:Inert-Condition-1} and \eqref{eq:Inert-Condition-2}] are satisfied.
Here, at some benchmark points, some quartic couplings are quite large, but they still satisfy a naive perturbation limit of $4\pi$.
From the above discussion, it can be said that our model tends to maintain the stability of the potential perturbatively while keeping the stability of dark matter by appropriately selecting the parameters of the potential.

\section{Bounds on Higgs trilinear self-coupling
\label{sec:trilinear-Higgs}}

\begin{figure}[t]
    \centering
    \includegraphics[width=0.5\textwidth]{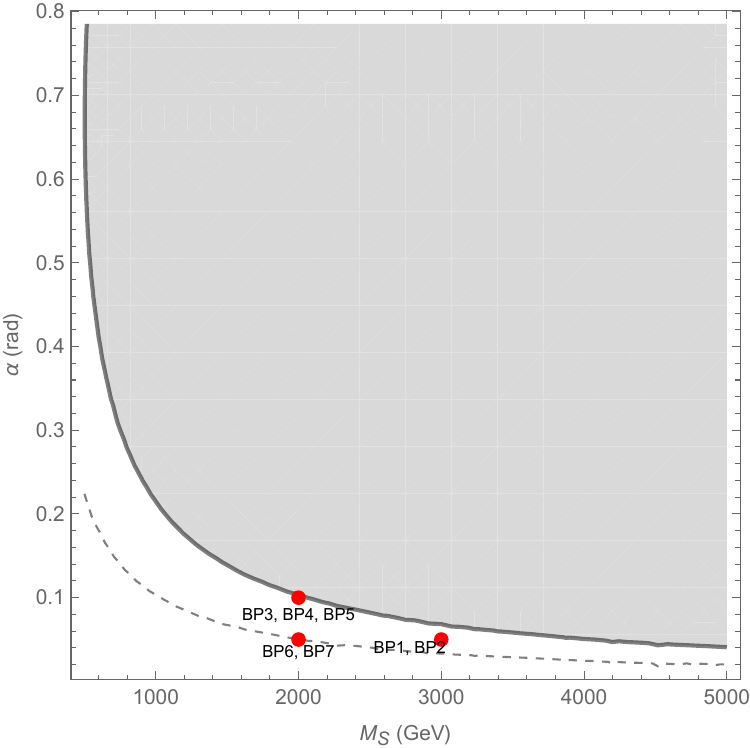}
\caption{
The grey-shaded region shows the {$95\%$-CL} disfavoured region on the $(M_s, \alpha)$ plane {(under the choice $v_S = 5\,\tx{TeV}$)} by the ATLAS constraints on the trilinear Higgs self-coupling~\cite{ATLAS:2022jtk}.
The dashed curve represents the {$95\%$-CL} future prospect of the ATLAS and CMS HL-LHC projection under the SM hypothesis~\cite{Cepeda:2019klc}.
Red points represent the benchmark points discussed in Section~\ref{sec:Analysis}
(see Figs.~\ref{fig:DM-result-1-updated} and \ref{fig:DM-result-2-updated}, and Table~\ref{tab:lepton_Higgs_portal_BP-updated} for details).
}
    \label{fig:kappa3}
\end{figure}

The ATLAS experiment places stringent constraints on the Higgs trilinear self-coupling \cite{ATLAS:2022jtk, Stylianou:2023tgg}. The parameter $\kappa_3$ quantifies the deviation from the Standard Model prediction for the trilinear Higgs self-coupling, defined as
\al{
\kappa_3 = {\lambda_{hhh}\ov\lambda_{hhh}^\tx{SM}}.
}
The current bound by ATLAS  on $\kappa_3$~\cite{ATLAS:2022jtk} is
\al{
-0.4 \le \kappa_3 \le 6.3 \quad \paren{\tx{at $95\%$ Confidence Level}}.
}
In the SM, the trilinear Higgs coupling arises from the scalar potential:
\als{
    \mathcal{V}^\tx{SM} &\supset {\lambda_H\ov 2}\paren{H^\dagger H}^2 \,,\\
                &\supset {\lambda_H v_H\ov 2} h'^3 = {1\ov 3!}{3\lambda_H v_H} h'^3 = {1\ov 3!}{\lambda_{hhh}^\tx{SM}} h'^3\,,
}
where the Higgs doublet is expanded around its VEV as shown in Eq.~\eqref{eq:multiplet-components} (see also Eq.~\eqref{eq:singlet-scalar-components}).
Therefore, the SM trilinear coupling is fully determined by the electroweak \cred{symmetry-breaking} scale $v_H=246\tx{ GeV}$ and the Higgs mass $M_h=125\,\tx{GeV}$,
\al{
\lambda_{hhh}^\tx{SM}=3\lambda_H v_H\,,
\quad
\lambda_H = {M_h^2 \ov v_H^2}\simeq0.26\,.
}

In our model, additional contributions to the trilinear coupling arise from the scalar singlet $S$, leading to the extended scalar potential:
\al{
{\cal V} 
	&\supset{\lambda_{HS} \ov 2} S^2 \paren{H^\dagger H} 
		+ {\lambda_{H} \ov 2} \paren{H^\dagger H}^2 + {\lambda_S \ov 4} S^4\,,\notag \\
        & \supset {1\ov 2} \br{v_H\paren{\lambda_H c_{\alpha}^3 + \lambda_{HS}\,c_{\alpha}\,s_\alpha^3}
      	  + v_S\paren{2\lambda_S\,s_\alpha^3 + \lambda_{HS}\, c_\alpha^2\, s_\alpha
        }}h^3\,, \notag \\
        &={1\ov 3!} \lambda_{hhh}h^3.
}
After spontaneous symmetry breaking and mixing between the Higgs and singlet scalars, the physical trilinear coupling becomes:
\al{
\lambda_{hhh}=3\br{v_H\paren{\lambda_H c_{\alpha}^3 + \lambda_{HS}\,c_{\alpha}\,s_\alpha^3}
        + v_S\paren{2\lambda_S\,s_\alpha^3 + \lambda_{HS}\, c_\alpha^2\, s_\alpha
        }}.
}
where we remind \cred{the reader of} the mixing angle, $s_\alpha=\sin\paren{\alpha}$, $c_\alpha=\cos\paren{\alpha}$, namely,
$s' = s_\alpha h + c_\alpha s$, $h' = c_\alpha h -  s_\alpha s$.
Here, $\lambda_{hhh}$ can be completely determined if we specify the $Z_4$ symmetry breaking $v_S$, mass of the new singlet $M_s$  and Higgs mixing angle $\alpha$ via,
\al{
\lambda_H &=\frac{M_{h}^2 c_\alpha^2 + M_{s}^2 s_\alpha^2}{v_H^2},&
\lambda_S &= \frac{M_{s}^2 c_\alpha^2 + M_{h}^2 s_\alpha^2}{2v_S^2},&
\lambda_{HS} &= -\frac{(M_{h}^2 - M_{s}^2) \sin(2 \alpha)}{2 v_H v_S}.&
	\label{eq:lambda-relations}
}

We pin down the corresponding points of the seven benchmark points discussed in Section~\ref{sec:Analysis} as a function of $\alpha$ and $M_s$ {under the choice $v_S = 5\,\tx{TeV}$}, and the result is summarised in Fig.~\ref{fig:kappa3}.\footnote{
{As recognised from Eq.~\eqref{eq:lambda-relations}, if $v_S$ is notably greater than $v_H$, the disfavoured region is not sensitive to the value of $v_S$.}
}
Note that all of the benchmark points are consistent with the latest ATLAS result~\cite{ATLAS:2022jtk}, {and} the five benchmark points (Nos. 1, 2, 3, 4, 5) are located near the current boundary.
The 95\% confidence-level combined ATLAS and CMS HL-LHC projection under the SM hypothesis was estimated as $\kappa_3 \in [0.1, 2.3]$~\cite{Cepeda:2019klc}, and the benchmark points will be indirectly investigated in the future.

\section{Comments on the domain wall problem
\label{sec:domain-wall}}

The authors of \cite{Zeldovich:1974uw} make the observation that when a discrete symmetry is spontaneously broken, multiple degenerate ground states become available. As there is no requirement for the \cred{Universe} to select a particular choice of ground state, we will have causally disconnected spatial regions having different ground states. The different ground state regions will be separated by stable domain walls~(DWs). The energy density of these domain walls scales as $\rho_\tx{DW} \propto 1/R$, where $R$ is the radius of the \cred{Universe}, $\rho_\tx{DW}$ decreases more slowly than normal radiation or matter energy density dominating the energy density of the \cred{Universe}, which can lead to disagreement with observation.

In our model, the $Z_4$ symmetry is spontaneously broken when the singlet field $S$ acquires a VEV, to a $Z_2$ subgroup.
We consider two possible mechanisms to address this issue in our model.\footnote{
{We may mention that while some works~\cite{delaVega:2024tuu, Ahriche:2020pwq,deAnda:2021jzc} have proposed Scotogenic scenarios based on $Z_4$ to a $Z_2$, no comments on the \cred{domain-wall} problem were found in them.}
}

\subsection{Domain wall production before Inflation}
The simplest way is to consider the symmetry-breaking scale to be higher than the inflation scale.
If the domain walls are produced before inflation, the resulting domain walls are inflated away.
Low-scale inflation models with $V^{1/4}\simeq 10$ -- $10^{4}\,\tx{TeV}$ \cite{Ghoshal:2022zwu, Caputo:2023ikd,Czerny:2014xja} offer a viable setting in which domain walls produced by discrete symmetry breaking can be inflated away.
Though our model benchmark points have $\tx{VEV}\sim \mathcal{O}(10 \tx{ TeV})$, we can increase the VEV above the inflation scale without altering the particle physics phenomenology discussed in this article to dilute away domain walls produced by discrete symmetry breaking.

\subsection{Domain wall decay in a $U(1)_X$-embedded scenario}
Another way to address this issue is the use of the gauged discrete symmetry prescription described in \cite{Vilenkin:1982ks} and \cite{Preskill:1991kd}. We embed our $Z_4$ symmetry into a continuous gauge symmetry such as $U(1)_X$, in this new scenario the \cred{Universe} first breaks the continuous gauge symmetry at some high scale $v_1$  to our $Z_4$ discrete symmetry forming strings and then subsequently $Z_4$ is broken to $Z_2$ at scale $v_S$ forming domain walls bounded by strings.
\als{
U(1)_{X}\xrightarrow{v_1}Z_4 \xrightarrow{v_S}Z_2.
}
Inflation must occur before the high \cred{symmetry-breaking} scale $v_1$, so that the strings produced are not diluted away.
These domain walls are metastable and collapse in time $t_c\sim {\mu/\sigma}\sim {v_1 ^2/ v_S^3}$
(from dimension analysis in~\cite{Dunsky:2021tih}), where $\mu$ and $\sigma$ respectively denote the string mass per unit length and wall mass per unit area.
If we assume scales of symmetry breaking $v_1\sim10^{16}\,\tx{GeV}$ and $v_S\sim 10\,\tx{TeV}$ or $\sim 1\,\tx{TeV}$, the domain walls collapse in time $t_c\sim 10^{-4}\,\tx{s}$ or $t_c\sim 10^{-2}\,\tx{s}$.\footnote{
Note that for a lower/higher $v_1$, we get a looser/tighter bound.
}
The homogeneity and isotropy of our \cred{Universe} require that the domain walls disappear quite early in the evolution of our \cred{Universe}. The most stringent bound is from \cred{the} big bang nucleosynthesis (BBN), the domain wall must disappear before BBN $T\sim 1\tx{ MeV}$ \cite{ParticleDataGroup:2024cfk} if we assume there is not \cred{a} domain wall dominated era in the \cred{Universe} evolution the temperature corresponds to $t\sim 1\,\tx{s}$, therefore, the bound on domain wall collapse is $t_c \lesssim 1\,\tx{s}~$\cite{Zeldovich:1974uw}.
This method was also used by the authors of \cite{Maji:2024pll} to analyse the gravitational signal resulting from the collapse of this metastable string-bounded \cred{domain-wall} system. While a detailed discussion of this process is not in the scope of our article, we show that we are able to overcome the \cred{domain-wall} issue within reasonable extensions of our model and its parameters while preserving our original result.\footnote{
While a description of the \cred{formulas} for $\mu$ and $\sigma$ is given in the original paper~\cite{Vilenkin:1982ks}, it contains a version of the \cred{formulas} in Eqs.~(12) and (13), which is a model-specific formula.
A more general case is studied in \cite{Dunsky:2021tih}, and the \cred{formulas} can be found in Eqs. (4) and (11).
}

\subsection{Explicit soft $Z_2$ breaking}

As a final mechanism to address the \cred{domain-wall} problem, we consider introducing a soft, explicit breaking of the discrete symmetry, which avoids the \cred{domain-wall} problem.
The softly broken $Z_4$ symmetry lifts the degeneracy of the VEV, triggering domain wall collapse.
Most studies illustrating this mechanism typically use $Z_2$ as a starting point~\cite{Larsson:1996sp,Saikawa:2017hiv,Hiramatsu:2010yz,Gelmini:1988sf}.  For $Z_N$ with $N > 2$, the breaking is typically considered in the context of complex scalar fields \cite{Wu:2022stu,Wu:2022tpe}, where the vacuum structure becomes elaborate.
Nonetheless, heuristically, the arguments discussed may still be valid for our model.
Let us begin with our potential for the $S$ field,
\al{
V_0\paren{S} = +{m_S^2\ov 2}S^2 +{\lambda_S\ov 4}S^4, 
}
with parameterisation $S =\paren{v_S + s'}$.
Though the $Z_4$ symmetry seems to allow four degenerate solutions for the VEV,
\al{
v_S=v_0 e^{N\pi\ov2}\,,\quad N=0,1,2,3.
}
These are only consistent if $S$ were complex, if not, the imaginary solutions of the VEV lead to a complex potential which is unphysical.
For a real scalar field $S$ with $Z_4$ charge $-1$, only the two real 
minima are allowed.
Therefore, in our setup, $S$ is effectively charged under a $Z_2$ symmetry and can only take VEV,
\al{
v_S = \pm v_0
}
An explicit $Z_4$ \cred{symmetry-breaking} term is added, which introduces a bias in the vacuum structure and generates a pressure on the domain walls,
\al{
\delta V = \epsilon S^3\xrightarrow{} \epsilon v_0^3e^{N\pi\ov2}\,,
}
Here $\epsilon$ is a dimensionful parameter controlling the strength of the explicit breaking, and $N=0, 2$ \cred{corresponds} to the real field configuration of our setup.
The energy difference between the two vacua \cred{reads}
\al{
\Delta V =V\paren{+v_S}-V\paren{-v_S} = 2\epsilon v_S^3.
}

This energy difference between the two vacuum states provides a pressure leading to \cred{domain-wall} collapse.
We require that the domain walls disappear before nucleosynthesis~\cite{Larsson:1996sp}; therefore, the pressure must satisfy,
\al{
{\Delta V\ov \sigma} > {1\ov 10^{10}\tx{cm}}\sim 10^{-24}\,\tx{GeV} \,,
}
which leads to the lower bound on $\epsilon$ 
\al{
\epsilon \gtrsim 5 \times 10^{-25}\,\tx{GeV}.
}
Here $\sigma$ is the \cred{domain-wall} tension $\sigma\sim v_{S}^3$.

On the other hand, the explicit soft-breaking term disables the absolute stability of the fermionic DM candidate,
which becomes a very long-lived particle. 
A possible decay channel of $\chi$ is $\chi \to \nu_\tx{SM} h h$ ($\nu_\tx{SM}$ is a SM neutrino), and the \cred{magnitude} of the corresponding total decay width of a decaying $\chi$ is very roughly estimated through naive dimensional analysis and the 3-body phase space factor (for massless final states) as
\al{
\Gamma_{\tx{decaying-}\chi}
	\sim
		{3 \ov 192 \pi^3} \times \paren{\epsilon c_{\cancel{\tx{inert}}}}^2 \times { M_\chi^3 \ov M_\tx{med}^4 },
}
where the factor $3$ comes from the three SM neutrinos,
the violation of the inert condition in the terms $\paren{H^\dagger H} \paren{\eta_A^\dagger \eta_A}$ and $\paren{H^\dagger H} \paren{\eta_B^\dagger \eta_B}$ is parametrised as $\epsilon c_{\cancel{\tx{inert}}}$ with a dimensionless constant $c_{\cancel{\tx{inert}}}$, $M_\tx{med}$ represents a mass of the scalar mediator into a pair of the SM Higgs boson, and here we assumed that all of the dimensionless coupling constants are ${\cal O}\fn{1}$.
The corresponding mean lifetime is estimated as
\al{
\tau_{\tx{decaying-}\chi}
	&=
		\Gamma_{\tx{decaying-}\chi}^{-1} \notag \\
	&\sim
		\paren{4 \times 10^{30} \, \tx{s}}
		\times
		\paren{5 \times 10^{-25}\,\tx{GeV} \ov \epsilon}^2
		\paren{1 \ov c_{\cancel{\tx{inert}}}}^2
		\paren{M_\tx{med} \ov 1\,\tx{TeV}}^4
		\paren{1\,\tx{TeV} \ov M_\chi}^3,
	\label{eq:tau-decaying-chi}
}
where, if this value is longer than the age of the \cred{Universe} $\sim 10^{17}\,\tx{s}$, this decaying DM can explain the relic abundance correctly.
Via Eq.~\eqref{eq:tau-decaying-chi}, we can conclude that, for a very wide region of $\br{ \epsilon, c_{\cancel{\tx{inert}}} }$, the effect of the soft explicit breaking of the $Z_2$ symmetry is harmless for the existence of a suitable DM candidate.
This means that the simplest option for evading the domain-wall problem works fine in our setup.

\section{Summary
\label{sec:Summary}}

We have constructed a new Scotogenic type model based on a global $Z_4$ symmetry involving dark matter candidates.
After the symmetry breaking of it as $Z_4$ to $Z_2$ via the singlet scalar VEV, the lightest Majorana fermion works as a viable WIMP DM candidate and the mass terms for active neutrinos are generated as a finite quantum correction.
A key point of realising our Scotogenic structure is to introduce two types of Majorana fermions (heavy right-handed neutrinos) and inert Higgs doublets with opposite $Z_4$ parities. 
Since a large VEV for the singlet scalar is not so harmful in an appropriate realisation of the Higgs mechanism for the SM gauge symmetry, we can naturally realise a TeV-scale fermionic DM candidate, where constraints via direct detection experiments are less than those for sub-TeV WIMPs.
For the case that $N_{B_3}$ is the DM candidate, we find a natural coannihilation partner $N_{A_3}$.
By taking `asymmetric' profiles in the Yukawa couplings to the singlet scalar $S$ as $y_{A_{33}} \sim {\cal O}\fn{1}$ but $y_{B_{33}} = 0$ and small but nonzero $\alpha$ and $\theta_{N_3}$, a `well-tempered' situation is realised, where the direct-detection cross section is relatively \cred{suppressed}, and the decent-sized Higgs-portal interaction (with coannihilation) naturally leads to a correct amount of the relic abundance of the fermionic thermal WIMP DM.
Our scenario has lots of parameter points where the observed DM relic abundance is explained under the correct realisation of the active neutrino masses and mixing angles (refer to Figs.~\ref{fig:DM-result-1-updated} and \ref{fig:DM-result-2-updated}).

The current scenario, above-TeV DM and its mediators, are favoured to evade the severe constraint via direct detection.
At the LHC, such cases are hard to investigate via events with large missing transverse energy, but the development in the measurement of the Higgs trilinear self-coupling enables us to survey the validity of our scenario indirectly.
As summarised in Table~\ref{tab:VstabilityCon-updated} and Fig.~\ref{fig:kappa3}, in our scenario, valid parameter points for the active neutrino profiles and the correct amount of WIMP DM tend to be located near or a bit away from the current boundary provided by the ATLAS experiment.
The future ATLAS-CMS investigation at high-luminosity LHC will survey such benchmark points.

Since our scenario predicts the transition of the global discrete symmetry $Z_4 \to Z_2$, the formation of domain walls at an early stage of the \cred{Universe} can lead to a mismatch with observations.
We proposed three possible solutions, where the two require some extension of the model, while the last, simplest one, introducing an explicit soft \cred{$Z_2$-symmetry-breaking} term, can be addressed within the present scenario. 
We checked that the simplest prescription works fine, where the lifetime of the decaying DM easily becomes longer than the age of the \cred{Universe}.

We focus on a significant property of our scenario.
As shown in Tables~\ref{tab:lepton_Higgs_portal_BP-updated} and \ref{tab:VstabilityCon-updated}, the Yukawa coupling to the singlet scalar $y_{A_{33}}$ and some of the scalar quartic couplings tend to be ${\cal O}\fn{1}$ at valid benchmarks.
If they follow the running through the renormalisation group equations, the running couplings may blow up or blow down at an energy scale that is not so far from the TeV scale.
Therefore, if we attempt to explain the properties of neutrinos and dark matter in this scenario, the scale of inflation must be quite low.
It looks like an interesting future study to investigate which low-energy inflation scenarios are consistent with our scenario.
Furthermore, our scenario includes numerous right-handed neutrinos and numerous new scalar fields, including $SU(2)_\tx{L}$ inert doublets.
It is an interesting challenge to consider various TeV-scale leptogenesis scenarios under such circumstances.
{Also, the feature of $Z_4$ to $Z_2$ symmetry breaking with a real scalar field can also be studied in the future through gravitational wave production.}


\section*{Acknowledgements}
We appreciate Igor Ivanov and Tetsuo Shindou for providing us with various fruitful comments during the workshop HPNP2025.
We thank Shiv Nadar Institution of Eminence for providing us with workstations for numerical calculations.
One of the authors (N.~J.~J.) is grateful to the Office of the Dean of Academics and the Department of Physics of Shiv Nadar Institution of Eminence for the kind financial support for the business trip to Japan in June 2025.


\section*{Appendix}
\appendix

\section{Decay width of $s$
\label{sec:decay-s}}

We begin with the scalar potential:
\al{
{\cal V}\supset{\lambda_{HS} \ov 2} S^2 \paren{H^\dagger H} 
		+ {\lambda_{H} \ov 2} \paren{H^\dagger H}^2 + {\lambda_S \ov 4} S^4.
}
After acquiring VEV, the SM Higgs and \cred{singlet} scalars are parametrised as follows,
\als{
H &= \bp 0 \\ {1 \ov \sqrt{2}} \paren{v_H + h'} \ep,&
S &= \paren{v_S + s'},&
	\\
s' &= s_\alpha h + c_\alpha s,&
h' &= c_\alpha h -  s_\alpha s.&
}

To compute the decay width $\Gamma_s$, we first isolate the interaction term in the potential that contributes to the decay of the singlet scalar $s$ into two Higgs bosons $hh$:\[
{\cal V}\supset {1\ov 2} \lambda_{shh}shh.
\]
The effective trilinear coupling coefficient $c_{shh}$ is
\al{
\lambda_{shh} = -3 \lambda_H v_{H}  c_\alpha^2  s_\alpha  + \frac{\lambda_{HS}}{4} \br{ v_S\paren{c_\alpha + 3c_{3\alpha}} + v_H\paren{- s_\alpha + 3s_{3\alpha}}} + 6 {\lambda_S} v_S c_\alpha  s_\alpha^2 .
}
The square amplitude for this process is,
\al{
\ab{M_{shh}}^2 = \lambda_{shh}^2\,.
}
The decay width of a particle of mass $M_s$ decaying into two identical Higgs particles is
\al{
\Gamma_{{s \to hh}} &={1\ov 2!} {p^* \ov 32 \pi^2 M_s^2} \int \ab{M_{shh}}^2 d\Omega,\\
p^* &={1\ov 2 M_s}\sqrt{\paren{M_s^2- 4M_h^2}M_s^2} \simeq {M_s \ov 2},
}
where we used the approximation $M_h\ll M_s$ for our focused parameter region.
\al{
\Gamma_{s \to hh} &= { \ab{M_{shh}}^2 \ov 32 \pi M_s},
}
where \cred{the} decay width of $s$ is dominantly described by $s\xrightarrow{}hh$ as $\Gamma_s \simeq \Gamma_{s \to hh}$.

\begin{figure}[t]
    \centering
    \begin{minipage}{0.48\textwidth}
        \centering
        \includegraphics[width=\linewidth]{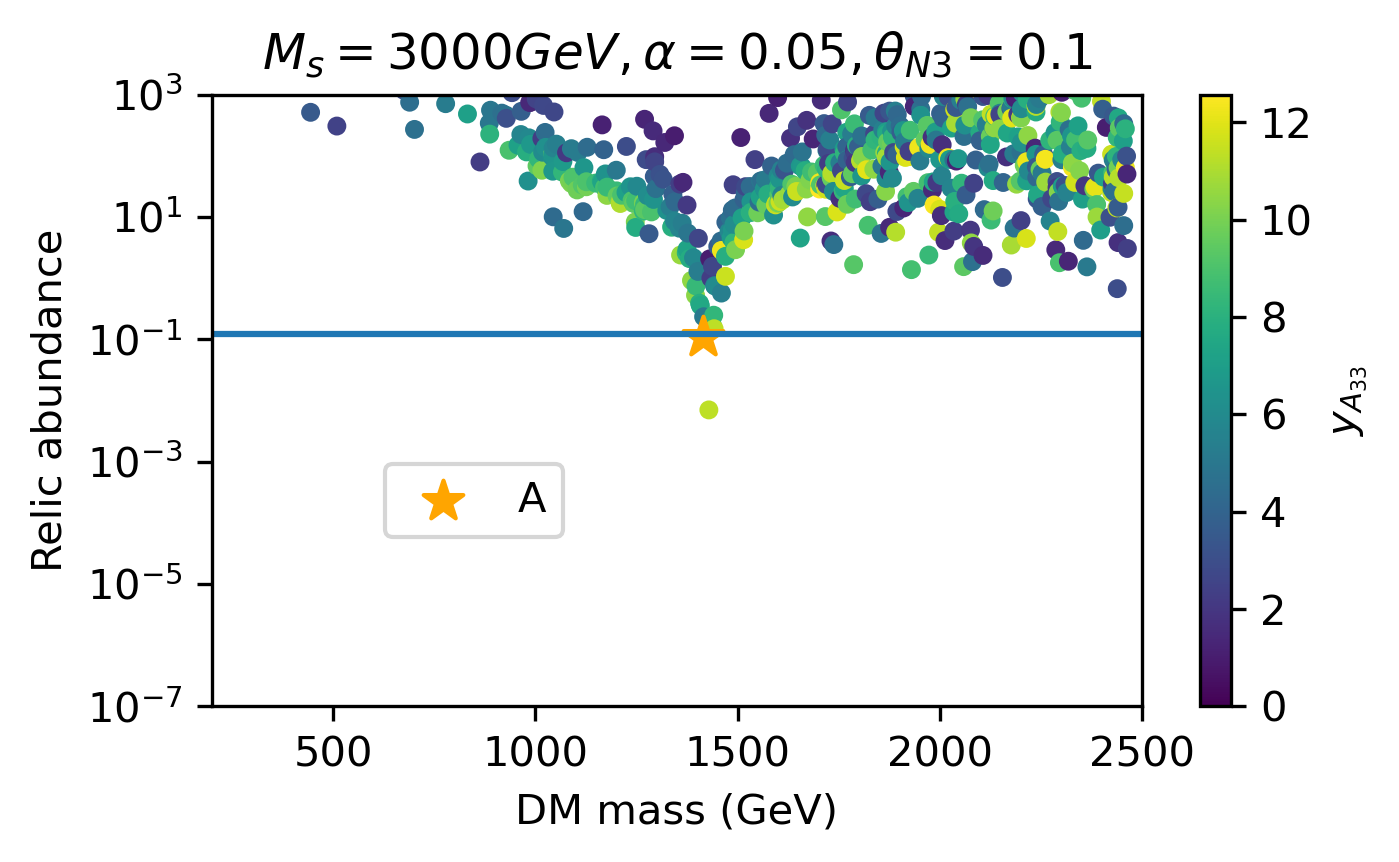} 
    \end{minipage}
    \hfill
    \begin{minipage}{0.48\textwidth}
        \centering
        \includegraphics[width=\linewidth]{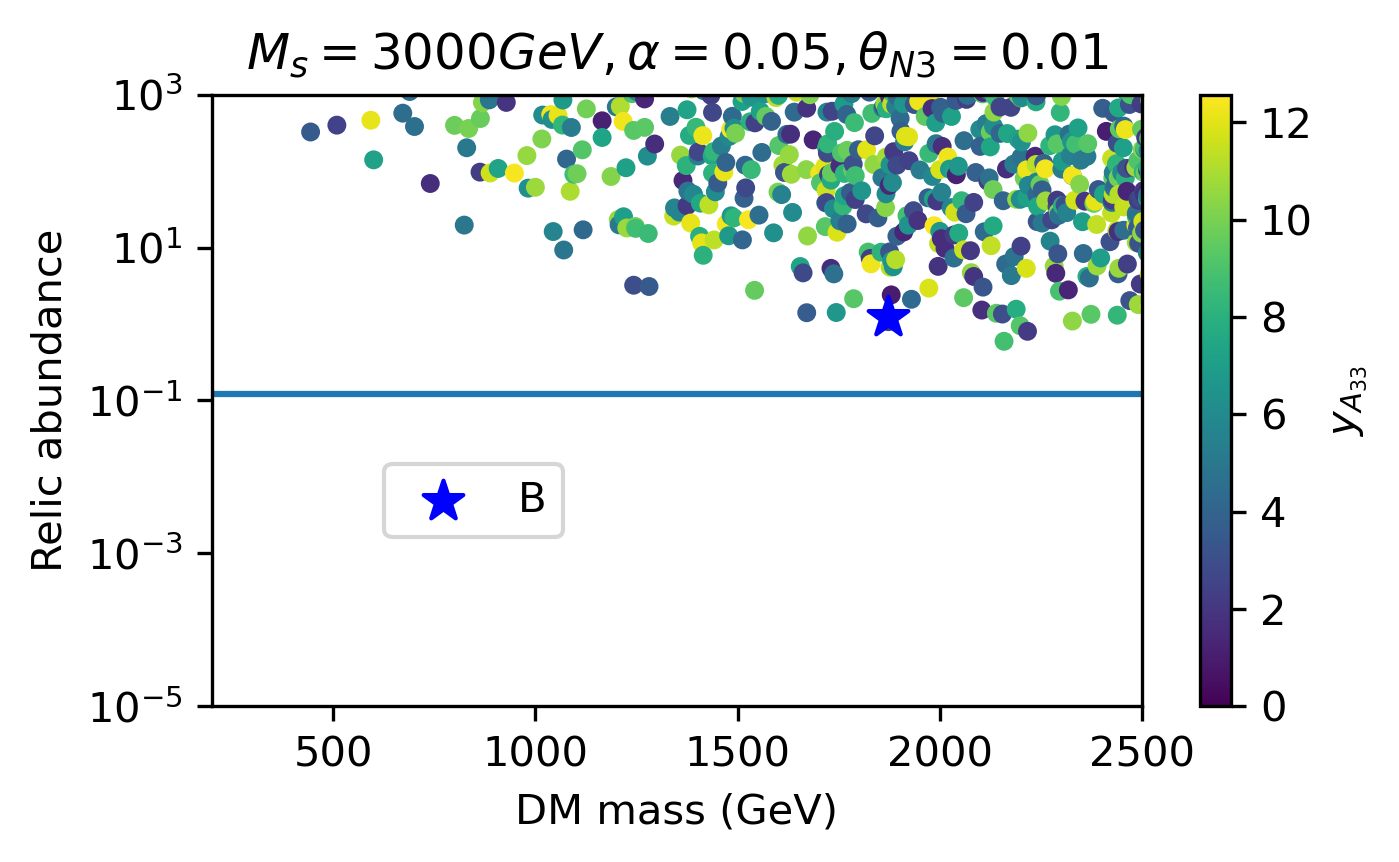} 
    \end{minipage}
    \caption{
Results of the relic abundance calculations under $M_s = 3\,\tx{TeV}$, $\alpha = 0.05$; $\theta_{N_3} = 0.1$ (Left panel) and $\theta_{N_3} = 0.01$ (Right panel), without coannihilation.
The horizontal blue line depicts the experimental observation of the DM relic abundance {($\Omega h^2 = 0.120 \pm 0.001$)~\cite{Planck:2018vyg}}.
Note that each point of the panels is consistent with {the observed neutrino profiles}, the constraints via the lepton flavour violation ($l_\alpha \to l_\beta + \gamma$) and  $T$ parameter.
The colours of the points indicate the magnitudes of $y_{A_{33}}$, while we take $y_{B_{33}} = 0$.
The details of the two benchmark points are available in Table~\ref{tab:BP-no-coannihilation}.
}
    \label{fig:DM-result-no-coannihilation}
\end{figure}

\begin{table}[t]
\centering
\begin{tabular}{|l||c|c|}
\hline
\textbf{Parameter} & \textbf{A} & \textbf{B}  \\
\hline
\hline
$\Omega h^2$ & 0.1107 & 1.2359  \\
\hline \hline
$M_s$ (TeV) & 3.0 & 3.0  \\
\hline
$M_{X_1} (= M_{N_{B_3}})$ (GeV) & 1415 & 1873 \\
\hline
$M_{X_2} (= M_{N_{A_3}})$ (GeV) & 1698 & 1883 \\
\hline
$M_{\eta_1^+}$ (GeV) & 3135.74 & 3645.31 \\
\hline
$M_{\eta_2^+}$ (GeV) & 3131.73 & 3478.74 \\
\hline
$M_{R_1}$ (GeV) & 3171.44 & 3611.76 \\
\hline
$M_{R_2}$ (GeV) & 3084.18 & 3469.54 \\
\hline
$M_{N_{A_1}}$ (GeV) & 1704 & 2253  \\
\hline
$M_{N_{A_2}}$ (GeV) & 1703 & 2252 \\
\hline
$M_{N_{B_1}}$ (GeV) & 1705 & 2254  \\
\hline
$M_{N_{B_2}}$ (GeV) & 1702 & 2251 \\
\hline
$\theta_R$ & 0.3529 & 0.3270  \\
\hline
$\theta_{N_1} (= \theta_{N_2}), \theta_{N_3}$ & 0.7, 0.1 & 0.7, 0.01 \\
\hline
$\theta_{cs}$ & 0.6 & 0.6  \\
\hline
$\alpha$ & 0.05 & 0.05  \\
\hline
$\log_{10}(\lambda_5)$ & -6.7201 & -7.2654 \\
\hline
$\lambda_6$ & 0 & 0  \\
\hline
$y_{A_{33}}$ & 8.0394 & 5.5746  \\
\hline
$y_{B_{33}}$ & 0 & 0  \\
\hline \hline
$M_{I_1}$ (GeV) & 3171.44 & 3611.76 \\
\hline
$M_{I_2}$ (GeV) & 3084.18 & 3469.54 \\
\hline
\end{tabular}
\qquad
\begin{tabular}{|l||c|c|}
        \hline
        \textbf{Parameter} & \textbf{A} & \textbf{B}\\
        \hline \hline
        $\lambda_{S\eta A}$ & 1.1854 & 1.1480 \\
        \hline
        $\lambda_{S\eta B}$ & 1.3617 & 1.1564 \\
        \hline
        $\lambda_{2_a}$ & 5 & 5\\
        \hline
        $\lambda_{2_b}$ & 7 & 5\\
        \hline
        $\lambda_{2_c}$ & 7 & 5\\
        \hline
        $\lambda_{2_d}$ & 8 & 8.5\\
        \hline
        $\lambda_{3A}$ & $-4.5508$ & -0.0226\\
        \hline
        $\lambda_{3B}$ & $8.8775$ & 12.1819\\
        \hline
        $\lambda_{4A}$ & $5.5508$ & 1.0226\\
        \hline
        $\lambda_{4B}$ & $-7.8775$ & -11.1819\\
        \hline
        $m_{\eta A}$ (GeV) & 2203.44\,$i$ & 1199.68\,$i$\\
        \hline
        $m_{\eta B}$ (GeV) & 2733.92\,$i$ & 1530.95\,$i$\\
        \hline
        $\lambda_{HS}$ & 0.36461 & 0.36461 \\
        \hline
        $\lambda_{H}$  & 0.629044 & 0.629044\\
        \hline
        $\lambda_{S}$  & 0.179551 & 0.179551  \\
        \hline
\end{tabular}
\caption{
[Left Table] The values of the relevant parameters for the two benchmark points without coannihilation in Fig.~\ref{fig:DM-result-no-coannihilation}, and realised relic abundances ($\Omega h^2$) are shown.
Note that they are consistent with {the observed neutrino profiles}, the lepton flavour violation ($l_\alpha \to l_\beta + \gamma$), $T$ parameter, and DM direct detection.
[Right Table] For each of the two benchmark points (A and B) of Fig.~\ref{fig:DM-result-no-coannihilation}, we provide a set of valid parameters describing the potential (also not being relevant for the phenomena considered in the numerical scans).
In any one of the sets, all of the conditions for 
the {bounded} from below [from Eq.~\eqref{eq:Det-condition-start} to Eq.~\eqref{eq:Det-condition-end}]
and the inert condition [from Eqs.~\eqref{eq:Inert-Condition-1} and \eqref{eq:Inert-Condition-2}] are satisfied.
}
    \label{tab:BP-no-coannihilation}
\end{table}

\section{DM results without coannihilation
\label{sec:DM-result-with-no-coannihilation}}

Here, we provide the numerical scans similar to those in Section~\ref{sec:Analysis}, but without coannihilation by taking
the mass of the coannihilation partner $M_{\chi_2} (= M_{N_{A_3}})$ as 20$\%$ larger than the DM mass $M_{\chi_1} (= M_{N_{B_3}})$, while the other schemes are the same.
For each scan, as before, we randomly selected the paramters $M_{\chi_1} (= M_{N_{B_3}})$, $M_{N_{A_1}}$, $M_{N_{B_1}}$, $M_{N_{A_2}}$, $M_{N_{B_2}}$, $M_{\eta^+_1}$,
$M_{\eta^+_2}$, $M_{R_1}$, $M_{R_2}$, $\theta_R$, $\lambda_5$, $y_{A_{33}}$, while the following other independent relevant parameters are fixed;
$M_s$, $M_{\chi_2} (= M_{N_{A_3}})$, $\theta_{N_1}$, $\theta_{N_2}$, $\theta_{N_3}$, $\theta_\tx{cs}$, $\alpha$, $\lambda_6 \, (=0)$, and $y_{B_{33}} \, (=0)$.
We chose the three significant parameters $M_s$, $\alpha$ and $\theta_{N_3}$ as $3\,\tx{TeV}$, $0.05$, and $0.1$ or $0.01$, respectively.
The results of the scans are available in Fig.~\ref{fig:DM-result-no-coannihilation}, where Left and Right panels correspond to $\theta_{N_3} = 0.1$ and $\theta_{N_3} = 0.01$.
It is apparent from the results, no valid points for addressing the DM relic and the active neutrino profiles are found, except for very near the narrow resonance of $s$.\footnote{
In our scan for $\theta_{N_3} = 0.01$, due to a suppressed DM interaction owing to the smaller value of $\theta_{N_3}$, we were not able to find a point near the resonance of $s$.
}
The details of the two focused points A and B are available in Table~\ref{tab:BP-no-coannihilation}.\footnote{
Note that the values of $M_{I_1}$ and $M_{I_2}$ are determined by the parameters above, but we show the corresponding digits for convenience.
Also, the two differences between $M_{R_i}$ and $M_{I_i}$ $(i=1,2)$ are not observed within the shown digits in Table~\ref{tab:lepton_Higgs_portal_BP-updated}, but they should be nonzero to produce the minuscule mass differences of the active neutrinos (see Section~\ref{sec:massterm}).
{Due to the same reason, $\theta_I$ (not shown in Table~\ref{tab:BP-no-coannihilation}) becomes extremely close to $\theta_R$.}
}
{While the choice of asymmetric Yukawa couplings is not optimal in the non-coannihilation regime, 
the inferences remain unchanged, i.e. only the narrow resonance region can produce the correct relic abundance.}

\bibliographystyle{utphys}
{\small \bibliography{Z4-Scotogenic,Scotogenic-Varieties,ref_Noel,ref_Noel-2}}

\providecommand{\href}[2]{#2}\begingroup\raggedright\begin{thebibliography}{100}

\bibitem{Super-Kamiokande:1998kpq}
{\bfseries Super-Kamiokande} Collaboration, Y.~Fukuda {\em et~al.}, ``{Evidence
  for oscillation of atmospheric neutrinos},''
  \href{http://dx.doi.org/10.1103/PhysRevLett.81.1562}{{\em Phys. Rev. Lett.}
  {\bfseries 81} (1998) 1562--1567},
  \href{http://arxiv.org/abs/hep-ex/9807003}{{\ttfamily arXiv:hep-ex/9807003}}.

\bibitem{SNO:2002tuh}
{\bfseries SNO} Collaboration, Q.~R. Ahmad {\em et~al.}, ``{Direct evidence for
  neutrino flavor transformation from neutral current interactions in the
  Sudbury Neutrino Observatory},''
  \href{http://dx.doi.org/10.1103/PhysRevLett.89.011301}{{\em Phys. Rev. Lett.}
  {\bfseries 89} (2002) 011301},
  \href{http://arxiv.org/abs/nucl-ex/0204008}{{\ttfamily
  arXiv:nucl-ex/0204008}}.

\bibitem{Giganti:2017fhf}
C.~Giganti, S.~Lavignac, and M.~Zito, ``{Neutrino oscillations: The rise of the
  PMNS paradigm},'' \href{http://dx.doi.org/10.1016/j.ppnp.2017.10.001}{{\em
  Prog. Part. Nucl. Phys.} {\bfseries 98} (2018) 1--54},
  \href{http://arxiv.org/abs/1710.00715}{{\ttfamily arXiv:1710.00715
  [hep-ex]}}.

\bibitem{Pontecorvo:1957cp}
B.~Pontecorvo, ``{Mesonium and anti-mesonium},'' {\em Sov. Phys. JETP}
  {\bfseries 6} (1957) 429.

\bibitem{Pontecorvo:1957qd}
B.~Pontecorvo, ``{Inverse beta processes and nonconservation of lepton
  charge},'' {\em Zh. Eksp. Teor. Fiz.} {\bfseries 34} (1957) 247.

\bibitem{Maki:1962mu}
Z.~Maki, M.~Nakagawa, and S.~Sakata, ``{Remarks on the unified model of
  elementary particles},'' \href{http://dx.doi.org/10.1143/PTP.28.870}{{\em
  Prog. Theor. Phys.} {\bfseries 28} (1962) 870--880}.

\bibitem{DUNE:2020jqi}
{\bfseries DUNE} Collaboration, B.~Abi {\em et~al.}, ``{Long-baseline neutrino
  oscillation physics potential of the DUNE experiment},''
  \href{http://dx.doi.org/10.1140/epjc/s10052-020-08456-z}{{\em Eur. Phys. J.
  C} {\bfseries 80} no.~10, (2020) 978},
  \href{http://arxiv.org/abs/2006.16043}{{\ttfamily arXiv:2006.16043
  [hep-ex]}}.

\bibitem{Hyper-Kamiokande:2016srs}
{\bfseries Hyper-Kamiokande} Collaboration, K.~Abe {\em et~al.}, ``{Physics
  potentials with the second Hyper-Kamiokande detector in Korea},''
  \href{http://dx.doi.org/10.1093/ptep/pty044}{{\em PTEP} {\bfseries 2018}
  no.~6, (2018) 063C01}, \href{http://arxiv.org/abs/1611.06118}{{\ttfamily
  arXiv:1611.06118 [hep-ex]}}.

\bibitem{Xing:2020ijf}
Z.-z. Xing, ``{Flavor structures of charged fermions and massive neutrinos},''
  \href{http://dx.doi.org/10.1016/j.physrep.2020.02.001}{{\em Phys. Rept.}
  {\bfseries 854} (2020) 1--147},
  \href{http://arxiv.org/abs/1909.09610}{{\ttfamily arXiv:1909.09610
  [hep-ph]}}.

\bibitem{Minkowski:1977sc}
P.~Minkowski, ``{$\mu \to e\gamma$ at a Rate of One Out of $10^{9}$ Muon
  Decays?},'' \href{http://dx.doi.org/10.1016/0370-2693(77)90435-X}{{\em Phys.
  Lett. B} {\bfseries 67} (1977) 421--428}.

\bibitem{Yanagida:1979as}
T.~Yanagida, ``{Horizontal gauge symmetry and masses of neutrinos},'' {\em
  Conf. Proc. C} {\bfseries 7902131} (1979) 95--99.

\bibitem{Gell-Mann:1979vob}
M.~Gell-Mann, P.~Ramond, and R.~Slansky, ``{Complex Spinors and Unified
  Theories},'' {\em Conf. Proc. C} {\bfseries 790927} (1979) 315--321,
  \href{http://arxiv.org/abs/1306.4669}{{\ttfamily arXiv:1306.4669 [hep-th]}}.

\bibitem{Glashow:1979nm}
S.~L. Glashow, ``{The Future of Elementary Particle Physics},''
  \href{http://dx.doi.org/10.1007/978-1-4684-7197-7_15}{{\em NATO Sci. Ser. B}
  {\bfseries 61} (1980) 687}.

\bibitem{Mohapatra:1979ia}
R.~N. Mohapatra and G.~Senjanovic, ``{Neutrino Mass and Spontaneous Parity
  Nonconservation},'' \href{http://dx.doi.org/10.1103/PhysRevLett.44.912}{{\em
  Phys. Rev. Lett.} {\bfseries 44} (1980) 912}.

\bibitem{Ramond:1979py}
P.~Ramond, ``{The Family Group in Grand Unified Theories},'' in {\em
  {International Symposium on Fundamentals of Quantum Theory and Quantum Field
  Theory}}.
\newblock 2, 1979.
\newblock \href{http://arxiv.org/abs/hep-ph/9809459}{{\ttfamily
  arXiv:hep-ph/9809459}}.

\bibitem{Schechter:1980gr}
J.~Schechter and J.~W.~F. Valle, ``{Neutrino Masses in SU(2) x U(1)
  Theories},'' \href{http://dx.doi.org/10.1103/PhysRevD.22.2227}{{\em Phys.
  Rev. D} {\bfseries 22} (1980) 2227}.

\bibitem{Lazarides:1980nt}
G.~Lazarides, Q.~Shafi, and C.~Wetterich, ``{Proton Lifetime and Fermion Masses
  in an SO(10) Model},''
  \href{http://dx.doi.org/10.1016/0550-3213(81)90354-0}{{\em Nucl. Phys. B}
  {\bfseries 181} (1981) 287--300}.

\bibitem{Mohapatra:1980yp}
R.~N. Mohapatra and G.~Senjanovic, ``{Neutrino Masses and Mixings in Gauge
  Models with Spontaneous Parity Violation},''
  \href{http://dx.doi.org/10.1103/PhysRevD.23.165}{{\em Phys. Rev. D}
  {\bfseries 23} (1981) 165}.

\bibitem{Wetterich:1981bx}
C.~Wetterich, ``{Neutrino Masses and the Scale of B-L Violation},''
  \href{http://dx.doi.org/10.1016/0550-3213(81)90279-0}{{\em Nucl. Phys. B}
  {\bfseries 187} (1981) 343--375}.

\bibitem{Schechter:1981cv}
J.~Schechter and J.~W.~F. Valle, ``{Neutrino Decay and Spontaneous Violation of
  Lepton Number},'' \href{http://dx.doi.org/10.1103/PhysRevD.25.774}{{\em Phys.
  Rev. D} {\bfseries 25} (1982) 774}.

\bibitem{Foot:1988aq}
R.~Foot, H.~Lew, X.~G. He, and G.~C. Joshi, ``{Seesaw Neutrino Masses Induced
  by a Triplet of Leptons},'' \href{http://dx.doi.org/10.1007/BF01415558}{{\em
  Z. Phys. C} {\bfseries 44} (1989) 441}.

\bibitem{Mohapatra:1986bd}
R.~N. Mohapatra and J.~W.~F. Valle, ``{Neutrino Mass and Baryon Number
  Nonconservation in Superstring Models},''
  \href{http://dx.doi.org/10.1103/PhysRevD.34.1642}{{\em Phys. Rev. D}
  {\bfseries 34} (1986) 1642}.

\bibitem{Wyler:1982dd}
D.~Wyler and L.~Wolfenstein, ``{Massless Neutrinos in Left-Right Symmetric
  Models},'' \href{http://dx.doi.org/10.1016/0550-3213(83)90482-0}{{\em Nucl.
  Phys. B} {\bfseries 218} (1983) 205--214}.

\bibitem{Akhmedov:1995ip}
E.~K. Akhmedov, M.~Lindner, E.~Schnapka, and J.~W.~F. Valle, ``{Left-right
  symmetry breaking in NJL approach},''
  \href{http://dx.doi.org/10.1016/0370-2693(95)01504-3}{{\em Phys. Lett. B}
  {\bfseries 368} (1996) 270--280},
  \href{http://arxiv.org/abs/hep-ph/9507275}{{\ttfamily arXiv:hep-ph/9507275}}.

\bibitem{Akhmedov:1995vm}
E.~K. Akhmedov, M.~Lindner, E.~Schnapka, and J.~W.~F. Valle, ``{Dynamical
  left-right symmetry breaking},''
  \href{http://dx.doi.org/10.1103/PhysRevD.53.2752}{{\em Phys. Rev. D}
  {\bfseries 53} (1996) 2752--2780},
  \href{http://arxiv.org/abs/hep-ph/9509255}{{\ttfamily arXiv:hep-ph/9509255}}.

\bibitem{Fukugita:1986hr}
M.~Fukugita and T.~Yanagida, ``{Baryogenesis Without Grand Unification},''
  \href{http://dx.doi.org/10.1016/0370-2693(86)91126-3}{{\em Phys. Lett. B}
  {\bfseries 174} (1986) 45--47}.

\bibitem{Weinberg:1979sa}
S.~Weinberg, ``{Baryon and Lepton Nonconserving Processes},''
  \href{http://dx.doi.org/10.1103/PhysRevLett.43.1566}{{\em Phys. Rev. Lett.}
  {\bfseries 43} (1979) 1566--1570}.

\bibitem{Bonnet:2012kz}
F.~Bonnet, M.~Hirsch, T.~Ota, and W.~Winter, ``{Systematic study of the d=5
  Weinberg operator at one-loop order},''
  \href{http://dx.doi.org/10.1007/JHEP07(2012)153}{{\em JHEP} {\bfseries 07}
  (2012) 153}, \href{http://arxiv.org/abs/1204.5862}{{\ttfamily arXiv:1204.5862
  [hep-ph]}}.

\bibitem{Zee:1980ai}
A.~Zee, ``{A Theory of Lepton Number Violation, Neutrino Majorana Mass, and
  Oscillation},'' \href{http://dx.doi.org/10.1016/0370-2693(80)90349-4}{{\em
  Phys. Lett. B} {\bfseries 93} (1980) 389}. [Erratum: Phys.Lett.B 95, 461
  (1980)].

\bibitem{Cheng:1980qt}
T.~P. Cheng and L.-F. Li, ``{Neutrino Masses, Mixings and Oscillations in SU(2)
  x U(1) Models of Electroweak Interactions},''
  \href{http://dx.doi.org/10.1103/PhysRevD.22.2860}{{\em Phys. Rev. D}
  {\bfseries 22} (1980) 2860}.

\bibitem{Zee:1985id}
A.~Zee, ``{Quantum Numbers of Majorana Neutrino Masses},''
  \href{http://dx.doi.org/10.1016/0550-3213(86)90475-X}{{\em Nucl. Phys. B}
  {\bfseries 264} (1986) 99--110}.

\bibitem{Babu:1988ki}
K.~S. Babu, ``{Model of 'Calculable' Majorana Neutrino Masses},''
  \href{http://dx.doi.org/10.1016/0370-2693(88)91584-5}{{\em Phys. Lett. B}
  {\bfseries 203} (1988) 132--136}.

\bibitem{Ma:1998dn}
E.~Ma, ``{Pathways to naturally small neutrino masses},''
  \href{http://dx.doi.org/10.1103/PhysRevLett.81.1171}{{\em Phys. Rev. Lett.}
  {\bfseries 81} (1998) 1171--1174},
  \href{http://arxiv.org/abs/hep-ph/9805219}{{\ttfamily arXiv:hep-ph/9805219}}.

\bibitem{Krauss:2002px}
L.~M. Krauss, S.~Nasri, and M.~Trodden, ``{A Model for neutrino masses and dark
  matter},'' \href{http://dx.doi.org/10.1103/PhysRevD.67.085002}{{\em Phys.
  Rev. D} {\bfseries 67} (2003) 085002},
  \href{http://arxiv.org/abs/hep-ph/0210389}{{\ttfamily arXiv:hep-ph/0210389}}.

\bibitem{Cai:2017jrq}
Y.~Cai, J.~Herrero-Garc\'\i{}a, M.~A. Schmidt, A.~Vicente, and R.~R. Volkas,
  ``{From the trees to the forest: a review of radiative neutrino mass
  models},'' \href{http://dx.doi.org/10.3389/fphy.2017.00063}{{\em Front. in
  Phys.} {\bfseries 5} (2017) 63},
  \href{http://arxiv.org/abs/1706.08524}{{\ttfamily arXiv:1706.08524
  [hep-ph]}}.

\bibitem{Ma:2006km}
E.~Ma, ``{Verifiable radiative seesaw mechanism of neutrino mass and dark
  matter},'' \href{http://dx.doi.org/10.1103/PhysRevD.73.077301}{{\em Phys.
  Rev. D} {\bfseries 73} (2006) 077301},
  \href{http://arxiv.org/abs/hep-ph/0601225}{{\ttfamily arXiv:hep-ph/0601225}}.

\bibitem{Ma:2008cu}
E.~Ma and D.~Suematsu, ``{Fermion Triplet Dark Matter and Radiative Neutrino
  Mass},'' \href{http://dx.doi.org/10.1142/S021773230903059X}{{\em Mod. Phys.
  Lett. A} {\bfseries 24} (2009) 583--589},
  \href{http://arxiv.org/abs/0809.0942}{{\ttfamily arXiv:0809.0942 [hep-ph]}}.

\bibitem{Farzan:2009ji}
Y.~Farzan, ``{A Minimal model linking two great mysteries: neutrino mass and
  dark matter},'' \href{http://dx.doi.org/10.1103/PhysRevD.80.073009}{{\em
  Phys. Rev. D} {\bfseries 80} (2009) 073009},
  \href{http://arxiv.org/abs/0908.3729}{{\ttfamily arXiv:0908.3729 [hep-ph]}}.

\bibitem{Chen:2009gd}
C.-H. Chen, C.-Q. Geng, and D.~V. Zhuridov, ``{Neutrino Masses, Leptogenesis
  and Decaying Dark Matter},''
  \href{http://dx.doi.org/10.1088/1475-7516/2009/10/001}{{\em JCAP} {\bfseries
  10} (2009) 001}, \href{http://arxiv.org/abs/0906.1646}{{\ttfamily
  arXiv:0906.1646 [hep-ph]}}.

\bibitem{Farzan:2010mr}
Y.~Farzan, S.~Pascoli, and M.~A. Schmidt, ``{AMEND: A model explaining neutrino
  masses and dark matter testable at the LHC and MEG},''
  \href{http://dx.doi.org/10.1007/JHEP10(2010)111}{{\em JHEP} {\bfseries 10}
  (2010) 111}, \href{http://arxiv.org/abs/1005.5323}{{\ttfamily arXiv:1005.5323
  [hep-ph]}}.

\bibitem{Aoki:2011yk}
M.~Aoki, S.~Kanemura, and K.~Yagyu, ``{Doubly-charged scalar bosons from the
  doublet},'' \href{http://dx.doi.org/10.1016/j.physletb.2011.07.017}{{\em
  Phys. Lett. B} {\bfseries 702} (2011) 355--358},
  \href{http://arxiv.org/abs/1105.2075}{{\ttfamily arXiv:1105.2075 [hep-ph]}}.
  [Erratum: Phys.Lett.B 706, 495--495 (2012)].

\bibitem{Parida:2011wh}
M.~K. Parida, ``{Radiative Seesaw in SO(10) with Dark Matter},''
  \href{http://dx.doi.org/10.1016/j.physletb.2011.09.016}{{\em Phys. Lett. B}
  {\bfseries 704} (2011) 206--210},
  \href{http://arxiv.org/abs/1106.4137}{{\ttfamily arXiv:1106.4137 [hep-ph]}}.

\bibitem{Cai:2011qr}
Y.~Cai, X.-G. He, M.~Ramsey-Musolf, and L.-H. Tsai, ``{R$\nu$MDM and Lepton
  Flavor Violation},'' \href{http://dx.doi.org/10.1007/JHEP12(2011)054}{{\em
  JHEP} {\bfseries 12} (2011) 054},
  \href{http://arxiv.org/abs/1108.0969}{{\ttfamily arXiv:1108.0969 [hep-ph]}}.

\bibitem{Chao:2012sz}
W.~Chao, ``{Dark matter, LFV and neutrino magnetic moment in the radiative
  seesaw model with fermion triplet},''
  \href{http://dx.doi.org/10.1142/S0217751X15500074}{{\em Int. J. Mod. Phys. A}
  {\bfseries 30} no.~01, (2015) 1550007},
  \href{http://arxiv.org/abs/1202.6394}{{\ttfamily arXiv:1202.6394 [hep-ph]}}.

\bibitem{Farzan:2012sa}
Y.~Farzan and E.~Ma, ``{Dirac neutrino mass generation from dark matter},''
  \href{http://dx.doi.org/10.1103/PhysRevD.86.033007}{{\em Phys. Rev. D}
  {\bfseries 86} (2012) 033007},
  \href{http://arxiv.org/abs/1204.4890}{{\ttfamily arXiv:1204.4890 [hep-ph]}}.

\bibitem{Okada:2012np}
H.~Okada and T.~Toma, ``{Fermionic Dark Matter in Radiative Inverse Seesaw
  Model with $U(1)_{B-L}$},''
  \href{http://dx.doi.org/10.1103/PhysRevD.86.033011}{{\em Phys. Rev. D}
  {\bfseries 86} (2012) 033011},
  \href{http://arxiv.org/abs/1207.0864}{{\ttfamily arXiv:1207.0864 [hep-ph]}}.

\bibitem{Hehn:2012kz}
D.~Hehn and A.~Ibarra, ``{A radiative model with a naturally mild neutrino mass
  hierarchy},'' \href{http://dx.doi.org/10.1016/j.physletb.2012.11.034}{{\em
  Phys. Lett. B} {\bfseries 718} (2013) 988--991},
  \href{http://arxiv.org/abs/1208.3162}{{\ttfamily arXiv:1208.3162 [hep-ph]}}.

\bibitem{Kajiyama:2013zla}
Y.~Kajiyama, H.~Okada, and K.~Yagyu, ``{Two Loop Radiative Seesaw Model with
  Inert Triplet Scalar Field},''
  \href{http://dx.doi.org/10.1016/j.nuclphysb.2013.05.020}{{\em Nucl. Phys. B}
  {\bfseries 874} (2013) 198--216},
  \href{http://arxiv.org/abs/1303.3463}{{\ttfamily arXiv:1303.3463 [hep-ph]}}.

\bibitem{Hirsch:2013ola}
M.~Hirsch, R.~A. Lineros, S.~Morisi, J.~Palacio, N.~Rojas, and J.~W.~F. Valle,
  ``{WIMP dark matter as radiative neutrino mass messenger},''
  \href{http://dx.doi.org/10.1007/JHEP10(2013)149}{{\em JHEP} {\bfseries 10}
  (2013) 149}, \href{http://arxiv.org/abs/1307.8134}{{\ttfamily arXiv:1307.8134
  [hep-ph]}}.

\bibitem{Ma:2013nga}
E.~Ma, ``{Unified Framework for Matter, Dark Matter, and Radiative Neutrino
  Mass},'' \href{http://dx.doi.org/10.1103/PhysRevD.88.117702}{{\em Phys. Rev.
  D} {\bfseries 88} no.~11, (2013) 117702},
  \href{http://arxiv.org/abs/1307.7064}{{\ttfamily arXiv:1307.7064 [hep-ph]}}.

\bibitem{Brdar:2013iea}
V.~Brdar, I.~Picek, and B.~Radovcic, ``{Radiative Neutrino Mass with Scotogenic
  Scalar Triplet},''
  \href{http://dx.doi.org/10.1016/j.physletb.2013.11.045}{{\em Phys. Lett. B}
  {\bfseries 728} (2014) 198--201},
  \href{http://arxiv.org/abs/1310.3183}{{\ttfamily arXiv:1310.3183 [hep-ph]}}.

\bibitem{Law:2013saa}
S.~S.~C. Law and K.~L. McDonald, ``{A Class of Inert N-tuplet Models with
  Radiative Neutrino Mass and Dark Matter},''
  \href{http://dx.doi.org/10.1007/JHEP09(2013)092}{{\em JHEP} {\bfseries 09}
  (2013) 092}, \href{http://arxiv.org/abs/1305.6467}{{\ttfamily arXiv:1305.6467
  [hep-ph]}}.

\bibitem{Okada:2014qsa}
H.~Okada, T.~Toma, and K.~Yagyu, ``{Inert Extension of the Zee-Babu Model},''
  \href{http://dx.doi.org/10.1103/PhysRevD.90.095005}{{\em Phys. Rev. D}
  {\bfseries 90} (2014) 095005},
  \href{http://arxiv.org/abs/1408.0961}{{\ttfamily arXiv:1408.0961 [hep-ph]}}.

\bibitem{Patra:2014sua}
S.~Patra, N.~Sahoo, and N.~Sahu, ``{Dipolar dark matter in light of the 3.5 keV
  x-ray line, neutrino mass, and LUX data},''
  \href{http://dx.doi.org/10.1103/PhysRevD.91.115013}{{\em Phys. Rev. D}
  {\bfseries 91} no.~11, (2015) 115013},
  \href{http://arxiv.org/abs/1412.4253}{{\ttfamily arXiv:1412.4253 [hep-ph]}}.

\bibitem{Fraser:2014yha}
S.~Fraser, E.~Ma, and O.~Popov, ``{Scotogenic Inverse Seesaw Model of Neutrino
  Mass},'' \href{http://dx.doi.org/10.1016/j.physletb.2014.08.069}{{\em Phys.
  Lett. B} {\bfseries 737} (2014) 280--282},
  \href{http://arxiv.org/abs/1408.4785}{{\ttfamily arXiv:1408.4785 [hep-ph]}}.

\bibitem{Okada:2014nea}
H.~Okada and Y.~Orikasa, ``{Classically conformal radiative neutrino model with
  gauged B \ensuremath{-} L symmetry},''
  \href{http://dx.doi.org/10.1016/j.physletb.2016.07.039}{{\em Phys. Lett. B}
  {\bfseries 760} (2016) 558--564},
  \href{http://arxiv.org/abs/1412.3616}{{\ttfamily arXiv:1412.3616 [hep-ph]}}.

\bibitem{Baek:2015mna}
S.~Baek, H.~Okada, and K.~Yagyu, ``{Flavour Dependent Gauged Radiative Neutrino
  Mass Model},'' \href{http://dx.doi.org/10.1007/JHEP04(2015)049}{{\em JHEP}
  {\bfseries 04} (2015) 049}, \href{http://arxiv.org/abs/1501.01530}{{\ttfamily
  arXiv:1501.01530 [hep-ph]}}.

\bibitem{Chowdhury:2015sla}
T.~A. Chowdhury and S.~Nasri, ``{Lepton Flavor Violation in the Inert Scalar
  Model with Higher Representations},''
  \href{http://dx.doi.org/10.1007/JHEP12(2015)040}{{\em JHEP} {\bfseries 12}
  (2015) 040}, \href{http://arxiv.org/abs/1506.00261}{{\ttfamily
  arXiv:1506.00261 [hep-ph]}}.

\bibitem{Diaz:2016udz}
M.~A. D\'\i{}az, N.~Rojas, S.~Urrutia-Quiroga, and J.~W.~F. Valle, ``{Heavy
  Higgs Boson Production at Colliders in the Singlet-Triplet Scotogenic Dark
  Matter Model},'' \href{http://dx.doi.org/10.1007/JHEP08(2017)017}{{\em JHEP}
  {\bfseries 08} (2017) 017}, \href{http://arxiv.org/abs/1612.06569}{{\ttfamily
  arXiv:1612.06569 [hep-ph]}}.

\bibitem{Ferreira:2016sbb}
P.~M. Ferreira, W.~Grimus, D.~Jurciukonis, and L.~Lavoura, ``{Scotogenic model
  for co-bimaximal mixing},''
  \href{http://dx.doi.org/10.1007/JHEP07(2016)010}{{\em JHEP} {\bfseries 07}
  (2016) 010}, \href{http://arxiv.org/abs/1604.07777}{{\ttfamily
  arXiv:1604.07777 [hep-ph]}}.

\bibitem{Ahriche:2016cio}
A.~Ahriche, K.~L. McDonald, and S.~Nasri, ``{The Scale-Invariant Scotogenic
  Model},'' \href{http://dx.doi.org/10.1007/JHEP06(2016)182}{{\em JHEP}
  {\bfseries 06} (2016) 182}, \href{http://arxiv.org/abs/1604.05569}{{\ttfamily
  arXiv:1604.05569 [hep-ph]}}.

\bibitem{vonderPahlen:2016cbw}
F.~von~der Pahlen, G.~Palacio, D.~Restrepo, and O.~Zapata, ``{Radiative Type
  III Seesaw Model and its collider phenomenology},''
  \href{http://dx.doi.org/10.1103/PhysRevD.94.033005}{{\em Phys. Rev. D}
  {\bfseries 94} no.~3, (2016) 033005},
  \href{http://arxiv.org/abs/1605.01129}{{\ttfamily arXiv:1605.01129
  [hep-ph]}}.

\bibitem{Lu:2016dbc}
W.-B. Lu and P.-H. Gu, ``{Mixed Inert Scalar Triplet Dark Matter, Radiative
  Neutrino Masses and Leptogenesis},''
  \href{http://dx.doi.org/10.1016/j.nuclphysb.2017.09.005}{{\em Nucl. Phys. B}
  {\bfseries 924} (2017) 279--311},
  \href{http://arxiv.org/abs/1611.02106}{{\ttfamily arXiv:1611.02106
  [hep-ph]}}.

\bibitem{Merle:2016scw}
A.~Merle, M.~Platscher, N.~Rojas, J.~W.~F. Valle, and A.~Vicente,
  ``{Consistency of WIMP Dark Matter as radiative neutrino mass messenger},''
  \href{http://dx.doi.org/10.1007/JHEP07(2016)013}{{\em JHEP} {\bfseries 07}
  (2016) 013}, \href{http://arxiv.org/abs/1603.05685}{{\ttfamily
  arXiv:1603.05685 [hep-ph]}}.

\bibitem{Rocha-Moran:2016enp}
P.~Rocha-Moran and A.~Vicente, ``{Lepton Flavor Violation in the
  singlet-triplet scotogenic model},''
  \href{http://dx.doi.org/10.1007/JHEP07(2016)078}{{\em JHEP} {\bfseries 07}
  (2016) 078}, \href{http://arxiv.org/abs/1605.01915}{{\ttfamily
  arXiv:1605.01915 [hep-ph]}}.

\bibitem{Nomura:2016dnf}
T.~Nomura, H.~Okada, and Y.~Orikasa, ``{Radiative neutrino model with $SU(2)_L$
  triplet fields},'' \href{http://dx.doi.org/10.1103/PhysRevD.94.115018}{{\em
  Phys. Rev. D} {\bfseries 94} no.~11, (2016) 115018},
  \href{http://arxiv.org/abs/1610.04729}{{\ttfamily arXiv:1610.04729
  [hep-ph]}}.

\bibitem{Cheung:2016frv}
K.~Cheung, T.~Nomura, and H.~Okada, ``{Three-loop neutrino mass model with a
  colored triplet scalar},''
  \href{http://dx.doi.org/10.1103/PhysRevD.95.015026}{{\em Phys. Rev. D}
  {\bfseries 95} no.~1, (2017) 015026},
  \href{http://arxiv.org/abs/1610.04986}{{\ttfamily arXiv:1610.04986
  [hep-ph]}}.

\bibitem{Chowdhury:2016mtl}
T.~A. Chowdhury and S.~Nasri, ``{The Sommerfeld Enhancement in the Scotogenic
  Model with Large Electroweak Scalar Multiplets},''
  \href{http://dx.doi.org/10.1088/1475-7516/2017/01/041}{{\em JCAP} {\bfseries
  01} (2017) 041}, \href{http://arxiv.org/abs/1611.06590}{{\ttfamily
  arXiv:1611.06590 [hep-ph]}}.

\bibitem{Cheung:2017efc}
K.~Cheung, T.~Nomura, and H.~Okada, ``{A Three-loop Neutrino Model with
  Leptoquark Triplet Scalars},''
  \href{http://dx.doi.org/10.1016/j.physletb.2017.03.021}{{\em Phys. Lett. B}
  {\bfseries 768} (2017) 359--364},
  \href{http://arxiv.org/abs/1701.01080}{{\ttfamily arXiv:1701.01080
  [hep-ph]}}.

\bibitem{Lee:2017ekw}
S.~Lee, T.~Nomura, and H.~Okada, ``{Radiatively Induced Neutrino Mass Model
  with Flavor Dependent Gauge Symmetry},''
  \href{http://dx.doi.org/10.1016/j.nuclphysb.2018.04.010}{{\em Nucl. Phys. B}
  {\bfseries 931} (2018) 179--191},
  \href{http://arxiv.org/abs/1702.03733}{{\ttfamily arXiv:1702.03733
  [hep-ph]}}.

\bibitem{Fortes:2017ndr}
E.~C. F.~S. Fortes, A.~C.~B. Machado, J.~Monta\~no, and V.~Pleitez, ``{Lepton
  masses and mixing in a scotogenic model},''
  \href{http://dx.doi.org/10.1016/j.physletb.2020.135289}{{\em Phys. Lett. B}
  {\bfseries 803} (2020) 135289},
  \href{http://arxiv.org/abs/1705.09414}{{\ttfamily arXiv:1705.09414
  [hep-ph]}}.

\bibitem{Tang:2017rhv}
Y.-L. Tang, ``{Some Phenomenologies of a Simple Scotogenic Inverse Seesaw
  Model},'' \href{http://dx.doi.org/10.1103/PhysRevD.97.035020}{{\em Phys. Rev.
  D} {\bfseries 97} no.~3, (2018) 035020},
  \href{http://arxiv.org/abs/1709.07735}{{\ttfamily arXiv:1709.07735
  [hep-ph]}}.

\bibitem{Guo:2018iix}
C.~Guo, S.-Y. Guo, and Y.~Liao, ``{Dark matter and LHC phenomenology of a scale
  invariant scotogenic model},''
  \href{http://dx.doi.org/10.1088/1674-1137/43/10/103102}{{\em Chin. Phys. C}
  {\bfseries 43} no.~10, (2019) 103102},
  \href{http://arxiv.org/abs/1811.01180}{{\ttfamily arXiv:1811.01180
  [hep-ph]}}.

\bibitem{Rojas:2018wym}
N.~Rojas, R.~Srivastava, and J.~W.~F. Valle, ``{Simplest Scoto-Seesaw
  Mechanism},'' \href{http://dx.doi.org/10.1016/j.physletb.2018.12.014}{{\em
  Phys. Lett. B} {\bfseries 789} (2019) 132--136},
  \href{http://arxiv.org/abs/1807.11447}{{\ttfamily arXiv:1807.11447
  [hep-ph]}}.

\bibitem{Aranda:2018lif}
A.~Aranda, C.~Bonilla, and E.~Peinado, ``{Dynamical generation of neutrino mass
  scales},'' \href{http://dx.doi.org/10.1016/j.physletb.2019.01.068}{{\em Phys.
  Lett. B} {\bfseries 792} (2019) 40--42},
  \href{http://arxiv.org/abs/1808.07727}{{\ttfamily arXiv:1808.07727
  [hep-ph]}}.

\bibitem{Han:2019lux}
Z.-L. Han and W.~Wang, ``{Predictive Scotogenic Model with Flavor Dependent
  Symmetry},'' \href{http://dx.doi.org/10.1140/epjc/s10052-019-7033-8}{{\em
  Eur. Phys. J. C} {\bfseries 79} no.~6, (2019) 522},
  \href{http://arxiv.org/abs/1901.07798}{{\ttfamily arXiv:1901.07798
  [hep-ph]}}.

\bibitem{Suematsu:2019kst}
D.~Suematsu, ``{Low scale leptogenesis in a hybrid model of the scotogenic type
  I and III seesaw models},''
  \href{http://dx.doi.org/10.1103/PhysRevD.100.055008}{{\em Phys. Rev. D}
  {\bfseries 100} no.~5, (2019) 055008},
  \href{http://arxiv.org/abs/1906.12008}{{\ttfamily arXiv:1906.12008
  [hep-ph]}}.

\bibitem{Pramanick:2019oxb}
S.~Pramanick, ``{Scotogenic S3 symmetric generation of realistic neutrino
  mixing},'' \href{http://dx.doi.org/10.1103/PhysRevD.100.035009}{{\em Phys.
  Rev. D} {\bfseries 100} no.~3, (2019) 035009},
  \href{http://arxiv.org/abs/1904.07558}{{\ttfamily arXiv:1904.07558
  [hep-ph]}}.

\bibitem{Restrepo:2019ilz}
D.~Restrepo and A.~Rivera, ``{Phenomenological consistency of the
  singlet-triplet scotogenic model},''
  \href{http://dx.doi.org/10.1007/JHEP04(2020)134}{{\em JHEP} {\bfseries 04}
  (2020) 134}, \href{http://arxiv.org/abs/1907.11938}{{\ttfamily
  arXiv:1907.11938 [hep-ph]}}.

\bibitem{Mandal:2019oth}
S.~Mandal, N.~Rojas, R.~Srivastava, and J.~W.~F. Valle, ``{Dark matter as the
  origin of neutrino mass in the inverse seesaw mechanism},''
  \href{http://dx.doi.org/10.1016/j.physletb.2021.136609}{{\em Phys. Lett. B}
  {\bfseries 821} (2021) 136609},
  \href{http://arxiv.org/abs/1907.07728}{{\ttfamily arXiv:1907.07728
  [hep-ph]}}.

\bibitem{Avila:2019hhv}
I.~M. \'Avila, V.~De~Romeri, L.~Duarte, and J.~W.~F. Valle, ``{Phenomenology of
  scotogenic scalar dark matter},''
  \href{http://dx.doi.org/10.1140/epjc/s10052-020-08480-z}{{\em Eur. Phys. J.
  C} {\bfseries 80} no.~10, (2020) 908},
  \href{http://arxiv.org/abs/1910.08422}{{\ttfamily arXiv:1910.08422
  [hep-ph]}}.

\bibitem{Escribano:2020iqq}
P.~Escribano, M.~Reig, and A.~Vicente, ``{Generalizing the Scotogenic model},''
  \href{http://dx.doi.org/10.1007/JHEP07(2020)097}{{\em JHEP} {\bfseries 07}
  (2020) 097}, \href{http://arxiv.org/abs/2004.05172}{{\ttfamily
  arXiv:2004.05172 [hep-ph]}}.

\bibitem{Nomura:2020dzw}
T.~Nomura, H.~Okada, and Y.~Uesaka, ``{A two-loop induced neutrino mass model,
  dark matter, and LFV processes $\ell_i \to \ell_j \gamma$, and $\mu e \to e
  e$ in a hidden local $U(1)$ symmetry},''
  \href{http://dx.doi.org/10.1016/j.nuclphysb.2020.115236}{{\em Nucl. Phys. B}
  {\bfseries 962} (2021) 115236},
  \href{http://arxiv.org/abs/2008.02673}{{\ttfamily arXiv:2008.02673
  [hep-ph]}}.

\bibitem{Beniwal:2020hjc}
A.~Beniwal, J.~Herrero-Garc\'\i{}a, N.~Leerdam, M.~White, and A.~G. Williams,
  ``{The ScotoSinglet Model: a scalar singlet extension of the Scotogenic
  Model},'' \href{http://dx.doi.org/10.1007/JHEP06(2021)136}{{\em JHEP}
  {\bfseries 21} (2020) 136}, \href{http://arxiv.org/abs/2010.05937}{{\ttfamily
  arXiv:2010.05937 [hep-ph]}}.

\bibitem{DeRomeri:2021yjo}
V.~De~Romeri, M.~Puerta, and A.~Vicente, ``{Dark matter in a charged variant of
  the Scotogenic model},''
  \href{http://dx.doi.org/10.1140/epjc/s10052-022-10532-5}{{\em Eur. Phys. J.
  C} {\bfseries 82} no.~7, (2022) 623},
  \href{http://arxiv.org/abs/2106.00481}{{\ttfamily arXiv:2106.00481
  [hep-ph]}}.

\bibitem{De:2021crr}
B.~De, D.~Das, M.~Mitra, and N.~Sahoo, ``{Magnetic moments of leptons, charged
  lepton flavor violations and dark matter phenomenology of a minimal radiative
  Dirac neutrino mass model},''
  \href{http://dx.doi.org/10.1007/JHEP08(2022)202}{{\em JHEP} {\bfseries 08}
  (2022) 202}, \href{http://arxiv.org/abs/2106.00979}{{\ttfamily
  arXiv:2106.00979 [hep-ph]}}.

\bibitem{Kang:2021jmi}
D.~W. Kang, J.~Kim, and H.~Okada, ``{Muon $g-2$ in $U(1)_{\mu - \tau}$
  symmetric gauged radiative neutrino mass model},''
  \href{http://dx.doi.org/10.1016/j.physletb.2021.136666}{{\em Phys. Lett. B}
  {\bfseries 822} (2021) 136666},
  \href{http://arxiv.org/abs/2107.09960}{{\ttfamily arXiv:2107.09960
  [hep-ph]}}.

\bibitem{Sarazin:2021nwo}
M.~Sarazin, J.~Bernigaud, and B.~Herrmann, ``{Dark matter and lepton flavour
  phenomenology in a singlet-doublet scotogenic model},''
  \href{http://dx.doi.org/10.1007/JHEP12(2021)116}{{\em JHEP} {\bfseries 12}
  (2021) 116}, \href{http://arxiv.org/abs/2107.04613}{{\ttfamily
  arXiv:2107.04613 [hep-ph]}}.

\bibitem{Nagao:2022oin}
K.~I. Nagao, T.~Nomura, and H.~Okada, ``{A model explaining the new CDF II W
  boson mass linking to muon $g-2$ and dark matter},''
  \href{http://dx.doi.org/10.1140/epjp/s13360-023-03992-5}{{\em Eur. Phys. J.
  Plus} {\bfseries 138} no.~4, (2023) 365},
  \href{http://arxiv.org/abs/2204.07411}{{\ttfamily arXiv:2204.07411
  [hep-ph]}}.

\bibitem{Ahriche:2022bpx}
A.~Ahriche, ``{A scotogenic model with two inert doublets},''
  \href{http://dx.doi.org/10.1007/JHEP02(2023)028}{{\em JHEP} {\bfseries 02}
  (2023) 028}, \href{http://arxiv.org/abs/2208.00500}{{\ttfamily
  arXiv:2208.00500 [hep-ph]}}.

\bibitem{Cepedello:2022xgb}
R.~Cepedello, P.~Escribano, and A.~Vicente, ``{Neutrino masses, flavor
  anomalies, and muon g-2 from dark loops},''
  \href{http://dx.doi.org/10.1103/PhysRevD.107.035034}{{\em Phys. Rev. D}
  {\bfseries 107} no.~3, (2023) 035034},
  \href{http://arxiv.org/abs/2209.02730}{{\ttfamily arXiv:2209.02730
  [hep-ph]}}.

\bibitem{Chun:2023vbh}
E.~J. Chun, A.~Roy, S.~Mandal, and M.~Mitra, ``{Fermionic dark matter in
  Dynamical Scotogenic Model},''
  \href{http://dx.doi.org/10.1007/JHEP08(2023)130}{{\em JHEP} {\bfseries 08}
  (2023) 130}, \href{http://arxiv.org/abs/2303.02681}{{\ttfamily
  arXiv:2303.02681 [hep-ph]}}.

\bibitem{Escribano:2023hxj}
P.~Escribano, V.~M. Lozano, and A.~Vicente, ``{Scotogenic explanation for the
  95~GeV excesses},'' \href{http://dx.doi.org/10.1103/PhysRevD.108.115001}{{\em
  Phys. Rev. D} {\bfseries 108} no.~11, (2023) 115001},
  \href{http://arxiv.org/abs/2306.03735}{{\ttfamily arXiv:2306.03735
  [hep-ph]}}.

\bibitem{Borah:2023hqw}
D.~Borah, S.~Mahapatra, P.~K. Paul, and N.~Sahu, ``{Scotogenic
  U(1)L\ensuremath{\mu}-L\ensuremath{\tau} origin of (g-2)\ensuremath{\mu},
  W-mass anomaly and 95~GeV excess},''
  \href{http://dx.doi.org/10.1103/PhysRevD.109.055021}{{\em Phys. Rev. D}
  {\bfseries 109} no.~5, (2024) 055021},
  \href{http://arxiv.org/abs/2310.11953}{{\ttfamily arXiv:2310.11953
  [hep-ph]}}.

\bibitem{Garbrecht:2024bbo}
B.~Garbrecht and E.~Wang, ``{A Scotogenic Model as a Prototype for Leptogenesis
  with One Single Gauge Singlet},''
  \href{http://arxiv.org/abs/2404.02207}{{\ttfamily arXiv:2404.02207
  [hep-ph]}}.

\bibitem{Nomura:2024zca}
T.~Nomura and O.~Popov, ``{Extended scotogenic model of neutrino mass and
  proton decay},'' \href{http://dx.doi.org/10.1103/PhysRevD.110.075035}{{\em
  Phys. Rev. D} {\bfseries 110} no.~7, (2024) 075035},
  \href{http://arxiv.org/abs/2406.00651}{{\ttfamily arXiv:2406.00651
  [hep-ph]}}.

\bibitem{Cardenas:2024ojd}
K.~M. C{\'a}rdenas, G.~Mohlabeng, and A.~C. Vincent, ``{Global fit to loopy
  dark matter and neutrino masses},''
  \href{http://dx.doi.org/10.1103/PhysRevD.111.055024}{{\em Phys. Rev. D}
  {\bfseries 111} no.~5, (2025) 055024},
  \href{http://arxiv.org/abs/2411.03470}{{\ttfamily arXiv:2411.03470
  [hep-ph]}}.

\bibitem{Singh:2025jtn}
L.~Singh, R.~Srivastava, S.~Verma, and S.~Yadav, ``{Type-III Scotogenic Model:
  Inflation, Dark Matter and Collider Phenomenology},''
  \href{http://arxiv.org/abs/2501.13171}{{\ttfamily arXiv:2501.13171
  [hep-ph]}}.

\bibitem{CarcamoHernandez:2025eyt}
A.~E. C{\'a}rcamo~Hern{\'a}ndez, J.~E. Puentes, R.~Pasechnik, and
  D.~Salinas-Arizmendi, ``{Strongly coupled inert scalar sector with radiative
  neutrino masses},'' \href{http://arxiv.org/abs/2504.07193}{{\ttfamily
  arXiv:2504.07193 [hep-ph]}}.

\bibitem{Escribano:2025kys}
P.~Escribano, V.~M. Lozano, S.~Norero, and A.~Vicente, ``{Exploring Dimuon
  Higgs Decay in an Extended Scotogenic Model},''
  \href{http://arxiv.org/abs/2501.17244}{{\ttfamily arXiv:2501.17244
  [hep-ph]}}.

\bibitem{AbuSiam:2025voc}
A.~AbuSiam and A.~Ahriche, ``{The Scotogenic Model with Two Inert Doublets:
  Parameters Space and Electroweak Precision Tests},''
  \href{http://arxiv.org/abs/2506.18051}{{\ttfamily arXiv:2506.18051
  [hep-ph]}}.

\bibitem{Ma:2008ym}
E.~Ma, ``{Dark Scalar Doublets and Neutrino Tribimaximal Mixing from A(4)
  Symmetry},'' \href{http://dx.doi.org/10.1016/j.physletb.2008.12.038}{{\em
  Phys. Lett. B} {\bfseries 671} (2009) 366--368},
  \href{http://arxiv.org/abs/0808.1729}{{\ttfamily arXiv:0808.1729 [hep-ph]}}.

\bibitem{Adulpravitchai:2009re}
A.~Adulpravitchai, M.~Lindner, A.~Merle, and R.~N. Mohapatra, ``{Radiative
  Transmission of Lepton Flavor Hierarchies},''
  \href{http://dx.doi.org/10.1016/j.physletb.2009.09.042}{{\em Phys. Lett. B}
  {\bfseries 680} (2009) 476--479},
  \href{http://arxiv.org/abs/0908.0470}{{\ttfamily arXiv:0908.0470 [hep-ph]}}.

\bibitem{Ma:2012ez}
E.~Ma, A.~Natale, and A.~Rashed, ``{Scotogenic $A_4$ Neutrino Model for Nonzero
  $\theta_{13}$ and Large $\delta_{CP}$},''
  \href{http://dx.doi.org/10.1142/S0217751X12501345}{{\em Int. J. Mod. Phys. A}
  {\bfseries 27} (2012) 1250134},
  \href{http://arxiv.org/abs/1206.1570}{{\ttfamily arXiv:1206.1570 [hep-ph]}}.

\bibitem{Bhattacharya:2013mpa}
S.~Bhattacharya, E.~Ma, A.~Natale, and A.~Rashed, ``{Radiative Scaling Neutrino
  Mass with $A_4$ Symmetry},''
  \href{http://dx.doi.org/10.1103/PhysRevD.87.097301}{{\em Phys. Rev. D}
  {\bfseries 87} (2013) 097301},
  \href{http://arxiv.org/abs/1302.6266}{{\ttfamily arXiv:1302.6266 [hep-ph]}}.

\bibitem{Kajiyama:2013rla}
Y.~Kajiyama, H.~Okada, and T.~Toma, ``{Multicomponent dark matter particles in
  a two-loop neutrino model},''
  \href{http://dx.doi.org/10.1103/PhysRevD.88.015029}{{\em Phys. Rev. D}
  {\bfseries 88} no.~1, (2013) 015029},
  \href{http://arxiv.org/abs/1303.7356}{{\ttfamily arXiv:1303.7356 [hep-ph]}}.

\bibitem{Ma:2013xqa}
E.~Ma, ``{Neutrino Mixing and Geometric CP Violation with Delta(27)
  Symmetry},'' \href{http://dx.doi.org/10.1016/j.physletb.2013.05.011}{{\em
  Phys. Lett. B} {\bfseries 723} (2013) 161--163},
  \href{http://arxiv.org/abs/1304.1603}{{\ttfamily arXiv:1304.1603 [hep-ph]}}.

\bibitem{Ma:2014eka}
E.~Ma and A.~Natale, ``{Scotogenic $Z_2$ or $U(1)_D$ Model of Neutrino Mass
  with $\Delta(27)$ Symmetry},''
  \href{http://dx.doi.org/10.1016/j.physletb.2014.05.070}{{\em Phys. Lett. B}
  {\bfseries 734} (2014) 403--405},
  \href{http://arxiv.org/abs/1403.6772}{{\ttfamily arXiv:1403.6772 [hep-ph]}}.

\bibitem{Okada:2015bxa}
H.~Okada, N.~Okada, and Y.~Orikasa, ``{Radiative seesaw mechanism in a minimal
  3-3-1 model},'' \href{http://dx.doi.org/10.1103/PhysRevD.93.073006}{{\em
  Phys. Rev. D} {\bfseries 93} no.~7, (2016) 073006},
  \href{http://arxiv.org/abs/1504.01204}{{\ttfamily arXiv:1504.01204
  [hep-ph]}}.

\bibitem{Baek:2017qos}
S.~Baek, H.~Okada, and Y.~Orikasa, ``{A Two Loop Radiative Neutrino Model},''
  \href{http://dx.doi.org/10.1016/j.nuclphysb.2019.03.002}{{\em Nucl. Phys. B}
  {\bfseries 941} (2019) 744--754},
  \href{http://arxiv.org/abs/1703.00685}{{\ttfamily arXiv:1703.00685
  [hep-ph]}}.

\bibitem{Ahriche:2020pwq}
A.~Ahriche, A.~Jueid, and S.~Nasri, ``{A natural scotogenic model for neutrino
  mass \& dark matter},''
  \href{http://dx.doi.org/10.1016/j.physletb.2021.136077}{{\em Phys. Lett. B}
  {\bfseries 814} (2021) 136077},
  \href{http://arxiv.org/abs/2007.05845}{{\ttfamily arXiv:2007.05845
  [hep-ph]}}.

\bibitem{Chen:2020ark}
S.-L. Chen, A.~Dutta~Banik, and Z.-K. Liu, ``{Common origin of radiative
  neutrino mass, dark matter and leptogenesis in scotogenic Georgi-Machacek
  model},'' \href{http://dx.doi.org/10.1016/j.nuclphysb.2021.115394}{{\em Nucl.
  Phys. B} {\bfseries 966} (2021) 115394},
  \href{http://arxiv.org/abs/2011.13551}{{\ttfamily arXiv:2011.13551
  [hep-ph]}}.

\bibitem{deAnda:2021jzc}
F.~J. de~Anda, O.~Medina, J.~W.~F. Valle, and C.~A. Vaquera-Araujo,
  ``{Scotogenic Majorana neutrino masses in a predictive orbifold theory of
  flavor},'' \href{http://dx.doi.org/10.1103/PhysRevD.105.055030}{{\em Phys.
  Rev. D} {\bfseries 105} no.~5, (2022) 055030},
  \href{http://arxiv.org/abs/2110.06810}{{\ttfamily arXiv:2110.06810
  [hep-ph]}}.

\bibitem{ChuliaCentelles:2022ogm}
S.~Chuli\'a~Centelles, R.~Cepedello, and O.~Medina, ``{Absolute neutrino mass
  scale and dark matter stability from flavour symmetry},''
  \href{http://dx.doi.org/10.1007/JHEP10(2022)080}{{\em JHEP} {\bfseries 10}
  (2022) 080}, \href{http://arxiv.org/abs/2204.12517}{{\ttfamily
  arXiv:2204.12517 [hep-ph]}}.

\bibitem{Barreiros:2022aqu}
D.~M. Barreiros, H.~B. Camara, and F.~R. Joaquim, ``{Flavour and dark matter in
  a scoto/type-II seesaw model},''
  \href{http://dx.doi.org/10.1007/JHEP08(2022)030}{{\em JHEP} {\bfseries 08}
  (2022) 030}, \href{http://arxiv.org/abs/2204.13605}{{\ttfamily
  arXiv:2204.13605 [hep-ph]}}.

\bibitem{Bonilla:2023pna}
C.~Bonilla, J.~Herms, O.~Medina, and E.~Peinado, ``{Discrete dark matter
  mechanism as the source of neutrino mass scales},''
  \href{http://dx.doi.org/10.1007/JHEP06(2023)078}{{\em JHEP} {\bfseries 06}
  (2023) 078}, \href{http://arxiv.org/abs/2301.10811}{{\ttfamily
  arXiv:2301.10811 [hep-ph]}}.

\bibitem{Kumar:2023moh}
R.~Kumar, P.~Mishra, M.~K. Behera, R.~Mohanta, and R.~Srivastava,
  ``{Predictions from scoto-seesaw with A4 modular symmetry},''
  \href{http://dx.doi.org/10.1016/j.physletb.2024.138635}{{\em Phys. Lett. B}
  {\bfseries 853} (2024) 138635},
  \href{http://arxiv.org/abs/2310.02363}{{\ttfamily arXiv:2310.02363
  [hep-ph]}}.

\bibitem{Ganguly:2023jml}
J.~Ganguly, J.~Gluza, B.~Karmakar, and S.~Mahapatra, ``{Phenomenology of the
  flavor symmetric scoto-seesaw model with dark matter and TM1 mixing},''
  \href{http://dx.doi.org/10.1103/PhysRevD.110.035012}{{\em Phys. Rev. D}
  {\bfseries 110} no.~3, (2024) 035012},
  \href{http://arxiv.org/abs/2311.15997}{{\ttfamily arXiv:2311.15997
  [hep-ph]}}.

\bibitem{Arora:2024mhg}
S.~Arora and B.~C. Chauhan, ``{Dark Matter and Muon (g {\ensuremath{-}} 2) from
  a Discrete Z4 Symmetric Model},''
  \href{http://dx.doi.org/10.31526/lhep.2024.512}{{\em LHEP} {\bfseries 2024}
  (2024) 512}.

\bibitem{Kumar:2024zfb}
R.~Kumar, N.~Nath, and R.~Srivastava, ``{Cutting the scotogenic loop: adding
  flavor to dark matter},''
  \href{http://dx.doi.org/10.1007/JHEP12(2024)036}{{\em JHEP} {\bfseries 12}
  (2024) 036}, \href{http://arxiv.org/abs/2406.00188}{{\ttfamily
  arXiv:2406.00188 [hep-ph]}}.

\bibitem{Kim:2024cwp}
J.~Kim, S.-S. Kim, H.~M. Lee, and R.~Padhan, ``{Small neutrino masses from a
  decoupled singlet scalar field},''
  \href{http://dx.doi.org/10.1016/j.physletb.2025.139243}{{\em Phys. Lett. B}
  {\bfseries 861} (2025) 139243},
  \href{http://arxiv.org/abs/2407.13595}{{\ttfamily arXiv:2407.13595
  [hep-ph]}}.

\bibitem{delaVega:2024tuu}
L.~M.~G. de~la Vega, P.~J. Fitzpatrick, R.~Martinez-Ramirez, and E.~Peinado,
  ``{Dark matter for Majorana neutrinos in a Z4 symmetry},''
  \href{http://dx.doi.org/10.1103/PhysRevD.110.115024}{{\em Phys. Rev. D}
  {\bfseries 110} no.~11, (2024) 115024},
  \href{http://arxiv.org/abs/2407.14447}{{\ttfamily arXiv:2407.14447
  [hep-ph]}}.

\bibitem{CarcamoHernandez:2024ycd}
A.~E. C{\'a}rcamo~Hern{\'a}ndez, D.~Salinas-Arizmendi, J.~Vignatti, and
  A.~Zerwekh, ``{Phenomenology of an Extended $1+2$ Higgs Doublet Model with
  $S_3$ Family Symmetry},''
  \href{http://dx.doi.org/10.1140/epjc/s10052-024-13501-2}{{\em Eur. Phys. J.
  C} {\bfseries 84} (2024) 1135},
  \href{http://arxiv.org/abs/2408.01497}{{\ttfamily arXiv:2408.01497
  [hep-ph]}}.

\bibitem{Luong:2025pjj}
D.~M. Luong and P.~Van~Dong, ``{Scotoseesaw mechanism from a $Z_3$ symmetry of
  matter},'' \href{http://dx.doi.org/10.1140/epjc/s10052-025-14206-w}{{\em Eur.
  Phys. J. C} {\bfseries 85} no.~4, (2025) 473},
  \href{http://arxiv.org/abs/2501.08711}{{\ttfamily arXiv:2501.08711
  [hep-ph]}}.

\bibitem{Kubo:2006rm}
J.~Kubo and D.~Suematsu, ``{Neutrino masses and CDM in a non-supersymmetric
  model},'' \href{http://dx.doi.org/10.1016/j.physletb.2006.11.005}{{\em Phys.
  Lett. B} {\bfseries 643} (2006) 336--341},
  \href{http://arxiv.org/abs/hep-ph/0610006}{{\ttfamily arXiv:hep-ph/0610006}}.

\bibitem{Chang:2011kv}
W.-F. Chang and C.-F. Wong, ``{A Model for Neutrino Masses and Dark Matter with
  the Discrete Gauge Symmetry},''
  \href{http://dx.doi.org/10.1103/PhysRevD.85.013018}{{\em Phys. Rev. D}
  {\bfseries 85} (2012) 013018},
  \href{http://arxiv.org/abs/1104.3934}{{\ttfamily arXiv:1104.3934 [hep-ph]}}.

\bibitem{Kajiyama:2013lja}
Y.~Kajiyama, H.~Okada, and K.~Yagyu, ``{$T_7$ Flavor Model in Three Loop Seesaw
  and Higgs Phenomenology},''
  \href{http://dx.doi.org/10.1007/JHEP10(2013)196}{{\em JHEP} {\bfseries 10}
  (2013) 196}, \href{http://arxiv.org/abs/1307.0480}{{\ttfamily arXiv:1307.0480
  [hep-ph]}}.

\bibitem{Ma:2013yga}
E.~Ma, I.~Picek, and B.~Radov\v{c}i\'c, ``{New Scotogenic Model of Neutrino
  Mass with $U(1)_D$ Gauge Interaction},''
  \href{http://dx.doi.org/10.1016/j.physletb.2013.09.049}{{\em Phys. Lett. B}
  {\bfseries 726} (2013) 744--746},
  \href{http://arxiv.org/abs/1308.5313}{{\ttfamily arXiv:1308.5313 [hep-ph]}}.

\bibitem{AristizabalSierra:2014irc}
D.~Aristizabal~Sierra, M.~Dhen, C.~S. Fong, and A.~Vicente, ``{Dynamical flavor
  origin of $\mathbb{Z}_N$ symmetries},''
  \href{http://dx.doi.org/10.1103/PhysRevD.91.096004}{{\em Phys. Rev. D}
  {\bfseries 91} no.~9, (2015) 096004},
  \href{http://arxiv.org/abs/1412.5600}{{\ttfamily arXiv:1412.5600 [hep-ph]}}.

\bibitem{Hatanaka:2014tba}
H.~Hatanaka, K.~Nishiwaki, H.~Okada, and Y.~Orikasa, ``{A Three-Loop Neutrino
  Model with Global $U(1)$ Symmetry},''
  \href{http://dx.doi.org/10.1016/j.nuclphysb.2015.03.006}{{\em Nucl. Phys. B}
  {\bfseries 894} (2015) 268--283},
  \href{http://arxiv.org/abs/1412.8664}{{\ttfamily arXiv:1412.8664 [hep-ph]}}.

\bibitem{Nishiwaki:2015iqa}
K.~Nishiwaki, H.~Okada, and Y.~Orikasa, ``{Three loop neutrino model with
  isolated $k^{\pm\pm}$},''
  \href{http://dx.doi.org/10.1103/PhysRevD.92.093013}{{\em Phys. Rev. D}
  {\bfseries 92} no.~9, (2015) 093013},
  \href{http://arxiv.org/abs/1507.02412}{{\ttfamily arXiv:1507.02412
  [hep-ph]}}.

\bibitem{Okada:2015vwh}
H.~Okada and Y.~Orikasa, ``{Radiative neutrino model with an inert triplet
  scalar},'' \href{http://dx.doi.org/10.1103/PhysRevD.94.055002}{{\em Phys.
  Rev. D} {\bfseries 94} no.~5, (2016) 055002},
  \href{http://arxiv.org/abs/1512.06687}{{\ttfamily arXiv:1512.06687
  [hep-ph]}}.

\bibitem{Kanemura:2015bli}
S.~Kanemura, K.~Nishiwaki, H.~Okada, Y.~Orikasa, S.~C. Park, and R.~Watanabe,
  ``{LHC 750 GeV diphoton excess in a radiative seesaw model},''
  \href{http://dx.doi.org/10.1093/ptep/ptw164}{{\em PTEP} {\bfseries 2016}
  no.~12, (2016) 123B04}, \href{http://arxiv.org/abs/1512.09048}{{\ttfamily
  arXiv:1512.09048 [hep-ph]}}.

\bibitem{Yu:2016lof}
J.-H. Yu, ``{Hidden Gauged U(1) Model: Unifying Scotogenic Neutrino and Flavor
  Dark Matter},'' \href{http://dx.doi.org/10.1103/PhysRevD.93.113007}{{\em
  Phys. Rev. D} {\bfseries 93} no.~11, (2016) 113007},
  \href{http://arxiv.org/abs/1601.02609}{{\ttfamily arXiv:1601.02609
  [hep-ph]}}.

\bibitem{Nomura:2016seu}
T.~Nomura and H.~Okada, ``{A four-loop Radiative Seesaw Model},''
  \href{http://dx.doi.org/10.1016/j.physletb.2017.05.002}{{\em Phys. Lett. B}
  {\bfseries 770} (2017) 307--313},
  \href{http://arxiv.org/abs/1601.04516}{{\ttfamily arXiv:1601.04516
  [hep-ph]}}.

\bibitem{Wang:2017mcy}
W.~Wang, R.~Wang, Z.-L. Han, and J.-Z. Han, ``{The $B-L$ Scotogenic Models for
  Dirac Neutrino Masses},''
  \href{http://dx.doi.org/10.1140/epjc/s10052-017-5446-9}{{\em Eur. Phys. J. C}
  {\bfseries 77} no.~12, (2017) 889},
  \href{http://arxiv.org/abs/1705.00414}{{\ttfamily arXiv:1705.00414
  [hep-ph]}}.

\bibitem{Nomura:2017vzp}
T.~Nomura and H.~Okada, ``{Radiative neutrino mass in an alternative
  $U(1)_{B-L}$ gauge symmetry},''
  \href{http://dx.doi.org/10.1016/j.nuclphysb.2019.02.025}{{\em Nucl. Phys. B}
  {\bfseries 941} (2019) 586--599},
  \href{http://arxiv.org/abs/1705.08309}{{\ttfamily arXiv:1705.08309
  [hep-ph]}}.

\bibitem{Nomura:2017tzj}
T.~Nomura and H.~Okada, ``{A model with isospin doublet $U(1)_D$ gauge
  symmetry},'' \href{http://dx.doi.org/10.1142/S0217751X18500896}{{\em Int. J.
  Mod. Phys. A} {\bfseries 33} no.~14n15, (2018) 1850089},
  \href{http://arxiv.org/abs/1706.05268}{{\ttfamily arXiv:1706.05268
  [hep-ph]}}.

\bibitem{Ma:2017zyb}
E.~Ma, D.~Restrepo, and O.~Zapata, ``{Anomalous leptonic U(1) symmetry:
  Syndetic origin of the QCD axion, weak-scale dark matter, and radiative
  neutrino mass},'' \href{http://dx.doi.org/10.1142/S0217732318500244}{{\em
  Mod. Phys. Lett. A} {\bfseries 33} (2018) 1850024},
  \href{http://arxiv.org/abs/1706.08240}{{\ttfamily arXiv:1706.08240
  [hep-ph]}}.

\bibitem{Nomura:2017psk}
T.~Nomura and H.~Okada, ``{Neutrino mass in flavor dependent gauged lepton
  model},'' \href{http://dx.doi.org/10.1103/PhysRevD.97.055044}{{\em Phys. Rev.
  D} {\bfseries 97} no.~5, (2018) 055044},
  \href{http://arxiv.org/abs/1707.06083}{{\ttfamily arXiv:1707.06083
  [hep-ph]}}.

\bibitem{Hagedorn:2018spx}
C.~Hagedorn, J.~Herrero-Garc\'\i{}a, E.~Molinaro, and M.~A. Schmidt,
  ``{Phenomenology of the Generalised Scotogenic Model with Fermionic Dark
  Matter},'' \href{http://dx.doi.org/10.1007/JHEP11(2018)103}{{\em JHEP}
  {\bfseries 11} (2018) 103}, \href{http://arxiv.org/abs/1804.04117}{{\ttfamily
  arXiv:1804.04117 [hep-ph]}}.

\bibitem{Han:2018zcn}
Z.-L. Han and W.~Wang, ``{$Z'$ Portal Dark Matter in $B-L$ Scotogenic Dirac
  Model},'' \href{http://dx.doi.org/10.1140/epjc/s10052-018-6308-9}{{\em Eur.
  Phys. J. C} {\bfseries 78} no.~10, (2018) 839},
  \href{http://arxiv.org/abs/1805.02025}{{\ttfamily arXiv:1805.02025
  [hep-ph]}}.

\bibitem{Calle:2018ovc}
J.~Calle, D.~Restrepo, C.~E. Yaguna, and O.~Zapata, ``{Minimal radiative Dirac
  neutrino mass models},''
  \href{http://dx.doi.org/10.1103/PhysRevD.99.075008}{{\em Phys. Rev. D}
  {\bfseries 99} no.~7, (2019) 075008},
  \href{http://arxiv.org/abs/1812.05523}{{\ttfamily arXiv:1812.05523
  [hep-ph]}}.

\bibitem{Carvajal:2018ohk}
C.~D.~R. Carvajal and O.~Zapata, ``{One-loop Dirac neutrino mass and mixed
  axion-WIMP dark matter},''
  \href{http://dx.doi.org/10.1103/PhysRevD.99.075009}{{\em Phys. Rev. D}
  {\bfseries 99} no.~7, (2019) 075009},
  \href{http://arxiv.org/abs/1812.06364}{{\ttfamily arXiv:1812.06364
  [hep-ph]}}.

\bibitem{CentellesChulia:2019gic}
S.~Centelles~Chuli{\'a}, R.~Cepedello, E.~Peinado, and R.~Srivastava,
  ``{Scotogenic dark symmetry as a residual subgroup of Standard Model
  symmetries},'' \href{http://dx.doi.org/10.1088/1674-1137/44/8/083110}{{\em
  Chin. Phys. C} {\bfseries 44} no.~8, (2020) 083110},
  \href{http://arxiv.org/abs/1901.06402}{{\ttfamily arXiv:1901.06402
  [hep-ph]}}.

\bibitem{Ma:2019yfo}
E.~Ma, ``{Scotogenic $U(1)_\chi$ Dirac neutrinos},''
  \href{http://dx.doi.org/10.1016/j.physletb.2019.05.006}{{\em Phys. Lett. B}
  {\bfseries 793} (2019) 411--414},
  \href{http://arxiv.org/abs/1901.09091}{{\ttfamily arXiv:1901.09091
  [hep-ph]}}.

\bibitem{Kang:2019sab}
S.~K. Kang, O.~Popov, R.~Srivastava, J.~W.~F. Valle, and C.~A. Vaquera-Araujo,
  ``{Scotogenic dark matter stability from gauged matter parity},''
  \href{http://dx.doi.org/10.1016/j.physletb.2019.135013}{{\em Phys. Lett. B}
  {\bfseries 798} (2019) 135013},
  \href{http://arxiv.org/abs/1902.05966}{{\ttfamily arXiv:1902.05966
  [hep-ph]}}.

\bibitem{Nomura:2019jxj}
T.~Nomura and H.~Okada, ``{A modular $A_4$ symmetric model of dark matter and
  neutrino},'' \href{http://dx.doi.org/10.1016/j.physletb.2019.134799}{{\em
  Phys. Lett. B} {\bfseries 797} (2019) 134799},
  \href{http://arxiv.org/abs/1904.03937}{{\ttfamily arXiv:1904.03937
  [hep-ph]}}.

\bibitem{Jana:2019mez}
S.~Jana, P.~K. Vishnu, and S.~Saad, ``{Minimal dirac neutrino mass models from
  $\hbox {U}(1)_{\mathrm{R}}$ gauge symmetry and left\textendash{}right
  asymmetry at colliders},''
  \href{http://dx.doi.org/10.1140/epjc/s10052-019-7441-9}{{\em Eur. Phys. J. C}
  {\bfseries 79} no.~11, (2019) 916},
  \href{http://arxiv.org/abs/1904.07407}{{\ttfamily arXiv:1904.07407
  [hep-ph]}}.

\bibitem{Ma:2019iwj}
E.~Ma, ``{Scotogenic cobimaximal Dirac neutrino mixing from $\Delta (27)$ and
  $U(1)_\chi $},'' \href{http://dx.doi.org/10.1140/epjc/s10052-019-7440-x}{{\em
  Eur. Phys. J. C} {\bfseries 79} no.~11, (2019) 903},
  \href{http://arxiv.org/abs/1905.01535}{{\ttfamily arXiv:1905.01535
  [hep-ph]}}.

\bibitem{Nomura:2019yft}
T.~Nomura and H.~Okada, ``{A two loop induced neutrino mass model with modular
  $A_4$ symmetry},''
  \href{http://dx.doi.org/10.1016/j.nuclphysb.2021.115372}{{\em Nucl. Phys. B}
  {\bfseries 966} (2021) 115372},
  \href{http://arxiv.org/abs/1906.03927}{{\ttfamily arXiv:1906.03927
  [hep-ph]}}.

\bibitem{Fuentes-Martin:2019bue}
J.~Fuentes-Mart\'\i{}n, M.~Reig, and A.~Vicente, ``{Strong $CP$ problem with
  low-energy emergent QCD: The 4321 case},''
  \href{http://dx.doi.org/10.1103/PhysRevD.100.115028}{{\em Phys. Rev. D}
  {\bfseries 100} no.~11, (2019) 115028},
  \href{http://arxiv.org/abs/1907.02550}{{\ttfamily arXiv:1907.02550
  [hep-ph]}}.

\bibitem{Okada:2019xqk}
H.~Okada and Y.~Orikasa, ``{Modular $S_3$ symmetric radiative seesaw model},''
  \href{http://dx.doi.org/10.1103/PhysRevD.100.115037}{{\em Phys. Rev. D}
  {\bfseries 100} no.~11, (2019) 115037},
  \href{http://arxiv.org/abs/1907.04716}{{\ttfamily arXiv:1907.04716
  [hep-ph]}}.

\bibitem{Bonilla:2019ipe}
C.~Bonilla, L.~M.~G. de~la Vega, J.~M. Lamprea, R.~A. Lineros, and E.~Peinado,
  ``{Fermion Dark Matter and Radiative Neutrino Masses from Spontaneous Lepton
  Number Breaking},'' \href{http://dx.doi.org/10.1088/1367-2630/ab7254}{{\em
  New J. Phys.} {\bfseries 22} no.~3, (2020) 033009},
  \href{http://arxiv.org/abs/1908.04276}{{\ttfamily arXiv:1908.04276
  [hep-ph]}}.

\bibitem{Han:2019diw}
Z.-L. Han, R.~Ding, S.-J. Lin, and B.~Zhu, ``{Gauged $U(1)_{L_\mu -L_\tau }$
  scotogenic model in light of $R_{K^{(*)}}$ anomaly and AMS-02 positron
  excess},'' \href{http://dx.doi.org/10.1140/epjc/s10052-019-7526-5}{{\em Eur.
  Phys. J. C} {\bfseries 79} no.~12, (2019) 1007},
  \href{http://arxiv.org/abs/1908.07192}{{\ttfamily arXiv:1908.07192
  [hep-ph]}}.

\bibitem{Nomura:2019lnr}
T.~Nomura, H.~Okada, and O.~Popov, ``{A modular $A_4$ symmetric scotogenic
  model},'' \href{http://dx.doi.org/10.1016/j.physletb.2020.135294}{{\em Phys.
  Lett. B} {\bfseries 803} (2020) 135294},
  \href{http://arxiv.org/abs/1908.07457}{{\ttfamily arXiv:1908.07457
  [hep-ph]}}.

\bibitem{Leite:2019grf}
J.~Leite, O.~Popov, R.~Srivastava, and J.~W.~F. Valle, ``{A theory for
  scotogenic dark matter stabilised by residual gauge symmetry},''
  \href{http://dx.doi.org/10.1016/j.physletb.2020.135254}{{\em Phys. Lett. B}
  {\bfseries 802} (2020) 135254},
  \href{http://arxiv.org/abs/1909.06386}{{\ttfamily arXiv:1909.06386
  [hep-ph]}}.

\bibitem{Jana:2019mgj}
S.~Jana, P.~K. Vishnu, and S.~Saad, ``{Minimal realizations of Dirac neutrino
  mass from generic one-loop and two-loop topologies at $d = 5$},''
  \href{http://dx.doi.org/10.1088/1475-7516/2020/04/018}{{\em JCAP} {\bfseries
  04} (2020) 018}, \href{http://arxiv.org/abs/1910.09537}{{\ttfamily
  arXiv:1910.09537 [hep-ph]}}.

\bibitem{Wang:2019byi}
W.~Wang and Z.-L. Han, ``{$U(1)_{B-3L_{\alpha}}$ extended scotogenic models and
  single-zero textures of neutrino mass matrices},''
  \href{http://dx.doi.org/10.1103/PhysRevD.101.115040}{{\em Phys. Rev. D}
  {\bfseries 101} no.~11, (2020) 115040},
  \href{http://arxiv.org/abs/1911.00819}{{\ttfamily arXiv:1911.00819
  [hep-ph]}}.

\bibitem{delaVega:2020jcp}
L.~M.~G. de~la Vega, N.~Nath, and E.~Peinado, ``{Dirac neutrinos from
  Peccei-Quinn symmetry: two examples},''
  \href{http://dx.doi.org/10.1016/j.nuclphysb.2020.115099}{{\em Nucl. Phys. B}
  {\bfseries 957} (2020) 115099},
  \href{http://arxiv.org/abs/2001.01846}{{\ttfamily arXiv:2001.01846
  [hep-ph]}}.

\bibitem{Okada:2020dmb}
H.~Okada and Y.~Shoji, ``{A radiative seesaw model with three Higgs doublets in
  modular $A_4$ symmetry},''
  \href{http://dx.doi.org/10.1016/j.nuclphysb.2020.115216}{{\em Nucl. Phys. B}
  {\bfseries 961} (2020) 115216},
  \href{http://arxiv.org/abs/2003.13219}{{\ttfamily arXiv:2003.13219
  [hep-ph]}}.

\bibitem{Kim:2020aua}
J.~Kim, T.~Nomura, and H.~Okada, ``{A radiative seesaw model linking to XENON1T
  anomaly},'' \href{http://dx.doi.org/10.1016/j.physletb.2020.135862}{{\em
  Phys. Lett. B} {\bfseries 811} (2020) 135862},
  \href{http://arxiv.org/abs/2007.09894}{{\ttfamily arXiv:2007.09894
  [hep-ph]}}.

\bibitem{Wong:2020obo}
C.-F. Wong, ``{Anomaly-free chiral $U(1)_D$ and its scotogenic implication},''
  \href{http://dx.doi.org/10.1016/j.dark.2021.100818}{{\em Phys. Dark Univ.}
  {\bfseries 32} (2021) 100818},
  \href{http://arxiv.org/abs/2008.08573}{{\ttfamily arXiv:2008.08573
  [hep-ph]}}.

\bibitem{Behera:2020lpd}
M.~K. Behera, S.~Singirala, S.~Mishra, and R.~Mohanta, ``{A modular A $_{4}$
  symmetric scotogenic model for neutrino mass and dark matter},''
  \href{http://dx.doi.org/10.1088/1361-6471/ac3cc2}{{\em J. Phys. G} {\bfseries
  49} no.~3, (2022) 035002}, \href{http://arxiv.org/abs/2009.01806}{{\ttfamily
  arXiv:2009.01806 [hep-ph]}}.

\bibitem{Bernal:2021ezl}
N.~Bernal, J.~Calle, and D.~Restrepo, ``{Anomaly-free Abelian gauge symmetries
  with Dirac scotogenic models},''
  \href{http://dx.doi.org/10.1103/PhysRevD.103.095032}{{\em Phys. Rev. D}
  {\bfseries 103} no.~9, (2021) 095032},
  \href{http://arxiv.org/abs/2102.06211}{{\ttfamily arXiv:2102.06211
  [hep-ph]}}.

\bibitem{Okada:2021nwo}
H.~Okada, Y.~Orikasa, and Y.~Shoji, ``{Radiative dark matter and neutrino
  masses from an alternative U(1) B-L gauge symmetry},''
  \href{http://dx.doi.org/10.1088/1475-7516/2021/07/006}{{\em JCAP} {\bfseries
  07} (2021) 006}, \href{http://arxiv.org/abs/2102.10944}{{\ttfamily
  arXiv:2102.10944 [hep-ph]}}.

\bibitem{Nomura:2021aep}
T.~Nomura and H.~Okada, ``{Radiative neutrino mass model in dark non-Abelian
  gauge symmetry},'' \href{http://dx.doi.org/10.1103/PhysRevD.105.075010}{{\em
  Phys. Rev. D} {\bfseries 105} no.~7, (2022) 075010},
  \href{http://arxiv.org/abs/2106.10451}{{\ttfamily arXiv:2106.10451
  [hep-ph]}}.

\bibitem{Escribano:2021ymx}
P.~Escribano and A.~Vicente, ``{An ultraviolet completion for the Scotogenic
  model},'' \href{http://dx.doi.org/10.1016/j.physletb.2021.136717}{{\em Phys.
  Lett. B} {\bfseries 823} (2021) 136717},
  \href{http://arxiv.org/abs/2107.10265}{{\ttfamily arXiv:2107.10265
  [hep-ph]}}.

\bibitem{Dasgupta:2021ggp}
A.~Dasgupta, T.~Nomura, H.~Okada, O.~Popov, and M.~Tanimoto, ``{Dirac Radiative
  Neutrino Mass with Modular Symmetry and Leptogenesis},''
  \href{http://arxiv.org/abs/2111.06898}{{\ttfamily arXiv:2111.06898
  [hep-ph]}}.

\bibitem{Berbig:2022nre}
M.~Berbig, ``{Freeze-In of radiative keV-scale neutrino dark matter from a new
  U(1)$_{B-L}$},'' \href{http://dx.doi.org/10.1007/JHEP09(2022)101}{{\em JHEP}
  {\bfseries 09} (2022) 101}, \href{http://arxiv.org/abs/2203.04276}{{\ttfamily
  arXiv:2203.04276 [hep-ph]}}.

\bibitem{Portillo-Sanchez:2023kbz}
D.~Portillo-S\'anchez, P.~Escribano, and A.~Vicente, ``{Ultraviolet extensions
  of the Scotogenic model},''
  \href{http://dx.doi.org/10.1007/JHEP08(2023)023}{{\em JHEP} {\bfseries 08}
  (2023) 023}, \href{http://arxiv.org/abs/2301.05249}{{\ttfamily
  arXiv:2301.05249 [hep-ph]}}.

\bibitem{deBoer:2023phz}
T.~de~Boer, M.~Klasen, and S.~Zeinstra, ``{Anomaly-free dark matter models with
  one-loop neutrino masses and a gauged U(1) symmetry},''
  \href{http://dx.doi.org/10.1007/JHEP01(2024)013}{{\em JHEP} {\bfseries 01}
  (2024) 013}, \href{http://arxiv.org/abs/2309.06920}{{\ttfamily
  arXiv:2309.06920 [hep-ph]}}.

\bibitem{Nomura:2023vmh}
T.~Nomura and H.~Okada, ``{Scotogenic models with a general lepton flavor
  dependent U(1) gauge symmetry},''
  \href{http://dx.doi.org/10.1016/j.physletb.2023.138393}{{\em Phys. Lett. B}
  {\bfseries 848} (2024) 138393},
  \href{http://arxiv.org/abs/2304.03905}{{\ttfamily arXiv:2304.03905
  [hep-ph]}}.

\bibitem{Leite:2023gzl}
J.~Leite, S.~Sadhukhan, and J.~W.~F. Valle, ``{Dynamical scoto-seesaw mechanism
  with gauged B-L symmetry},''
  \href{http://dx.doi.org/10.1103/PhysRevD.109.035023}{{\em Phys. Rev. D}
  {\bfseries 109} no.~3, (2024) 035023},
  \href{http://arxiv.org/abs/2307.04840}{{\ttfamily arXiv:2307.04840
  [hep-ph]}}.

\bibitem{Nomura:2024jxc}
T.~Nomura and H.~Okada, ``{Radiative inverse seesaw model with hidden $U(1)$
  gauge symmetry enhancing lepton $g-2$},''
  \href{http://arxiv.org/abs/2403.14193}{{\ttfamily arXiv:2403.14193
  [hep-ph]}}.

\bibitem{VanLoi:2024ptt}
D.~Van~Loi, N.~T. Duy, C.~H. Nam, and P.~Van~Dong, ``{Scoto-seesaw model
  implied by flavor-dependent Abelian gauge charge},''
  \href{http://dx.doi.org/10.1140/epjc/s10052-025-13743-8}{{\em Eur. Phys. J.
  C} {\bfseries 85} no.~1, (2025) 109},
  \href{http://arxiv.org/abs/2409.06393}{{\ttfamily arXiv:2409.06393
  [hep-ph]}}.

\bibitem{Nomura:2024vzw}
T.~Nomura, H.~Okada, and O.~Popov, ``{Non-holomorphic modular A4 symmetric
  scotogenic model},''
  \href{http://dx.doi.org/10.1016/j.physletb.2024.139171}{{\em Phys. Lett. B}
  {\bfseries 860} (2025) 139171},
  \href{http://arxiv.org/abs/2409.12547}{{\ttfamily arXiv:2409.12547
  [hep-ph]}}.

\bibitem{CentellesChulia:2024iom}
S.~Centelles~Chuli{\'a}, R.~Srivastava, and S.~Yadav, ``{Comprehensive
  phenomenology of the Dirac Scotogenic Model: Novel low-mass dark matter},''
  \href{http://dx.doi.org/10.1007/JHEP04(2025)038}{{\em JHEP} {\bfseries 04}
  (2025) 038}, \href{http://arxiv.org/abs/2409.18513}{{\ttfamily
  arXiv:2409.18513 [hep-ph]}}.

\bibitem{Garnica:2024wur}
Y.~Garnica, A.~Morales, and C.~A. Vaquera-Araujo, ``{Scotogenic dark matter
  from gauged B{\ensuremath{-}}L},''
  \href{http://dx.doi.org/10.1016/j.physletb.2025.139790}{{\em Phys. Lett. B}
  {\bfseries 868} (2025) 139790},
  \href{http://arxiv.org/abs/2411.13756}{{\ttfamily arXiv:2411.13756
  [hep-ph]}}.

\bibitem{Pathak:2024sei}
G.~Pathak, P.~Das, and M.~K. Das, ``{Neutrino mass genesis in scoto-inverse
  seesaw with modular $A_4$},''
  \href{http://dx.doi.org/10.1140/epjc/s10052-025-14263-1}{{\em Eur. Phys. J.
  C} {\bfseries 85} no.~5, (2025) 569},
  \href{http://arxiv.org/abs/2411.13895}{{\ttfamily arXiv:2411.13895
  [hep-ph]}}.

\bibitem{Agudelo:2024luc}
K.~Agudelo, D.~Restrepo, A.~Rivera, and D.~Suarez, ``{Multicomponent secluded
  WIMP dark matter and Dirac neutrino masses with an extra Abelian gauge
  symmetry},'' \href{http://dx.doi.org/10.1103/PhysRevD.111.095018}{{\em Phys.
  Rev. D} {\bfseries 111} no.~9, (2025) 095018},
  \href{http://arxiv.org/abs/2412.02027}{{\ttfamily arXiv:2412.02027
  [hep-ph]}}.

\bibitem{Babu:2024jdw}
K.~S. Babu and S.~Saad, ``{Ultraviolet completion of a two-loop neutrino mass
  model},'' \href{http://dx.doi.org/10.1007/JHEP03(2025)132}{{\em JHEP}
  {\bfseries 03} (2025) 132}, \href{http://arxiv.org/abs/2412.14562}{{\ttfamily
  arXiv:2412.14562 [hep-ph]}}.

\bibitem{Gola:2024tfx}
S.~Gola, ``{Dark matter from axions and small neutrino masses},''
  \href{http://dx.doi.org/10.1103/v9ky-jfnb}{{\em Phys. Rev. D} {\bfseries 112}
  no.~3, (2025) 035006}, \href{http://arxiv.org/abs/2412.19094}{{\ttfamily
  arXiv:2412.19094 [hep-ph]}}.

\bibitem{Batra:2025gzy}
A.~Batra, H.~B. C{\^a}mara, F.~R. Joaquim, N.~Nath, R.~Srivastava, and J.~W.~F.
  Valle, ``{Axion framework with color-mediated Dirac neutrino masses},''
  \href{http://dx.doi.org/10.1016/j.physletb.2025.139629}{{\em Phys. Lett. B}
  {\bfseries 868} (2025) 139629},
  \href{http://arxiv.org/abs/2501.13156}{{\ttfamily arXiv:2501.13156
  [hep-ph]}}.

\bibitem{Borboruah:2025bwx}
Z.~A. Borboruah, L.~Malhotra, U.~Patel, S.~Patra, and S.~U. Sankar,
  ``{Left-Right Symmetric Neutrino Mass Model without Scalar Bi-doublet},''
  \href{http://arxiv.org/abs/2504.08267}{{\ttfamily arXiv:2504.08267
  [hep-ph]}}.

\bibitem{Dorsner:2025rzj}
I.~Dor{\v{s}}ner, M.~Matkovi{\'c}, and S.~Saad, ``{Nonrenormalizable SU(5)
  GUTs: Leptoquark-induced neutrino masses},''
  \href{http://dx.doi.org/10.1103/3plx-vwrb}{{\em Phys. Rev. D} {\bfseries 111}
  no.~11, (2025) 115039}, \href{http://arxiv.org/abs/2504.16022}{{\ttfamily
  arXiv:2504.16022 [hep-ph]}}.

\bibitem{Kumar:2025cte}
R.~Kumar, N.~Nath, R.~Srivastava, and S.~Yadav, ``{Dirac Scoto Inverse-Seesaw
  from $A_4$ Flavor Symmetry},''
  \href{http://arxiv.org/abs/2505.01407}{{\ttfamily arXiv:2505.01407
  [hep-ph]}}.

\bibitem{Nomura:2025raf}
T.~Nomura and H.~Okada, ``{Neutrino mass model at a three-loop level from a
  non-holomorphic modular $A_4$ symmetry},''
  \href{http://arxiv.org/abs/2506.02639}{{\ttfamily arXiv:2506.02639
  [hep-ph]}}.

\bibitem{Ma:2025ymy}
E.~Ma, ``{MonoHiggsology},'' \href{http://arxiv.org/abs/2507.04563}{{\ttfamily
  arXiv:2507.04563 [hep-ph]}}.

\bibitem{Leite:2025yuq}
J.~Leite, J.~Perez-Soler, and A.~Vicente, ``{Scotogenic mechanism from an
  extended $\boldsymbol{SU(2)_1 \times SU(2)_2 \times U(1)_Y}$ electroweak
  symmetry},'' \href{http://arxiv.org/abs/2507.21223}{{\ttfamily
  arXiv:2507.21223 [hep-ph]}}.

\bibitem{Reig:2018mdk}
M.~Reig, D.~Restrepo, J.~W.~F. Valle, and O.~Zapata, ``{Bound-state dark matter
  and Dirac neutrino masses},''
  \href{http://dx.doi.org/10.1103/PhysRevD.97.115032}{{\em Phys. Rev. D}
  {\bfseries 97} no.~11, (2018) 115032},
  \href{http://arxiv.org/abs/1803.08528}{{\ttfamily arXiv:1803.08528
  [hep-ph]}}.

\bibitem{Reig:2018ztc}
M.~Reig, D.~Restrepo, J.~W.~F. Valle, and O.~Zapata, ``{Bound-state dark matter
  with Majorana neutrinos},''
  \href{http://dx.doi.org/10.1016/j.physletb.2019.01.023}{{\em Phys. Lett. B}
  {\bfseries 790} (2019) 303--307},
  \href{http://arxiv.org/abs/1806.09977}{{\ttfamily arXiv:1806.09977
  [hep-ph]}}.

\bibitem{AristizabalSierra:2007nf}
D.~Aristizabal~Sierra, M.~Hirsch, and S.~G. Kovalenko, ``{Leptoquarks: Neutrino
  masses and accelerator phenomenology},''
  \href{http://dx.doi.org/10.1103/PhysRevD.77.055011}{{\em Phys. Rev. D}
  {\bfseries 77} (2008) 055011},
  \href{http://arxiv.org/abs/0710.5699}{{\ttfamily arXiv:0710.5699 [hep-ph]}}.

\bibitem{FileviezPerez:2009ud}
P.~Fileviez~Perez and M.~B. Wise, ``{On the Origin of Neutrino Masses},''
  \href{http://dx.doi.org/10.1103/PhysRevD.80.053006}{{\em Phys. Rev. D}
  {\bfseries 80} (2009) 053006},
  \href{http://arxiv.org/abs/0906.2950}{{\ttfamily arXiv:0906.2950 [hep-ph]}}.

\bibitem{Kohda:2012sr}
M.~Kohda, H.~Sugiyama, and K.~Tsumura, ``{Lepton number violation at the LHC
  with leptoquark and diquark},''
  \href{http://dx.doi.org/10.1016/j.physletb.2012.12.048}{{\em Phys. Lett. B}
  {\bfseries 718} (2013) 1436--1440},
  \href{http://arxiv.org/abs/1210.5622}{{\ttfamily arXiv:1210.5622 [hep-ph]}}.

\bibitem{Cordova:2022fhg}
C.~Cordova, S.~Hong, S.~Koren, and K.~Ohmori, ``{Neutrino Masses from
  Generalized Symmetry Breaking},''
  \href{http://dx.doi.org/10.1103/PhysRevX.14.031033}{{\em Phys. Rev. X}
  {\bfseries 14} no.~3, (2024) 031033},
  \href{http://arxiv.org/abs/2211.07639}{{\ttfamily arXiv:2211.07639
  [hep-ph]}}.

\bibitem{Kobayashi:2025cwx}
T.~Kobayashi, H.~Okada, and H.~Otsuka, ``{Radiative neutrino mass models from
  non-invertible selection rules},''
  \href{http://arxiv.org/abs/2505.14878}{{\ttfamily arXiv:2505.14878
  [hep-ph]}}.

\bibitem{Nomura:2025yoa}
T.~Nomura and O.~Popov, ``{No-group Scotogenic Model},''
  \href{http://arxiv.org/abs/2507.10299}{{\ttfamily arXiv:2507.10299
  [hep-ph]}}.

\bibitem{Suzuki:2025bxg}
M.~Suzuki, L.-X. Xu, and H.~Y. Zhang, ``{Spurion Analysis for Non-Invertible
  Selection Rules from Near-Group Fusions},''
  \href{http://arxiv.org/abs/2508.14970}{{\ttfamily arXiv:2508.14970
  [hep-ph]}}.

\bibitem{Darricau:2025vcs}
A.~Darricau, H.~Lee, J.~Orloff, and A.~M. Teixeira, ``{Flavour and precision
  probes of a class of scotogenic models},''
  \href{http://arxiv.org/abs/2506.23383}{{\ttfamily arXiv:2506.23383
  [hep-ph]}}.

\bibitem{Cacciapaglia:2020psm}
G.~Cacciapaglia and M.~Rosenlyst, ``{Loop-generated neutrino masses in
  composite Higgs models},''
  \href{http://dx.doi.org/10.1007/JHEP09(2021)167}{{\em JHEP} {\bfseries 09}
  (2021) 167}, \href{http://arxiv.org/abs/2010.01437}{{\ttfamily
  arXiv:2010.01437 [hep-ph]}}.

\bibitem{LZ:2022lsv}
{\bfseries LZ} Collaboration, J.~Aalbers {\em et~al.}, ``{First Dark Matter
  Search Results from the LUX-ZEPLIN (LZ) Experiment},''
  \href{http://dx.doi.org/10.1103/PhysRevLett.131.041002}{{\em Phys. Rev.
  Lett.} {\bfseries 131} no.~4, (2023) 041002},
  \href{http://arxiv.org/abs/2207.03764}{{\ttfamily arXiv:2207.03764
  [hep-ex]}}.

\bibitem{LZ:2024zvo}
{\bfseries LZ} Collaboration, J.~Aalbers {\em et~al.}, ``{Dark Matter Search
  Results from 4.2 Tonne-Years of Exposure of the LUX-ZEPLIN (LZ)
  Experiment},'' \href{http://arxiv.org/abs/2410.17036}{{\ttfamily
  arXiv:2410.17036 [hep-ex]}}.

\bibitem{Freytsis:2010ne}
M.~Freytsis and Z.~Ligeti, ``{On dark matter models with uniquely
  spin-dependent detection possibilities},''
  \href{http://dx.doi.org/10.1103/PhysRevD.83.115009}{{\em Phys. Rev. D}
  {\bfseries 83} (2011) 115009},
  \href{http://arxiv.org/abs/1012.5317}{{\ttfamily arXiv:1012.5317 [hep-ph]}}.

\bibitem{Ipek:2014gua}
S.~Ipek, D.~McKeen, and A.~E. Nelson, ``{A Renormalizable Model for the
  Galactic Center Gamma Ray Excess from Dark Matter Annihilation},''
  \href{http://dx.doi.org/10.1103/PhysRevD.90.055021}{{\em Phys. Rev. D}
  {\bfseries 90} no.~5, (2014) 055021},
  \href{http://arxiv.org/abs/1404.3716}{{\ttfamily arXiv:1404.3716 [hep-ph]}}.

\bibitem{Arcadi:2017wqi}
G.~Arcadi, M.~Lindner, F.~S. Queiroz, W.~Rodejohann, and S.~Vogl,
  ``{Pseudoscalar Mediators: A WIMP model at the Neutrino Floor},''
  \href{http://dx.doi.org/10.1088/1475-7516/2018/03/042}{{\em JCAP} {\bfseries
  03} (2018) 042}, \href{http://arxiv.org/abs/1711.02110}{{\ttfamily
  arXiv:1711.02110 [hep-ph]}}.

\bibitem{Bell:2018zra}
N.~F. Bell, G.~Busoni, and I.~W. Sanderson, ``{Loop Effects in Direct
  Detection},'' \href{http://dx.doi.org/10.1088/1475-7516/2018/08/017}{{\em
  JCAP} {\bfseries 08} (2018) 017},
  \href{http://arxiv.org/abs/1803.01574}{{\ttfamily arXiv:1803.01574
  [hep-ph]}}. [Erratum: JCAP 01, E01 (2019)].

\bibitem{Abe:2018emu}
T.~Abe, M.~Fujiwara, and J.~Hisano, ``{Loop corrections to dark matter direct
  detection in a pseudoscalar mediator dark matter model},''
  \href{http://dx.doi.org/10.1007/JHEP02(2019)028}{{\em JHEP} {\bfseries 02}
  (2019) 028}, \href{http://arxiv.org/abs/1810.01039}{{\ttfamily
  arXiv:1810.01039 [hep-ph]}}.

\bibitem{Abe:2019wjw}
T.~Abe, M.~Fujiwara, J.~Hisano, and Y.~Shoji, ``{Maximum value of the
  spin-independent cross section in the 2HDM+a},''
  \href{http://dx.doi.org/10.1007/JHEP01(2020)114}{{\em JHEP} {\bfseries 01}
  (2020) 114}, \href{http://arxiv.org/abs/1910.09771}{{\ttfamily
  arXiv:1910.09771 [hep-ph]}}.

\bibitem{Barger:2010yn}
V.~Barger, M.~McCaskey, and G.~Shaughnessy, ``{Complex Scalar Dark Matter
  vis-\textbackslash{}`{a}-vis CoGeNT, DAMA/LIBRA and XENON100},''
  \href{http://dx.doi.org/10.1103/PhysRevD.82.035019}{{\em Phys. Rev. D}
  {\bfseries 82} (2010) 035019},
  \href{http://arxiv.org/abs/1005.3328}{{\ttfamily arXiv:1005.3328 [hep-ph]}}.

\bibitem{Barducci:2016fue}
D.~Barducci, A.~Bharucha, N.~Desai, M.~Frigerio, B.~Fuks, A.~Goudelis,
  S.~Kulkarni, G.~Polesello, and D.~Sengupta, ``{Monojet searches for
  momentum-dependent dark matter interactions},''
  \href{http://dx.doi.org/10.1007/JHEP01(2017)078}{{\em JHEP} {\bfseries 01}
  (2017) 078}, \href{http://arxiv.org/abs/1609.07490}{{\ttfamily
  arXiv:1609.07490 [hep-ph]}}.

\bibitem{Gross:2017dan}
C.~Gross, O.~Lebedev, and T.~Toma, ``{Cancellation Mechanism for
  Dark-Matter\textendash{}Nucleon Interaction},''
  \href{http://dx.doi.org/10.1103/PhysRevLett.119.191801}{{\em Phys. Rev.
  Lett.} {\bfseries 119} no.~19, (2017) 191801},
  \href{http://arxiv.org/abs/1708.02253}{{\ttfamily arXiv:1708.02253
  [hep-ph]}}.

\bibitem{Balkin:2017aep}
R.~Balkin, M.~Ruhdorfer, E.~Salvioni, and A.~Weiler, ``{Charged Composite
  Scalar Dark Matter},'' \href{http://dx.doi.org/10.1007/JHEP11(2017)094}{{\em
  JHEP} {\bfseries 11} (2017) 094},
  \href{http://arxiv.org/abs/1707.07685}{{\ttfamily arXiv:1707.07685
  [hep-ph]}}.

\bibitem{Ishiwata:2018sdi}
K.~Ishiwata and T.~Toma, ``{Probing pseudo Nambu-Goldstone boson dark matter at
  loop level},'' \href{http://dx.doi.org/10.1007/JHEP12(2018)089}{{\em JHEP}
  {\bfseries 12} (2018) 089}, \href{http://arxiv.org/abs/1810.08139}{{\ttfamily
  arXiv:1810.08139 [hep-ph]}}.

\bibitem{Abe:2021byq}
Y.~Abe, T.~Toma, K.~Tsumura, and N.~Yamatsu, ``{Pseudo-Nambu-Goldstone dark
  matter model inspired by grand unification},''
  \href{http://dx.doi.org/10.1103/PhysRevD.104.035011}{{\em Phys. Rev. D}
  {\bfseries 104} no.~3, (2021) 035011},
  \href{http://arxiv.org/abs/2104.13523}{{\ttfamily arXiv:2104.13523
  [hep-ph]}}.

\bibitem{Okada:2021qmi}
N.~Okada, D.~Raut, Q.~Shafi, and A.~Thapa, ``{Pseudo-Goldstone dark matter in
  SO(10)},'' \href{http://dx.doi.org/10.1103/PhysRevD.104.095002}{{\em Phys.
  Rev. D} {\bfseries 104} no.~9, (2021) 095002},
  \href{http://arxiv.org/abs/2105.03419}{{\ttfamily arXiv:2105.03419
  [hep-ph]}}.

\bibitem{Chiang:2023omu}
C.-W. Chiang, K.~Tsumura, Y.~Uchida, and N.~Yamatsu, ``{Pseudo-Nambu-Goldstone
  dark matter in SU(7) grand unification},''
  \href{http://dx.doi.org/10.1103/PhysRevD.109.055040}{{\em Phys. Rev. D}
  {\bfseries 109} no.~5, (2024) 055040},
  \href{http://arxiv.org/abs/2311.13753}{{\ttfamily arXiv:2311.13753
  [hep-ph]}}.

\bibitem{Bhattacharya:2024ohh}
S.~Bhattacharya, D.~Mahanta, N.~Mondal, and D.~Pradhan, ``{Two-component dark
  matter and low scale thermal Leptogenesis},''
  \href{http://dx.doi.org/10.1088/1475-7516/2025/09/032}{{\em JCAP} {\bfseries
  09} (2025) 032}, \href{http://arxiv.org/abs/2412.21202}{{\ttfamily
  arXiv:2412.21202 [hep-ph]}}.

\bibitem{Wang:2024qhe}
Z.~Wang, Y.~Reyimuaji, and N.~Yalikun, ``{A $Z_4$ symmetric inverse seesaw
  model for neutrino masses and FIMP dark matter},''
  \href{http://arxiv.org/abs/2412.15672}{{\ttfamily arXiv:2412.15672
  [hep-ph]}}.

\bibitem{Ellis:2016jkw}
J.~Ellis, ``{TikZ-Feynman: Feynman diagrams with TikZ},''
  \href{http://dx.doi.org/10.1016/j.cpc.2016.08.019}{{\em Comput. Phys.
  Commun.} {\bfseries 210} (2017) 103--123},
  \href{http://arxiv.org/abs/1601.05437}{{\ttfamily arXiv:1601.05437
  [hep-ph]}}.

\bibitem{Casas:2001sr}
J.~A. Casas and A.~Ibarra, ``{Oscillating neutrinos and $\mu \to e, \gamma$},''
  \href{http://dx.doi.org/10.1016/S0550-3213(01)00475-8}{{\em Nucl. Phys. B}
  {\bfseries 618} (2001) 171--204},
  \href{http://arxiv.org/abs/hep-ph/0103065}{{\ttfamily arXiv:hep-ph/0103065}}.

\bibitem{Cordero-Carrion:2019qtu}
I.~Cordero-Carri{\'o}n, M.~Hirsch, and A.~Vicente, ``{General parametrization
  of Majorana neutrino mass models},''
  \href{http://dx.doi.org/10.1103/PhysRevD.101.075032}{{\em Phys. Rev. D}
  {\bfseries 101} no.~7, (2020) 075032},
  \href{http://arxiv.org/abs/1912.08858}{{\ttfamily arXiv:1912.08858
  [hep-ph]}}.

\bibitem{ParticleDataGroup:2024cfk}
{\bfseries Particle Data Group} Collaboration, S.~Navas {\em et~al.}, ``{Review
  of particle physics},''
  \href{http://dx.doi.org/10.1103/PhysRevD.110.030001}{{\em Phys. Rev. D}
  {\bfseries 110} no.~3, (2024) 030001}.

\bibitem{Esteban:2024eli}
I.~Esteban, M.~C. Gonzalez-Garcia, M.~Maltoni, I.~Martinez-Soler, J.~P.
  Pinheiro, and T.~Schwetz, ``{NuFit-6.0: updated global analysis of
  three-flavor neutrino oscillations},''
  \href{http://dx.doi.org/10.1007/JHEP12(2024)216}{{\em JHEP} {\bfseries 12}
  (2024) 216}, \href{http://arxiv.org/abs/2410.05380}{{\ttfamily
  arXiv:2410.05380 [hep-ph]}}.

\bibitem{Thomson:2013zua}
M.~Thomson, \href{http://dx.doi.org/10.1017/CBO9781139525367}{{\em {Modern
  particle physics}}}.
\newblock Cambridge University Press, New York, 10, 2013.

\bibitem{MEGII:2023ltw}
{\bfseries MEG II} Collaboration, K.~Afanaciev {\em et~al.}, ``{A search for
  $\mu^+ \rightarrow e^+ \gamma$ with the first dataset of the MEG~II
  experiment},'' \href{http://dx.doi.org/10.1140/epjc/s10052-024-12416-2}{{\em
  Eur. Phys. J. C} {\bfseries 84} no.~3, (2024) 216},
  \href{http://arxiv.org/abs/2310.12614}{{\ttfamily arXiv:2310.12614
  [hep-ex]}}. [Erratum: Eur.Phys.J.C 84, 1042 (2024)].

\bibitem{BaBar:2009hkt}
{\bfseries BaBar} Collaboration, B.~Aubert {\em et~al.}, ``{Searches for Lepton
  Flavor Violation in the Decays $\tau^\pm \to e^\pm \gamma$ and $\tau^\pm \to
  \mu^\pm \gamma$},''
  \href{http://dx.doi.org/10.1103/PhysRevLett.104.021802}{{\em Phys. Rev.
  Lett.} {\bfseries 104} (2010) 021802},
  \href{http://arxiv.org/abs/0908.2381}{{\ttfamily arXiv:0908.2381 [hep-ex]}}.

\bibitem{Grimus:2007if}
W.~Grimus, L.~Lavoura, O.~M. Ogreid, and P.~Osland, ``{A Precision constraint
  on multi-Higgs-doublet models},''
  \href{http://dx.doi.org/10.1088/0954-3899/35/7/075001}{{\em J. Phys. G}
  {\bfseries 35} (2008) 075001},
  \href{http://arxiv.org/abs/0711.4022}{{\ttfamily arXiv:0711.4022 [hep-ph]}}.

\bibitem{particle2022review}
P.~D. Group, R.~Workman, V.~Burkert, V.~Crede, E.~Klempt, U.~Thoma, L.~Tiator,
  K.~Agashe, G.~Aielli, B.~Allanach, {\em et~al.}, ``Review of particle
  physics,'' {\em Progress of theoretical and experimental physics} {\bfseries
  2022} no.~8, (2022) 083C01.

\bibitem{Kannike:2016fmd}
K.~Kannike, ``{Vacuum Stability of a General Scalar Potential of a Few
  Fields},'' \href{http://dx.doi.org/10.1140/epjc/s10052-016-4160-3}{{\em Eur.
  Phys. J. C} {\bfseries 76} no.~6, (2016) 324},
  \href{http://arxiv.org/abs/1603.02680}{{\ttfamily arXiv:1603.02680
  [hep-ph]}}. [Erratum: Eur.Phys.J.C 78, 355 (2018)].

\bibitem{Keus:2013hya}
V.~Keus, S.~F. King, and S.~Moretti, ``{Three-Higgs-doublet models: symmetries,
  potentials and Higgs boson masses},''
  \href{http://dx.doi.org/10.1007/JHEP01(2014)052}{{\em JHEP} {\bfseries 01}
  (2014) 052}, \href{http://arxiv.org/abs/1310.8253}{{\ttfamily arXiv:1310.8253
  [hep-ph]}}.

\bibitem{Edsjo:1997bg}
J.~Edsjo and P.~Gondolo, ``{Neutralino relic density including
  coannihilations},'' \href{http://dx.doi.org/10.1103/PhysRevD.56.1879}{{\em
  Phys. Rev. D} {\bfseries 56} (1997) 1879--1894},
  \href{http://arxiv.org/abs/hep-ph/9704361}{{\ttfamily arXiv:hep-ph/9704361}}.

\bibitem{Srednicki:1988ce}
M.~Srednicki, R.~Watkins, and K.~A. Olive, ``{Calculations of Relic Densities
  in the Early Universe},''
  \href{http://dx.doi.org/10.1016/0550-3213(88)90099-5}{{\em Nucl. Phys. B}
  {\bfseries 310} (1988) 693}.

\bibitem{Planck:2018vyg}
{\bfseries Planck} Collaboration, N.~Aghanim {\em et~al.}, ``{Planck 2018
  results. VI. Cosmological parameters},''
  \href{http://dx.doi.org/10.1051/0004-6361/201833910}{{\em Astron. Astrophys.}
  {\bfseries 641} (2020) A6}, \href{http://arxiv.org/abs/1807.06209}{{\ttfamily
  arXiv:1807.06209 [astro-ph.CO]}}. [Erratum: Astron.Astrophys. 652, C4
  (2021)].

\bibitem{tanedo2011defense}
F.~Tanedo, ``Defense against the dark arts,'' {\em Notes on dark matter and
  particle physics} (2011) .

\bibitem{Husdal:2016haj}
L.~Husdal, ``{On Effective Degrees of Freedom in the Early Universe},''
  \href{http://dx.doi.org/10.3390/galaxies4040078}{{\em Galaxies} {\bfseries 4}
  no.~4, (2016) 78}, \href{http://arxiv.org/abs/1609.04979}{{\ttfamily
  arXiv:1609.04979 [astro-ph.CO]}}.

\bibitem{Kopp:2014tsa}
J.~Kopp, L.~Michaels, and J.~Smirnov, ``{Loopy Constraints on Leptophilic Dark
  Matter and Internal Bremsstrahlung},''
  \href{http://dx.doi.org/10.1088/1475-7516/2014/04/022}{{\em JCAP} {\bfseries
  04} (2014) 022}, \href{http://arxiv.org/abs/1401.6457}{{\ttfamily
  arXiv:1401.6457 [hep-ph]}}.

\bibitem{Bai:2014osa}
Y.~Bai and J.~Berger, ``{Lepton Portal Dark Matter},''
  \href{http://dx.doi.org/10.1007/JHEP08(2014)153}{{\em JHEP} {\bfseries 08}
  (2014) 153}, \href{http://arxiv.org/abs/1402.6696}{{\ttfamily arXiv:1402.6696
  [hep-ph]}}.

\bibitem{Arcadi:2021mag}
G.~Arcadi, A.~Djouadi, and M.~Kado, ``{The Higgs-portal for dark matter:
  effective field theories versus concrete realizations},''
  \href{http://dx.doi.org/10.1140/epjc/s10052-021-09411-2}{{\em Eur. Phys. J.
  C} {\bfseries 81} no.~7, (2021) 653},
  \href{http://arxiv.org/abs/2101.02507}{{\ttfamily arXiv:2101.02507
  [hep-ph]}}.

\bibitem{10.5555/1593511}
G.~Van~Rossum and F.~L. Drake, {\em Python 3 Reference Manual}.
\newblock CreateSpace, Scotts Valley, CA, 2009.

\bibitem{van1995python}
G.~Van~Rossum, {\em The Python Library Reference, release 3.8.2}.
\newblock Python Software Foundation, 2020.

\bibitem{harris2020array}
C.~R. Harris, K.~J. Millman, S.~J. van~der Walt, R.~Gommers, P.~Virtanen,
  D.~Cournapeau, E.~Wieser, J.~Taylor, S.~Berg, N.~J. Smith, R.~Kern, M.~Picus,
  S.~Hoyer, M.~H. van Kerkwijk, M.~Brett, A.~Haldane, J.~F. del R{\'{i}}o,
  M.~Wiebe, P.~Peterson, P.~G{\'{e}}rard-Marchant, K.~Sheppard, T.~Reddy,
  W.~Weckesser, H.~Abbasi, C.~Gohlke, and T.~E. Oliphant, ``Array programming
  with {NumPy},'' \href{http://dx.doi.org/10.1038/s41586-020-2649-2}{{\em
  Nature} {\bfseries 585} no.~7825, (Sept., 2020) 357--362}.
  \url{https://doi.org/10.1038/s41586-020-2649-2}.

\bibitem{reback2020pandas}
T.~pandas~development team, ``pandas-dev/pandas: Pandas,'' Feb., 2020.
\newblock \url{https://doi.org/10.5281/zenodo.3509134}.

\bibitem{2020SciPy-NMeth}
P.~Virtanen, R.~Gommers, T.~E. Oliphant, M.~Haberland, T.~Reddy, D.~Cournapeau,
  E.~Burovski, P.~Peterson, W.~Weckesser, J.~Bright, S.~J. {van der Walt},
  M.~Brett, J.~Wilson, K.~J. Millman, N.~Mayorov, A.~R.~J. Nelson, E.~Jones,
  R.~Kern, E.~Larson, C.~J. Carey, {\.I}.~Polat, Y.~Feng, E.~W. Moore,
  J.~{VanderPlas}, D.~Laxalde, J.~Perktold, R.~Cimrman, I.~Henriksen, E.~A.
  Quintero, C.~R. Harris, A.~M. Archibald, A.~H. Ribeiro, F.~Pedregosa, P.~{van
  Mulbregt}, and {SciPy 1.0 Contributors}, ``{{SciPy} 1.0: Fundamental
  Algorithms for Scientific Computing in Python},''
  \href{http://dx.doi.org/10.1038/s41592-019-0686-2}{{\em Nature Methods}
  {\bfseries 17} (2020) 261--272}.

\bibitem{Hunter:2007}
J.~D. Hunter, ``Matplotlib: A 2d graphics environment,''
  \href{http://dx.doi.org/10.1109/MCSE.2007.55}{{\em Computing in Science \&
  Engineering} {\bfseries 9} no.~3, (2007) 90--95}.

\bibitem{Gondolo:1990dk}
P.~Gondolo and G.~Gelmini, ``{Cosmic abundances of stable particles: Improved
  analysis},'' \href{http://dx.doi.org/10.1016/0550-3213(91)90438-4}{{\em Nucl.
  Phys. B} {\bfseries 360} (1991) 145--179}.

\bibitem{ATLAS:2022jtk}
{\bfseries ATLAS} Collaboration, G.~Aad {\em et~al.}, ``{Constraints on the
  Higgs boson self-coupling from single- and double-Higgs production with the
  ATLAS detector using pp collisions at s=13 TeV},''
  \href{http://dx.doi.org/10.1016/j.physletb.2023.137745}{{\em Phys. Lett. B}
  {\bfseries 843} (2023) 137745},
  \href{http://arxiv.org/abs/2211.01216}{{\ttfamily arXiv:2211.01216
  [hep-ex]}}.

\bibitem{Cepeda:2019klc}
M.~Cepeda {\em et~al.}, ``{Report from Working Group 2}: {Higgs Physics at the
  HL-LHC and HE-LHC},''
  \href{http://dx.doi.org/10.23731/CYRM-2019-007.221}{{\em CERN Yellow Rep.
  Monogr.} {\bfseries 7} (2019) 221--584},
  \href{http://arxiv.org/abs/1902.00134}{{\ttfamily arXiv:1902.00134
  [hep-ph]}}.

\bibitem{Stylianou:2023tgg}
P.~Stylianou and G.~Weiglein, ``{Constraints on the trilinear and quartic Higgs
  couplings from triple Higgs production at the LHC and beyond},''
  \href{http://dx.doi.org/10.1140/epjc/s10052-024-12722-9}{{\em Eur. Phys. J.
  C} {\bfseries 84} no.~4, (2024) 366},
  \href{http://arxiv.org/abs/2312.04646}{{\ttfamily arXiv:2312.04646
  [hep-ph]}}.

\bibitem{Zeldovich:1974uw}
Y.~B. Zeldovich, I.~Y. Kobzarev, and L.~B. Okun, ``{Cosmological Consequences
  of the Spontaneous Breakdown of Discrete Symmetry},'' {\em Zh. Eksp. Teor.
  Fiz.} {\bfseries 67} (1974) 3--11.

\bibitem{Ghoshal:2022zwu}
A.~Ghoshal, N.~Okada, and A.~Paul, ``{eV Hubble scale inflation with a
  radiative plateau: Very light inflaton, reheating, and dark matter in B-L
  extensions},'' \href{http://dx.doi.org/10.1103/PhysRevD.106.095021}{{\em
  Phys. Rev. D} {\bfseries 106} no.~9, (2022) 095021},
  \href{http://arxiv.org/abs/2203.03670}{{\ttfamily arXiv:2203.03670
  [hep-ph]}}.

\bibitem{Caputo:2023ikd}
A.~Caputo, M.~Geller, and G.~Rossi, ``{New source for light dark matter
  isocurvature in low scale inflation},''
  \href{http://dx.doi.org/10.1103/PhysRevD.110.055027}{{\em Phys. Rev. D}
  {\bfseries 110} no.~5, (2024) 055027},
  \href{http://arxiv.org/abs/2306.00056}{{\ttfamily arXiv:2306.00056
  [hep-ph]}}.

\bibitem{Czerny:2014xja}
M.~Czerny, T.~Higaki, and F.~Takahashi, ``{Multi-Natural Inflation in
  Supergravity},'' \href{http://dx.doi.org/10.1007/JHEP05(2014)144}{{\em JHEP}
  {\bfseries 05} (2014) 144}, \href{http://arxiv.org/abs/1403.0410}{{\ttfamily
  arXiv:1403.0410 [hep-ph]}}.

\bibitem{Vilenkin:1982ks}
A.~Vilenkin and A.~E. Everett, ``{Cosmic Strings and Domain Walls in Models
  with Goldstone and PseudoGoldstone Bosons},''
  \href{http://dx.doi.org/10.1103/PhysRevLett.48.1867}{{\em Phys. Rev. Lett.}
  {\bfseries 48} (1982) 1867--1870}.

\bibitem{Preskill:1991kd}
J.~Preskill, S.~P. Trivedi, F.~Wilczek, and M.~B. Wise, ``{Cosmology and broken
  discrete symmetry},''
  \href{http://dx.doi.org/10.1016/0550-3213(91)90241-O}{{\em Nucl. Phys. B}
  {\bfseries 363} (1991) 207--220}.

\bibitem{Dunsky:2021tih}
D.~I. Dunsky, A.~Ghoshal, H.~Murayama, Y.~Sakakihara, and G.~White, ``{GUTs,
  hybrid topological defects, and gravitational waves},''
  \href{http://dx.doi.org/10.1103/PhysRevD.106.075030}{{\em Phys. Rev. D}
  {\bfseries 106} no.~7, (2022) 075030},
  \href{http://arxiv.org/abs/2111.08750}{{\ttfamily arXiv:2111.08750
  [hep-ph]}}.

\bibitem{Maji:2024pll}
R.~Maji, Q.~Shafi, and A.~Tiwari, ``{Topological structures, dark matter and
  gravitational waves in E$_{6}$},''
  \href{http://dx.doi.org/10.1007/JHEP08(2024)060}{{\em JHEP} {\bfseries 08}
  (2024) 060}, \href{http://arxiv.org/abs/2406.06308}{{\ttfamily
  arXiv:2406.06308 [hep-ph]}}.

\bibitem{Larsson:1996sp}
S.~E. Larsson, S.~Sarkar, and P.~L. White, ``{Evading the cosmological domain
  wall problem},'' \href{http://dx.doi.org/10.1103/PhysRevD.55.5129}{{\em Phys.
  Rev. D} {\bfseries 55} (1997) 5129--5135},
  \href{http://arxiv.org/abs/hep-ph/9608319}{{\ttfamily arXiv:hep-ph/9608319}}.

\bibitem{Saikawa:2017hiv}
K.~Saikawa, ``{A review of gravitational waves from cosmic domain walls},''
  \href{http://dx.doi.org/10.3390/universe3020040}{{\em Universe} {\bfseries 3}
  no.~2, (2017) 40}, \href{http://arxiv.org/abs/1703.02576}{{\ttfamily
  arXiv:1703.02576 [hep-ph]}}.

\bibitem{Hiramatsu:2010yz}
T.~Hiramatsu, M.~Kawasaki, and K.~Saikawa, ``{Gravitational Waves from
  Collapsing Domain Walls},''
  \href{http://dx.doi.org/10.1088/1475-7516/2010/05/032}{{\em JCAP} {\bfseries
  05} (2010) 032}, \href{http://arxiv.org/abs/1002.1555}{{\ttfamily
  arXiv:1002.1555 [astro-ph.CO]}}.

\bibitem{Gelmini:1988sf}
G.~B. Gelmini, M.~Gleiser, and E.~W. Kolb, ``{Cosmology of Biased Discrete
  Symmetry Breaking},'' \href{http://dx.doi.org/10.1103/PhysRevD.39.1558}{{\em
  Phys. Rev. D} {\bfseries 39} (1989) 1558}.

\bibitem{Wu:2022stu}
Y.~Wu, K.-P. Xie, and Y.-L. Zhou, ``{Collapsing domain walls beyond Z2},''
  \href{http://dx.doi.org/10.1103/PhysRevD.105.095013}{{\em Phys. Rev. D}
  {\bfseries 105} no.~9, (2022) 095013},
  \href{http://arxiv.org/abs/2204.04374}{{\ttfamily arXiv:2204.04374
  [hep-ph]}}.

\bibitem{Wu:2022tpe}
Y.~Wu, K.-P. Xie, and Y.-L. Zhou, ``{Classification of Abelian domain walls},''
  \href{http://dx.doi.org/10.1103/PhysRevD.106.075019}{{\em Phys. Rev. D}
  {\bfseries 106} no.~7, (2022) 075019},
  \href{http://arxiv.org/abs/2205.11529}{{\ttfamily arXiv:2205.11529
  [hep-ph]}}.

\end{thebibliography}\endgroup

\end{document}